\documentclass[a4paper]{article}
\usepackage{a4, ams, amsmath, amssymb, graphics, subfigure, fullpage, color}%
\usepackage[pdftex]{graphicx}
\usepackage{fancyhdr}
\usepackage{multirow}
\usepackage{amsthm}
\usepackage[colorlinks]{hyperref}
\hypersetup{colorlinks,citecolor=green,filecolor=black,linkcolor=blue,urlcolor=blue}

\usepackage{upgreek} 

\newcommand{\MyBib}{./MyBib}

\usepackage[round]{natbib}

\usepackage{lscape}
\numberwithin{equation}{section}
\numberwithin{theorem}{section}
\numberwithin{corollary}{section}
\numberwithin{definition}{section}
\numberwithin{lemma}{section}
\numberwithin{remark}{section}
\numberwithin{example}{section}

\DeclareMathOperator{\E}{\mathbb{E}}
\renewcommand{\Q}{\mathbb{Q}}
\DeclareMathOperator{\cov}{\textrm{cov}}

\DeclareMathOperator{\e}{\textrm{e}}
\DeclareMathOperator{\ind}{1{\hskip -2.5 pt}\textrm{I}}
\DeclareMathOperator{\sign}{\textrm{sign}}

\newcommand{\beq}{\begin{equation}}
\newcommand{\eeq}{\end{equation}}
\newcommand{\beqn}{\begin{eqnarray}}
\newcommand{\eeqn}{\end{eqnarray}}
\newcommand{\bfig}{\begin{figure}}
\newcommand{\efig}{\end{figure}}
\newcommand{\btab}{\begin{table}}
\newcommand{\etab}{\end{table}}

\newcommand{\brho}{\bar{\rho}}
\newcommand{\talpha}{\tilde{\alpha}}

\newcommand{\tbeta}{\tilde{\beta}}

\newcommand{\tgamma}{\tilde{\gamma}}

\title{Wrong-Way Risk Models:\\ A Comparison of Analytical Exposures}
\author{Fr\'ed\'eric VRINS\thanks{Contact information: Chauss\'ee de Binche 151, Office A.212, B-7000 Mons, Belgium. E-mail: \href{mailto:frederic.vrins@uclouvain.be}{frederic.vrins@uclouvain.be}. We thank the CVA desk of ING Bank for providing us with data for calibrating exposure profiles.}\vspace{0.2cm}\\ Louvain School of Management \& CORE\\ Universit\'e catholique de Louvain}
\date\today

\begin{document}
\maketitle
\begin{abstract}
In this paper, we compare static and dynamic (reduced form) approaches for modeling wrong-way risk in the context of CVA. Although all these approaches potentially suffer from arbitrage problems, they are popular (respectively) in industry and academia, mainly due to analytical tractability reasons. We complete the stochastic intensity models with another dynamic approach, consisting in the straight modeling of the survival (Az\'ema supermartingale) process using the $\Phi$-martingale. Just like the other approaches, this method allows for automatic calibration to a given default probability curve. We derive analytically the positive exposures $V^+_t$ ``conditional upon default'' associated to prototypical market price processes of FRA and IRS in all cases. We further discuss the link between the ``default'' condition and change-of-measure techniques. The expectation of $V^+_t$ conditional upon $\tau=t$ is equal to the unconditional expectation of $V^+_t\zeta_t$. The process $\zeta$ is explicitly derived in the dynamic approaches: it is proven to be positive and to have unit expectation. Unfortunately however, it fails to be a martingale, so that Girsanov machinery cannot be used. Nevertheless, the expectation of $V^+_t\zeta_t$ can be computed explicitly, leading to analytical expected positive exposure profiles in the considered examples.
\end{abstract}

\section{Introduction}
\label{sec:Intro}

An important factor driving the price of derivatives is counterparty risk. The \textit{counterparty risk-free} price given by computing the risk-neutral expected value of discounted future cashflows needs to be adjusted by the so-called \textit{Credit Value Adjustment}, or CVA. The later aims at capturing the value of the losses resulting from the default of the transaction's counterparty. Its calculation involves conditional expectations. The associated \textit{condition} refers to the fact that we are interested in portfolio value \textit{given default}: it captures the hybrid credit/market dependency. This market-credit relationship is commonly referred to as wrong-way risk (WWR). Depending on the considered portfolio, WWR can be very significant; the self-speaking example being when one is long a call (right-way) or a put (wrong-way) on counterparty's own stock.

Although counterparty risk in general and CVA in particular is receiving much more attention since the financial crisis and the bankruptcy of Lehman Brothers, the modeling of the WWR effect remains a difficult problem. In the financial industry, most of the existing models rely on static or dynamic copulas (\cite{Grego10}, \cite{Sokol11b}, \cite{Cesp10}) while the problem is generally tackled quite differently by academics, who typically prefer a dynamic setup based on stochastic intensity approaches (\cite{Hull12}, \cite{Brigo13}). Specific methods have also been developed to handle CVA on Credit Default Swaps or bilateral counterparty risk management where the problem of simultaneous defaults become relevant (see \cite{Brigo13} and references therein, and in particular the work of Cr\'epey, Jeanblanc, Bielecki and co-authors in that respect, for example~\cite{CrepXX}). Surprisingly however, no model comparison have been proposed so far with respect to the WWR expected positive exposure (EPE) profiles and CVA levels. Moreover, existing continuous models focus on either of the above approaches, whilst other alternatives can be thought of. Another aspect that is worth investigating is whether the WWR condition in the expectation can be dealt with using change-of-measure techniques. Indeed, a conditional expectation is nothing but an expectation with respect to the conditional density, and one may wonder whether such techniques combined with Girsanov theorem could help to decrease the dimensionality of the problem.

The goal of this paper is to fill this gap by looking at the impact of hybrid correlation on WWR EPE profiles, as well as on CVA levels on prototypical exposure paths for both static and stochastic intensity models. We also derive a new dynamic modeling setup based on the $\Phi$-martingale, which is a tractable martingale evolving in $[0,1]$. We show that it behaves similarly than the copula approach in that the correlation impact is much more pronounced than that produced by stochastic intensity models. 

The paper is organized as follows. We first recall the pricing formula for CVA and stress the WWR effect. We then review the mechanics of the static (copula-based resampling) models in Section~\ref{sec:WWRCop}. In Section~\ref{sec:WWRDyn}, we introduce two dynamic approcahes for WWR. We first recall the intensity paradigm for credit risk modeling before introducing a new setup, that we call the \textit{martingale} approach. We then introduce in Section~\ref{sec:ExpMod} two schemes aiming at modeling in a prototypical way the exposures of a Foward Rate Agreement (FRA) and of an Interest Rate Swap (IRS). For each of them, the WWR exposures are derived in both static (Section \ref{sec:WWREPEGC}) and dynamic (Section \ref{sec:WWREPEDyn}) models. Finally, a comparison among resampling, stochastic intensity and martingale approaches is performed in Section~\ref{sec:Comp}.

\section{CVA and Wrong-Way Risk Models}
\label{sec:WWRCop}

Our starting point is the expression of the Credit Valuation Adjustment (CVA), expressed as the present value of the non-recovered losses of a derivative portfolio resulting from counterparty's default\footnote{We have implicitly assumed that counterparty's recovery rate is zero in order to keep notations simple, but this assumption can easily been relaxed by rescaling the above expression. Similarly, we have disregarded the possibility for the financial institution to default as well that is, focus on unilateral CVA.}~\cite{Grego10,Brigo13}:
\beq
CVA:=\E[V^+_\tau\ind_{\{\tau\leq T\}}]\;.\label{eq:CVAdef}
\eeq

In this expression, $\E$ stands for the expectation under the risk-neutral measure $\Q$, $V_t$ denotes the num\'eraire-adjusted (i.e. discounted) value of the derivative portfolio at time $t$, $T$ is the portfolio maturity, $\tau$ the random variable standing for the counterparty's default time and $\ind_{\{\omega\}}$ is the indicator function, which is 1 if $\omega$ is true and zero otherwise. We refer to the discounted price process $V_t$ as the \textit{unconditional exposure} process.

This expression can be rewritten by first conditioning on the default time of the counterparty and then integrate out the latter using counterparty's survival probability curve $G(t):=\Q\{\tau>t\}$, deterministically given by the prevailing available market information\footnote{We assumed that $(V_t,\tau)$ admits a joint density $f_{V_t,\tau}$ and the density $f_\tau$ of $\tau$ is strictly positive on $\mathbb{R}^+$ so that $\E[V^+_t|\tau=t]=\frac{1}{f_\tau(t)}\int_{0}^\infty xf_{V_t,\tau}(x,t)dx$ is well-defined}:
\beq
CVA=-\int_{0}^T\E[V^+_t|\tau=t]dG(t)=\int_{0}^T\E[V^+_t|\tau=t]h(t)G(t)dt\label{eq:CVAdefint}
\eeq
where, in the last expression, we have made the usual assumption that $\tau$ is triggered by the first jump of a time-inhomogeneous Poisson process with deterministic hazard rate function $h(t)>0$:
\beq
G(t):=\e^{-\int_0^t h(s) ds}=:1-\bar{G}(t)\;.\label{eq:Gt}
\eeq
In the particular case where the discounted exposure is independent of counterparty's credit worthiness, we get
\beq
CVA^\perp=\int_{0}^T\E[V^+_t]h(t)G(t)dt\;.
\eeq

As explained above, this credit-market independence assumption is often not realistic and one may distinguish the \textit{unconditional (discounted) expected positive exposure $\E[V^+_t]$} from the \textit{conditional (discounted) expected positive exposure $\E[V^+_t|\tau=t]$}. We naturally refer $\E[V^+_t|\tau=t]$ to as the \textit{wrong-way expected positive exposure} or WWR EPE, implicitly noting that we are considering \textit{discounted} exposures.

Modeling of credit contingent products can be handled in two main ways that we know review. 


\subsection{Resampling using static (Gaussian) copula}
\label{sec:WWRCop}

%

The static setup is very popular among practitioners (\cite{Sokol11b},\cite{Cesp10},\cite{Sokol11}). It relies on copulas and allows to uncouple credit and market risks adopting a computationally efficient two-steps procedure. 

In this approach, one first computes the risk-neutral distributions $F_{V_t}(x)$ of the discounted exposure at some future points in time $\{t_1,\ldots,t_n\}$ up to portfolio maturity. These distributions need to be computed at portfolio level. Therefore, in practice, one needs to draw sample paths for discounted portfolio prices (called \textit{exposure profiles}) disregarding counterparty risk, $V_t(\omega)$, and then compute the empirical distributions $\hat{F}_{V_t}(x)$. 

In a second phase, for each $t_i$, we compute $\E[V_{t_i}^+|\tau=t_i]$ by averaging samples drawn from the distribution of $V_{t_i}$. From the probability integral transform, if $U$ is a uniform random variable in $[0,1]$ independent from the risk factors driving the exposure and $\hat{F}^{-1}_{V_t}(x)$ stands for the generalized inverse of $\hat{F}_{V_t}(x)$, then we have (up to the estimation error resulting from $\hat{F}^{-1}_{V_t}\approx F_{V_t}^{-1}$):
\beq
\hat{F}^{-1}_{V_t}(U)\sim V_t\;.
\eeq

The WWR effect results from the fact that the sampling variable $U$ is correlated with the default time $\tau$: $U=U(\tau)$. Samples from $V_{t_i}$ given $\tau=t_i$ are obtained by sampling $F_{V_{t_i}}$ with $U(t_i)$. 

The idea behind the resampling technique is to couple $U$ and $\tau$ using two independent variables $(\tilde{Z},Z)$:  $U=f(\tilde{Z},Z;\rho)$, where $\tilde{Z}=g(\tau)$ and $\rho$ controls the impact of $\tilde{Z}$ on $U$. Therefore, a sample  $V_t(t,\omega)$ of today's portfolio value conditional upon $\tau=t$ is obtained by replacing $\tau$ by $t$ in $\tilde{Z}$:
\beq
V_t(t,\omega)=\hat{F}^{-1}_{V_t}(U(t,\omega))=\hat{F}^{-1}_{V_t}\circ f(g(t),Z(\omega);\rho)\;,
\eeq
where the time argument $t$ in the brackets of $V_t(t,\omega)$ emphasizes the default condition $\tau=t$, in contrasts with $V_t(\omega)$ that represents the unconditional exposure. We thus have samples for the \textit{conditional exposure}, from which one can compute the empirical expectations $\hat{\E}[V_t^+(t)]$. 

Finally, by the law of large numbers, we get $\E[V_t^+|\tau=t]\approx\hat{\E}[V_t^+(t)]$. Therefore, CVA is given by computing the weighted sum of the conditional expected values $\hat{\E}[V_{t_i}^+(t_i)]$ by the corresponding default probabilities. The CVA expression (including WWR) is given by replacing $\E[V^+_t|\tau=t]$ by $\E[V^+_t(t)]$ in eq.~(\ref{eq:CVAdefint}):
\beq
CVA=-\int_{0}^T\E[V_t^+(t)]dG(t)=\int_{0}^T\E[V_t^+(t)]h(t)G(t)dt\;.
\eeq

The resulting coupling scheme (that is, the copula) associated to this model depends on the choice of the set $\{f,g,Z\}$. Let $\Phi(x)$ be the standard Normal cumulative distribution function. Then, the procedure consisting in choosing $Z$ as a standard Normal variable, $f(x,y;\rho)=\Phi(\rho x +\sqrt{1-\rho^2} y) $ and $g(x)=\Phi^{-1}(G(x))$ (so that $\tilde{Z}=g(\tau)=\Phi^{-1}(G(\tau))$ is a standard Normal variable independent of $Z$) corresponds to the Gaussian copula. Other copulas can be considered as well. Nevertheless, we will use the Gaussian copula in this example as it is a market standard and allows for analytical results in the considered cases. In such a framework, the coupling between $\tau$ and $V_t(t)=V_t|\tau=t$ is introduced via a standard Normal latent variable $Z$ independent from the default time variable:

%
\beqn
V_t(\tau,\omega)&=&F_{V_t}^{-1}\circ\Phi\left(\rho \Phi^{-1}(G(\tau))+\sqrt{1-\rho^2}Z(\omega) \right)\;.
\eeqn

Because $G(\tau)$ is uniform in $[0,1]$, $\Phi^{-1}(G(\tau))$ is standard Normal independent of $Z$, $V_t(\tau)\sim V_t$:
\beqn
\Phi^{-1}\circ F_{V_t}(V_t(\tau))&\sim&\mathcal{N}(0,1)\;.
\eeqn

By contrast, the \textit{conditional} exposure (where $\tau$ is set to $t$) has both different mean and variance:
\beqn
V_t(t,\omega)&=&F_{V_t}^{-1}\circ\Phi\left(\rho \Phi^{-1}(G(t))+\sqrt{1-\rho^2}Z(\omega) \right)\label{eq:VtGC}\\
\Phi^{-1}\circ F_{V_t}(V_t(t))&\sim&\mathcal{N}\left(\rho \Phi^{-1}(G(t)),\sqrt{1-\rho^2}\right)\;.
\eeqn

Note that it is straightforward to see that in the limit where $\rho\to 1$ (resp. -1), $V_t(t)$ is given by its $G(t)$-quantile (resp. $1-G(t)$-quantile):
\beqn
V_t(t,\omega)&\stackrel{(\rho=1)}{=}&F_{V_t}^{-1}\circ G(t)=:q_+(t)\\
&\stackrel{(\rho=-1)}{=}&F_{V_t}^{-1}\circ (1-G(t))=:q_-(t)\;.
\eeqn
%

\subsection{Dynamic Wrong-Way risk modeling}
\label{sec:WWRDyn}

The alternative (dynamic) approach is to model the default and the exposure processes in a joint, dynamic way. The default process can be either firm-valued (structural models) or intensity-based (reduced form) models. Defaults are rare events that have an extreme impact. This means that in practice, a lot of simulations are required to reach convergence. And these simulations are costly since both default times and portfolio price paths need to be drawn. In order to circumvent this issue, one can work with \textit{survival probability processes} instead of actual default times. This setup requires a proper modeling of the information flow (filtration). The \textit{full} information is noted $\mathbb{G}=(\mathcal{G}_t)_{t\geq 0}$, and can be written as the \textit{default-free market information} (excluding defaults) $\mathbb{F}=(\mathcal{F}_t)_{t\geq 0}$ enlarged with the $\sigma$-field of the default indicator
\beq
\mathcal{G}_t=\mathcal{F}_t\vee\sigma\left(\ind_{\{\tau>s\}},0\leq s\leq t\right)\;.
\eeq
This setup allows us to work in the filtration $\mathbb{F}$, getting rid of the indicators featuring the $\mathbb{G}$-stopping times $\tau$. This approach features the $\mathbb{F}$-predictable survival process known as the \textit{Az\'ema supermartingale}:
\beq
S_t:=\E[\ind_{\{\tau>t\}}|\mathcal{F}_t]=\Q(\tau>t|\mathcal{F}_t)=:1-F_t\;.\label{eq:defazema}
\eeq

In dynamic approaches, the objective is to get rid of the $\mathbb{G}$-stopping times $\tau$ by working in a smaller filtration $\mathbb{F}$. Actual default simulations can be avoided, which is good news. Those methods rely on the Az\'ema supermartingale. Assuming $\tau>0$, straight application of a key lemma about filtration changes (see e.g. Lemma 3.1.3. of \cite{Biel11}) allows us to write the right-hand side of eq.~(\ref{eq:CVAdef}) as
\beq
\E\left[V^+_\tau\ind_{\{\tau\leq T\}}\right]=-\E\left[\int_{0}^TV^+_tdS_t\right]\;.\label{eq:EdS1}
\eeq
The interesting point here is that this expression features no default indicators or default times. They have been replaced by the $\mathbb{F}$-predictable survival process. The choice of the dynamics of $S$ specifies the WWR model. It can be shown (see e.g.~\cite{Biel11}) that $S$ admits the Doob-Meyer decomposition
\beq
dS_t=dA_t+dM_t\;,
\eeq
where $A$ is a $\mathbb{F}$-predictable decreasing process satisfying $A_0=1$, and $M$ a non-negative $\mathbb{F}$-martingale. In particular, there is no restriction on the decreasingness of $S$ for eq.~(\ref{eq:EdS1}) to hold. The only restriction is that $S\in[0,1]$ almost surely since it is a probability (see eq.~(\ref{eq:defazema})). In the continuous case, the dynamics of $S$ take the general form:
\beq
dS_t=\mu(.)dt+\sigma(.)dW_t\label{eq:dSMu}\;.
\eeq
Since $h(t)G(t)>0$ for all $t$, we can define the \textit{wrong-way process} $\zeta=(\zeta_t)_{t\geq 0}$:
\beq
\zeta_t:=-\frac{\mu(.)}{h(t)G(t)}\;.\label{eq:Zetat}
\eeq
Recalling that It\^o integrals have zero expectation and using Fubini's theorem, eq.~(\ref{eq:EdS1}) becomes
\beq
CVA
=-\int_{0}^T\E\left[V^+_t\zeta_t\right]dG(t)=\int_{0}^T\E\left[V^+_t\zeta_t\right]h(t)G(t)dt\;.\label{eq:EdS}
\eeq
In the sequel, we shall assume that the risk-neutral expected value $\E[S_t]$ is constrained to be given by the continuous positive non-increasing curve $G(t)$ satisfying $G(0)=1$ given in eq.~(\ref{eq:Gt}). In financial applications, the calibration equation 
\beq
\E[S_t]=G(t)\label{eq:cal}
\eeq
ensures that the stochastic default model is in line with market quotes. In particular, it corresponds to the calibration to CDS quotes under independence between rates and credit. This assumption is widely accepted because it is known to have little impact on the CDS spreads implied by the model, see e.g.\cite{Brigo05}. This equation implies that the wrong-way process $\zeta$ satisfies, for all $t\geq 0$
\beq
 \E[\zeta_t]=1\;.\label{eq:calzeta}
\eeq
Equation~(\ref{eq:EdS}) justifies the name \textit{wrong-way process} for $\zeta$ which dynamics control the WWR effect. Should $\zeta$ be independent of $V$ for example, $CVA=CVA^\perp$: there is no WWR effect.

In a dynamic, one-factor setup, wrong-way risk impact will be introduced by correlating the exposure and credit risk drivers. In the case of exposure and survival processes driven by a single Brownian motion $B$ and $W$ we can assume
\beq
d\langle B,W\rangle_t=\rho_t dt\;.\label{eq:rhodt}
\eeq
We shall use a time-independent correlation in the sequel, $\rho_t=\rho$. We now recall the stochastic intensity method for dynamic modeling of wrong-way risk, and introduce the so-called \textit{martingale approach}.

\subsubsection{Stochastic Intensity Approach}
\label{sec:StochInt}

The most popular approach for credit risk modeling is Cox setup. In this case, default events are triggered by the first jump of a Poisson process which intensity is a non-negative stochastic process $\lambda$ that is, $\tau$ is modeled as the first jump of a Cox process.  The resulting Az\'ema supermartingale $S_t$ is the stochastic version of eq.~(\ref{eq:Gt}) where the deterministic hazard rate function $h(t)$ is replaced by the stochastic process $\lambda_t$ being correlated with market factors:
\beq
S_t=\e^{-\int_{0}^t \lambda_s ds}\;.\label{eq:SCox}
\eeq
%

It is well-known and easy to show that under the calibration equation (\ref{eq:cal}), one can sample the default distribution $\bar{G}(t)$ by looking for the passage time $\tau$ satisfying $S_\tau=U$ where $U$ is an independent barrier with uniform distribution.

In such a framework, the dynamics of the Az\'ema supermatringale takes the form
\beq
dS_t=-\lambda_tS_tdt\;,\label{eq:dSCox}
\eeq
where $\lambda_t\geq 0$ $\Q$-a.s. Cox setup thus corresponds to the special case where the Az\'ema supermartingale $S_t$ is a decreasing $\mathbb{F}$-adapted process. In particular, the martingale part in the Doob-Meyer decomposition vanishes, $M\equiv 0$. The restriction of having a decreasing survival process is motivated by the Cox setup where $\lambda_t$ is restricted to be positive since it corresponds to a default intensity. From eq.~(\ref{eq:dSMu}), $\mu(\cdot)=-\lambda_tS_t$ so that the wrong-way process becomes
\beq
\zeta_t=\frac{\lambda_tS_t}{h(t)G(t)}=\frac{\lambda_t\e^{-\int_{0}^t \lambda_s ds}}{h(t)G(t)}\;.\label{eq:zetalambda}
\eeq

This approach has been originally proposed in~\cite{Lando98} and~\cite{Duffie99} in the context of defaultable bonds valuation and pricing of other credit risky securities. Here again, several stochastic processes can be used but in practice, one exploits relationship between short rate and default intensity (see e.g.~\cite{Brigo06}). Equation~(\ref{eq:SCox}) indeed corresponds to the stochastic discount factor computed from continuously compounded rate $r$:
\beq
D_t=\e^{-\int_{0}^t r_s ds}\;.
\eeq

All affine models lead to the following expression for zero-coupon bond prices (see e.g.~\cite{Brigo06})
\beq
P_\lambda(t,T)\doteq\E\left[\e^{-\int_{t}^Tr_s ds}|\mathcal{F}_t\right]=A(t,T)\e^{-B(t,T)r_t}\;.
\eeq

Replacing $r_t$ by $\lambda_t$, the above expression with $t\leftarrow 0$ is nothing but the calibration equation~(\ref{eq:cal}) with $P_r(0,t)=G(t)$. 

Analytical and exact calibration to any valid curve $G(t)$ can thus be achieved by modeling the intensity $\lambda_t$ ``as a short rate process $r$'', shifted by a deterministic function $\phi$ to allow for perfect fit to the initial survival probability curve:
\beqn
\lambda_t&=&r_t+\phi(t)\\
S_t&=&D_t\e^{-\int_{0}^t \phi(s) ds}\;.
\eeqn
The calibration equation eq.~(\ref{eq:cal}) yields the drift function:
\beq
G(t)=A(0,t)\e^{-B(0,t)r_0}\e^{-\int_{0}^t \phi(s) ds}\;.
\eeq

Let us now focus on the standard cases where $r_t$ is governed by Ornstein-Uhlenbeck (OU) and Square-Root Diffusion (SRD) dynamics. They are popular in industry, essentially for the same reason as the Gaussian copula: their tractability. The shifted version of the OU is known as the \textit{Hull-White} process. It is commonly used in interest rates modeling, but also to describe the dynamics of funding spread~\cite{PiterbargFunding}, that is for credit spread, or hazard rate~\cite{Lando04}, \cite{Ben11}. Practitioners are aware of the fact that negative intensities can be drawn, but they commonly agree to work with models that violate properties theoretically required, provided that one can keep some control on the number of ``wrong paths'' (thereby trying to make sure the aggregated results are trustworthy, for whatever it means); see e.g.~\cite{Ces09}. An appealing alternative to the Hull-White process for the sake of default intensity modeling is the square root diffusion (also known as {CIR}, \cite{Duffie99}) or, allowing for a perfect fit to market data, the shifted square root diffusion (SSRD, also known as CIR$^{++}$), with or without jumps (\cite{Brigo05},\cite{Brigo13}). Depending on the market data however (that is, function $G(t)$), the shift function may lead to non-zero probability of having negative intensities, and thus may violate the $S_t\leq 1$ constraint. We shall thus focus on the Hull-White process, which is Gaussian and allows for analytical results. However, we shall comment some results obtained for the SSRD as well (most of the results below can be found in \cite{Brigo06}). 

We model the intensity process as
\beq
\lambda_t=r_t+\phi(t;\kappa,\theta,\sigma)\;,
\eeq
where
\beq
dr_t=\kappa(\theta-r_t)dt+\sigma(r_t) dW_t\;.
\eeq

Noting 
\beq
\xi(x,t)=\frac{1-\e^{-x\kappa t}}{x\kappa}\;,
\eeq
the analytical expressions for the functions $A,B$ and $\phi$ are given in Table~\ref{tab:rproc}.
\begin{table}[h!]
\begin{tabular}{|c|c|c|}
\hline
&Vasicek&CIR\\
\hline
$\sigma(x)$&$\sigma$&$\sigma\sqrt{x}$\\
$\ln A(t,T)$&$\left(\theta-\frac{\sigma^2}{2\kappa^2}\right)\left(B(t,T)-T+t-\frac{\sigma^2}{4\kappa}B^2(t,T)\right)$&${2\kappa\theta/\sigma^2}\ln\left(\frac{2k\e^{(\kappa+k)(T-t)/2}}{2k+(\kappa+k)\left(\e^{(T-t)k}-1\right)}\right)$\\
$B(t,T)$&$\xi(1,T-t)$&$\frac{2\left(\e^{(T-t)k}-1\right)}{2k+(\kappa+k)\left(\e^{(T-t)k}-1\right)}$\\
$\phi(t;\kappa,\theta,\sigma)$&$h(t)+B(0,t)\left(\frac{\sigma^2}{2}B(0,t)-\theta\kappa\right)-r_0\e^{-\kappa t}$&$h(t)-\frac{2\kappa\theta\e^{tk}-1}{2k+(\kappa+k)(\e^{tk}-1)}+r_0\frac{4(k^2)\e^{tk}}{(2k+(\kappa+k)(\e^{tk}-1))^2}$\\
\hline
\end{tabular}
\label{tab:rproc}
\end{table}

In spite of the possible negative intensities in either short-rate models, this approach is probably the most popular candidate for modeling continuous ``positive intensity" processes in the Cox setup. 

Of course, other models for the stochastic intensity can be adopted (see e.g.~\cite{Hull12},\cite{Willemen14}), but they do not admit analytical calibration and require a numerical scheme. Out of the positivity property, they are expected to behave similarly to the above two approaches in terms of CVA when the later is substantially positive (see comparison below with SSRD).

\subsubsection{Martingale Approach}
\label{sec:Mart}

An alternative to Cox setup is to directly model the supermartingale $S$ without passing through the latent process $\lambda$. More explicitly, we follow Cesari's setup depicted in~\cite{Ces09} initially derived in the context of CVA on credit-linked options. 

In this section, we specify the dynamics of a family of $\mathbb{F}$-martingales
\beq
S_{t,T}:=\Q(\tau>T|\mathcal{F}_t)
\eeq
and define the associated Az\'ema supermartingale (or survival process) by letting $T\downarrow t$
\beq
S_t:=S_{t,t}\;.
\eeq

In~\cite{Ces09}, the authors propose to model the martingales $(S_{t,T})_{t\geq 0}$ on $t\in[0,T]$ using a Gaussian process
\beqn
S_{t,T}&=&G(T)+\int_{0}^t\eta(s,T)dW_s
\eeqn
leading to the dynamics
\beq
dS_{t,T}=\eta(t,T)dW_t\;.
\eeq

The volatility coefficient $\eta$ is supposed to meet the (Lipschitz or Holder-$1/2$ continuity together with the linear growth bound, see e.g. \cite{Kloed99}) regularity assumptions for $S_{t,T}$ to be a martingale (and not merely a local martingale). The corresponding Az\'ema supermartingale is thus given by
\beq
S_{t}=G(t)+\int_{0}^t\eta(s,t)dW_s\;.
\eeq

Observe that in our specific context, this setup is clearly inappropriate when $\eta(t,T)=\eta(t)$. Indeed, we have seen that the WWR feature is controlled by the wrong-way process $\zeta$. In this specific case however,
\beq
dS_{t}=-h(t)G(t)dt+\eta(t) dW_t\label{eq:dSGaussian}
\eeq
that is, $\mu(t)=-h(t)G(t)$ implying $\zeta\equiv 1$. In other words, there is no WWR effect, whatever the correlation $\rho$. 

Considering more complex functions $\eta(t,T)$ does not help fixing the problem of the range of $S_t$, which is $\mathbb{R}$ and not $[0,1]$ as it should. The authors argue that for practical cases, this is acceptable provided that one keeps control on probabilities $\Q(S_t>1)$ and $\Q(S_t<0)$. In a sense, this situation is indeed not much different from the above ``Cox models'' since the Hull-White model yields negative intensities, while CIR$^{++}$ can have the same drawback depending on the market data and the process parameters.

The range problem can easily be circumvented by considering martingales $S_{t,T}$ that belong to $[0,1]$, $\Q$-a.s. Although such processes received little attention in finance, analytically tractable processes sharing this property have been recently proposed, see e.g.~\cite{Vrins14,Vrins15a}. Let us apply the procedure depicted in~\cite{Vrins14} to our CVA context, named \textit{conic martingale} hereafter. We consider a family of latent processes $X_{t,T}$ with dynamics
\beq
dX_{t,T}=a(X_{t,T})dt+\sigma dW_t\label{eq:dXtT}
\eeq
and then choose a bijection $H:\mathbb{R}\to[0,1]$. Set $S_{t,T}=H(X_{t,T})$. We then use It\^o's lemma to determine the drift function $a(x)$ ensuring that $S_{t,T}$ is a local martingale provided that the solution $X_{t,T}$ to eq.~(\ref{eq:dXtT}) exists. Since any bounded local martingale is a martingale (see e.g. Th. 5.1 in ~\cite{Prott05}), $S_{t,T}$ would be a genuine martingale, too. In the specific case where we choose the standard Normal cumulative distribution $\Phi$ as mapping function $H$, It\^o's lemma combined with the local martingale condition uniquely determines the drift function to be
\beq
a(x)=\frac{\sigma^2}{2}x\;.
\eeq

With this specific drift, the stochastic differential equation~(\ref{eq:dXtT}) admits a unique strong solution for each $T$. This yields a family of Vasicek processes with zero long-term mean and negative mean-reversion speed (\cite{Vrins15a}):
\beq
X_{t,T}=X_{0,T}\e^{\sigma^{2}/2t}+\sigma\int_{0}^t\e^{\frac{\sigma^2}{2}(t-s)}dW_s\;.
\eeq
Noting $\varphi(x)=\Phi'(x)$ the standard Normal density and using $-\Phi''(x)/\Phi'(x)=-\varphi'(x)/\varphi(x)=x$, the dynamics of 
\beq
S_{t,T}=\Phi(X_{t,T})
\eeq
are given by
\beq
dS_{t,T}=\sigma\varphi(\Phi^{-1}(S_{t,T})) dW_t
\eeq
Because $S_{t,T}$ is given by mapping a process through a cumulative distribution function, it is constrained to evolve in $[0,1]$ (the diffusion coefficient $\sigma\varphi(\Phi^{-1}(x))$ vanishes at the boundaries $\{0,1\}$ of the unit interval). It is thus a bounded local martingale, and hence a martingale. More details about the statical properties of this stochastic process can be found in~\cite{Vrins15a}.

The interesting point here is that the solution, and thus the distribution of $S_{t,T}$ is readily given from that of $X_{t,T}$, which is Normal with mean $X_{0,T}\e^{\sigma^{2}/2t}$ and variance $\e^{\sigma^2 t}-1$. In the following, we set 
\beq
X_{0,T}:=\Phi^{-1}(G(T))\label{eq:calCM}
\eeq
and define the survival process as $S:=\Phi(X)$ where $X:=(X_t)_{t\geq 0}$, $X_{t}:= X_{t,t}$. It is easy to show that $S$ corresponds to the following It\^o process:
\beq
S_{t}=1+\int_{0}^t\e^{\sigma^2/2s}\frac{\varphi(\Phi^{-1}(S_s))}{\varphi(\Phi^{-1}(G(s)))}dG(s)+\sigma\int_{0}^t\varphi(\Phi^{-1}(S_{s}))dW_s\;.
\eeq

In this model, the wrong-way process $\zeta$ thus takes the form
\beq
\zeta_t=\e^{\sigma^2/2t}\frac{\varphi(\Phi^{-1}(S_t))}{\varphi(\Phi^{-1}(G(t)))}\;.\label{eq:ZetaConic}
\eeq

The initialization~(\ref{eq:calCM}) guarantees that~(\ref{eq:cal}) holds:
\beq
\E[S_t]=\E[\Phi(X_t)]=\E\left[\Phi\left(X_{0,t}\e^{\mu t} + \sqrt{\e^{\sigma^2 t}-1} Z\right)\right]=\Phi(X_{0,t})=G(t)\;.
\eeq

Intuitively, the process behaves as follows. Around $t=0$, $S_0\simeq G(t)\simeq 1$: the volatility $S_t$ is negligible and the survival process decreases from 1 as per the deterministic curve $G(t)$: $dS_t\simeq dG(t)$. When getting away from 1, the martingale part enters the picture, but $S_t$ is frozen at the boundaries of the unit interval $[0,1]$.

We shall now proceed to the analysis of WWR EPE profiles and CVA levels among the presented approaches. To that end, we introduce prototypical derivatives exposure profiles.

\section{Exposures Comparison}
\label{sec:ExpMod}

Our purposes here is to compare the above WWR models from both exposure profiles and CVA levels perspectives. We are therefore more interested in the impact of the credit/market coupling than in the precise modeling of portfolio dynamics. Instead, we would like to work out simplified processes easy to handle and reproduce. To that end, we shall use prototypical instruments. In particular, we assume a flat hazard rate process for prevailing survival probability curve ($h(t)=h$) and model the discounted portfolio process using a rescaled Brownian motion or a Brownian bridge, thereby attempting to mimic the profile of a plain vanilla forward contract and interest rate swap (IRS), respectively. This modeling setup has the advantage of leading to Gaussian processes, allowing us to derive closed form expressions for EPE profiles.

\subsection{Prototypical Exposures}

It is clear from risk-neutral pricing theory that the exposure (i.e. discounted price) process of a forward contract with expiry date $T$ must be a $\mathbb{F}$-martingale on $[0,T)$, with initial value zero. Therefore, we can model our exposure process as, with $0\leq s\leq t< T$ :
\beqn
V_t&=&\vartheta B_{t}\label{eq:FF}\\
V_t|V_s&=&V_s+\vartheta (B_{t}-B_{s})\\
&\sim&V_s+\vartheta\sqrt{t-s}Z
\eeqn
where $B_t$ is a Brownian motion, $Z\sim\mathcal{N}(0,1)$.


A IRS is paying a stream of cashflows, and thus exhibits a pull-to-zero effect. The discounted portfolio process can thus be modeled in $t\in [0,T)$ as a Brownian bridge shifted in a deterministic way. Assuming $V_0=0$:
\beqn
V_t&=&\gamma t (T-t)+\vartheta (T-t)\int_{0}^t \frac{1}{T-s}dB_{s}\label{eq:BB}\\
V_t|V_s&=&\frac{T-t}{T-s}V_s+\gamma (t-s)(T-t)+\vartheta(T-t)\int_{s}^t \frac{1}{T-u}dB_{u}\\
&\sim&\frac{T-t}{T-s}V_s+\gamma (t-s)(T-t)+\vartheta\sqrt{\frac{(t-s)(T-t)}{T-s}}Z
\eeqn
where $Z\sim\mathcal{N}(0,1)$ and using the terminal condition $V_T=0$.

In eq.~(\ref{eq:BB}), $B_t$ is a Brownian motion, the first term controls the expected exposure (EE) profile and the second part is a rescaled Brownian bridge that will drive expected positive exposures (EPE) and expected negative exposure (ENE) profiles away from the expected exposure (average, EE) profiles. The constant $\vartheta$ controls the exposure volatility and $\gamma$ controls the profile moneyness. For a payer swap, a positive (resp. negative) value for $\gamma$ mimics an increasing (resp. decreasing) forward curve. This specific formulation of the Brownian bridge is given in Theorem 4.7.6 of \cite{Shrev04}, and is particularly important as it leads to a $\mathcal{F}_t$-measurable process.
\begin{figure}
\centering
\subfigure[$(\gamma,\vartheta)=(0\%,2.2\%)$]{\includegraphics[width=0.45\columnwidth]{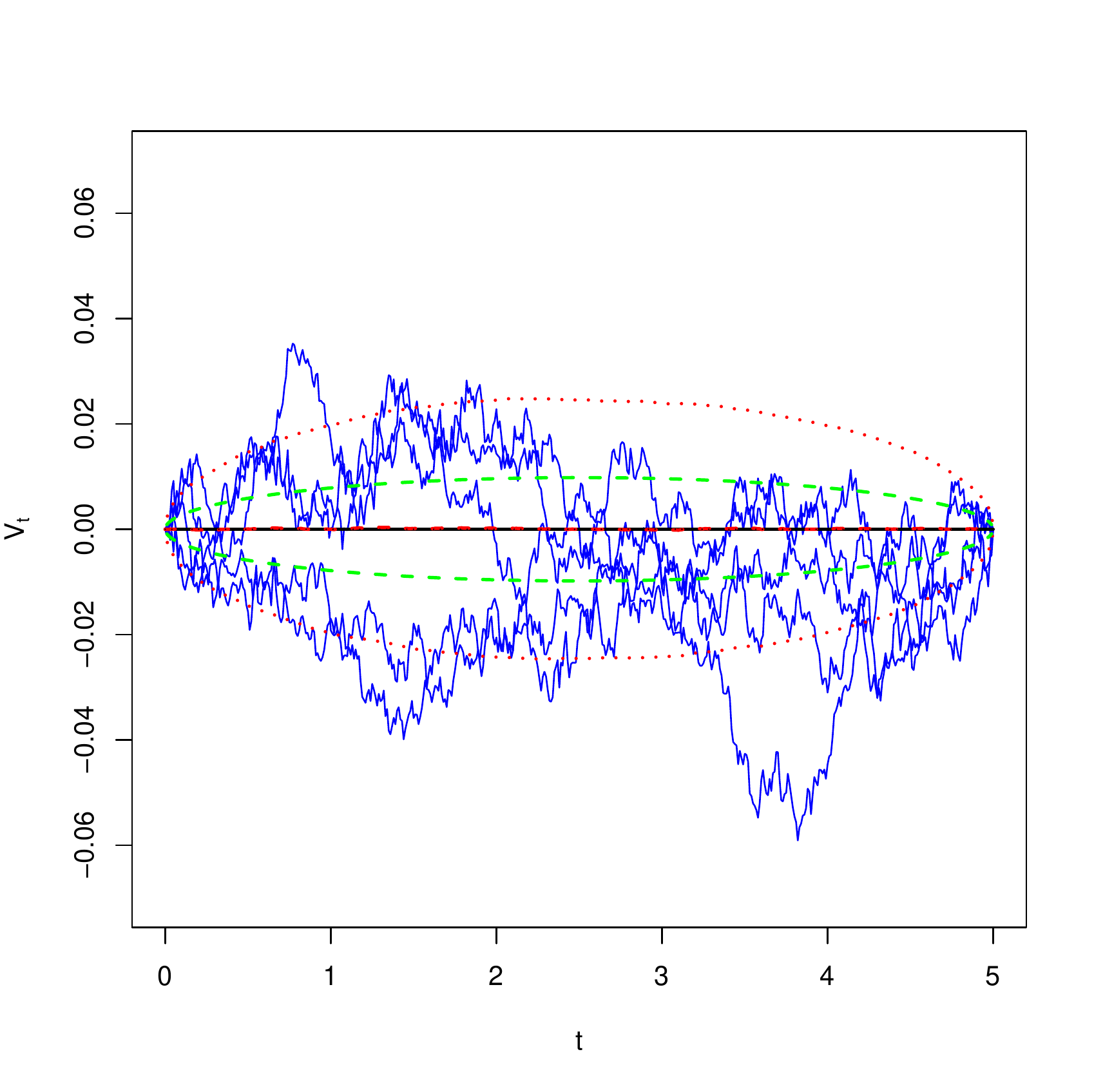}}\hspace{0.2cm}
\subfigure[$(\gamma,\vartheta)=(-0.1\%,2.2\%)$]{\includegraphics[width=0.45\columnwidth]{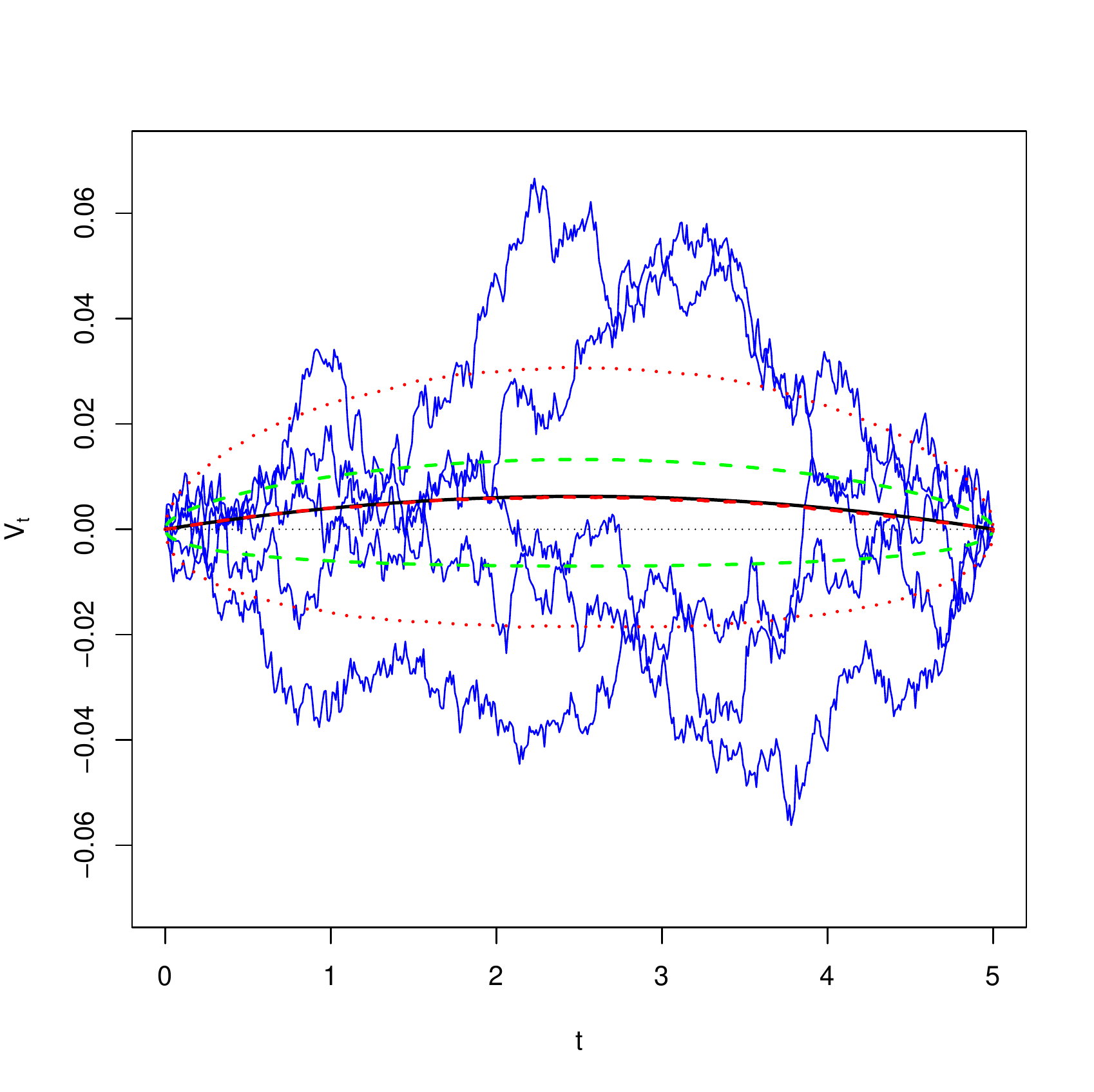}}\\
\subfigure[$(\gamma,\vartheta)=(+0.1\%,2.2\%)$]{\includegraphics[width=0.45\columnwidth]{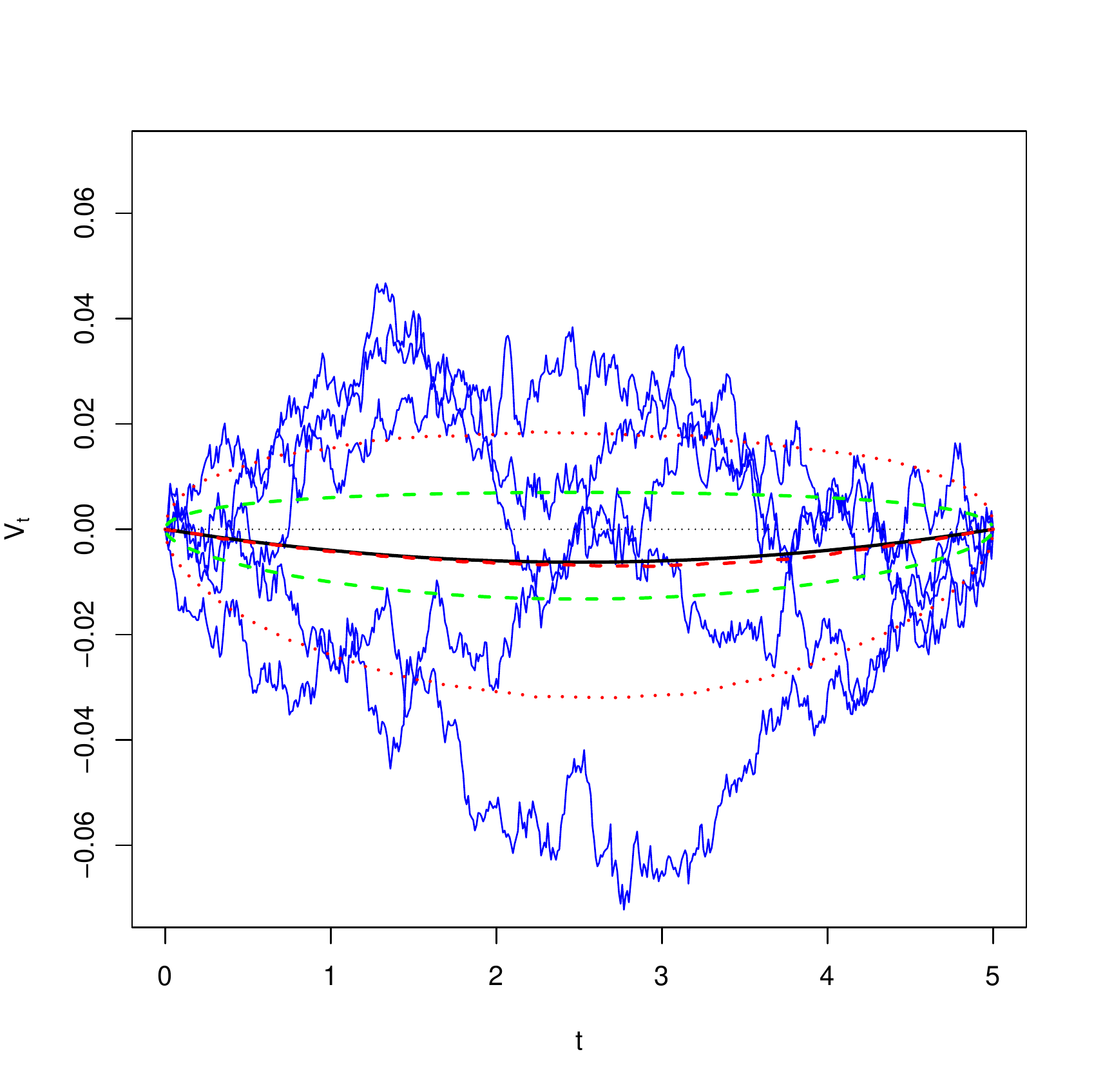}}\hspace{0.2cm}
\subfigure[$(\gamma,\vartheta)=(0.1\%,4.4\%)$]{\includegraphics[width=0.45\columnwidth]{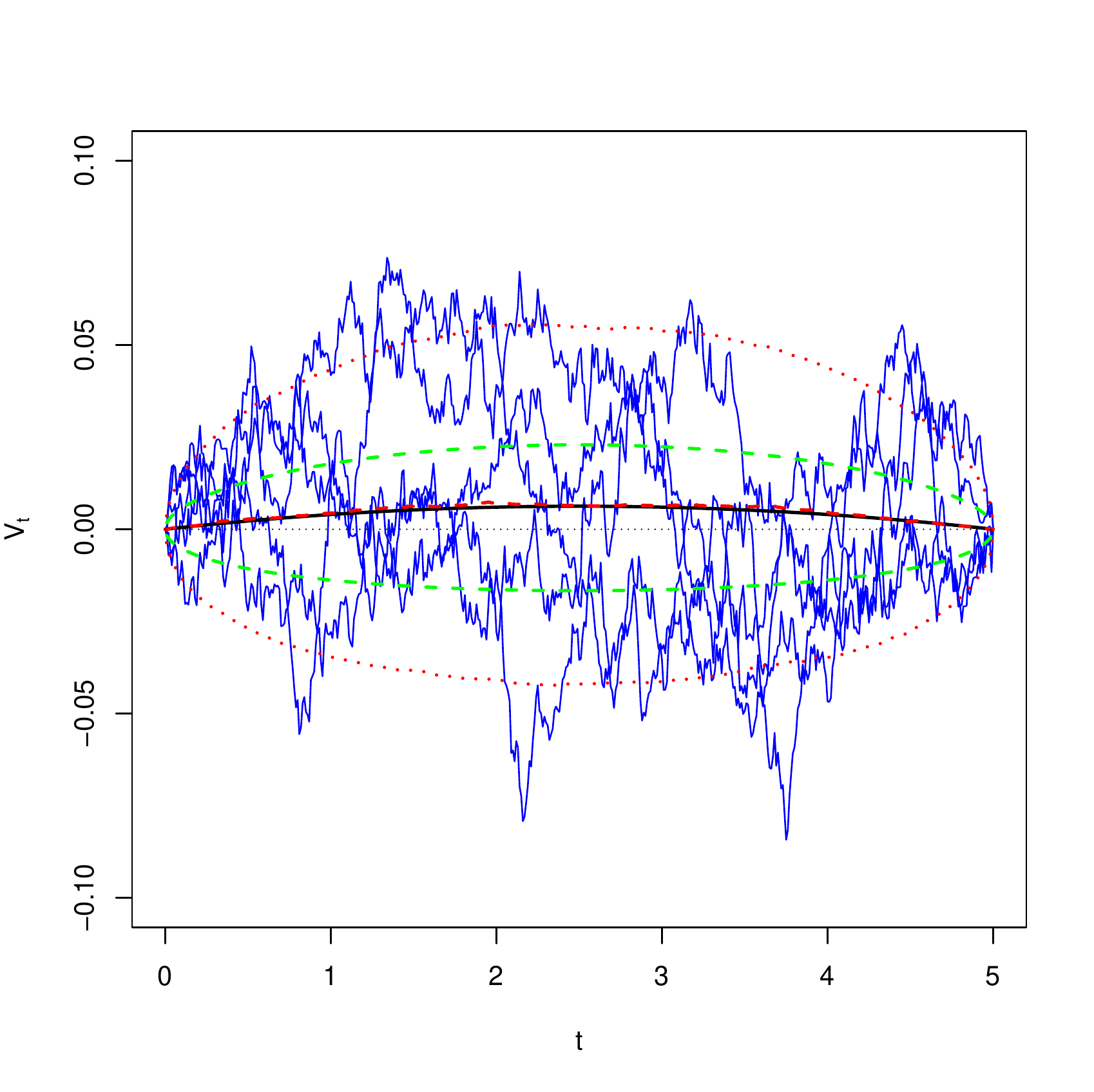}}\\
\caption{Discounted exposure profiles for IRS-type profile. Five sample paths (blue), average of 5k paths (solid red) average $\pm$ one standard deviation (dashed red), Expected exposure $\gamma T(T-t)$ (solid, black), EPE and ENE (expected positive and negative profiles, given by the positive and negative green dashed profiles)}\label{fig:BB}
\end{figure}
%

For a EURO 5Y payer swap, reasonable values that can be obtained from swaptions or cap/floor prices are $\gamma=0.5\%$ and $\vartheta=2.2\%$.
Since the profiles are Gaussian processes, they are Normally distributed so that generally speaking, 
\beq
V_t\sim\mathcal{N}(a(t),b(t))\;.\label{eq:GaussUncondExp}
\eeq
In the above cases, we respectively have
\begin{table}[h]
\centering
\begin{tabular}{|c|c|c|}
\hline
&$a(t)$&$b(t)$\\
\hline
Forward&$0$&$\vartheta\sqrt{t}$\\
IRS&$\gamma t(T-t)$&$\vartheta\sqrt{t(1-t/T)}$\\
\hline
\end{tabular}
\label{tab:abNoWWR}
\end{table}

The \textit{unconditional} expected exposure, positive expected exposure and negative expected exposure are given by
\beqn
\E[V_t]&=&a(t)\\
\E[V^+_t]&=&b(t)\varphi\left(\frac{a(t)}{b(t)}\right)+a(t)\Phi\left(\frac{a(t)}{b(t)}\right)\;.
\eeqn

We now proceed with the computation of the \textit{conditional} EPE profiles and CVA levels in static and dynamic approaches.

\subsection{Static WWR CVA (with Gaussian Copula)}
\label{sec:WWREPEGC}

In this section, we compute the WWR EPE profiles and the associated CVA for both prototypical instruments under the One-Factor Gaussian copula setup. One advantage of this prototypical modeling is that it allows for closed form expression of EPE under this setup, thereby avoiding the resampling step: analytical expressions are available and Monte Carlo simulations are not required. We now proceed to the derivation of the EPEs for both products.

Because the \textit{unconditional} exposures are Gaussian processes, eq.~(\ref{eq:GaussUncondExp}) yields
\beq
F_{V_t}^{-1}(u)=a(t)+b(t)\Phi^{-1}(u)\;.
\eeq

In the static setup, the WWR exposure are modelled according to
\beq
\E[V_t^+|\tau=t]=\E[V^+_t(t)]\;,
\eeq
where $V_t(t)$ is a resampled, conditional exposure. The Gaussian copula setup transforms \textit{unconditional} exposures $V_t$ with mean $a(t)$ and variance $b^2(t)$ into WWR (\textit{conditional}) exposures $V_t(t)$ according to
\beqn
V_t(t,\omega)&=&F_{V_t}^{-1}(\Phi(\rho\Phi^{-1}(G(t))+\sqrt{1-\rho^2}Z(\omega)))\\
&=&a(t)+b(t)\rho\Phi^{-1}(G(t))+b(t)\sqrt{1-\rho^2}Z(\omega)\;.
\eeqn

Therefore, WWR exposures are Normally distributed, too:
\beq
V_t(t)\sim\mathcal{N}(\tilde{a}(t),\tilde{b}(t))\;,
\eeq
where
\beqn
\tilde{a}(t)&:=&a(t)+\rho\Phi^{-1}(G(t))b(t)\\
\tilde{b}(t)&:=&b(t)\sqrt{1-\rho^2}
\eeqn
so that the WWR expected discounted positive exposure is given by
\beq
\E[V^+_t|\tau=t]=\E\left[V^+_t(t)\right]=\tilde{b}(t)\varphi\left(\frac{\tilde{a}(t)}{\tilde{b}(t)}\right)+\tilde{a}(t)\Phi\left(\frac{\tilde{a}(t)}{\tilde{b}(t)}\right)\;.
\eeq

In both cases, the Gaussian behavior of the price process allows to get the EPE profile in closed form, so that the CVA is given by the semi-analytical formula
\beqn
CVA&=&-\int_{0}^T\varphi\left(\frac{\tilde{a}(t)}{\tilde{b}(t)}\right)\tilde{b}(t)dG(t)-\int_{0}^T\Phi\left(\frac{\tilde{a}(t)}{\tilde{b}(t)}\right)\tilde{a}(t)dG(t)\;.\label{eq:CVASAGC}
\eeqn
\begin{figure}
\centering
\subfigure[Forward-type profile (h=1\%)]{\includegraphics[width=0.45\columnwidth]{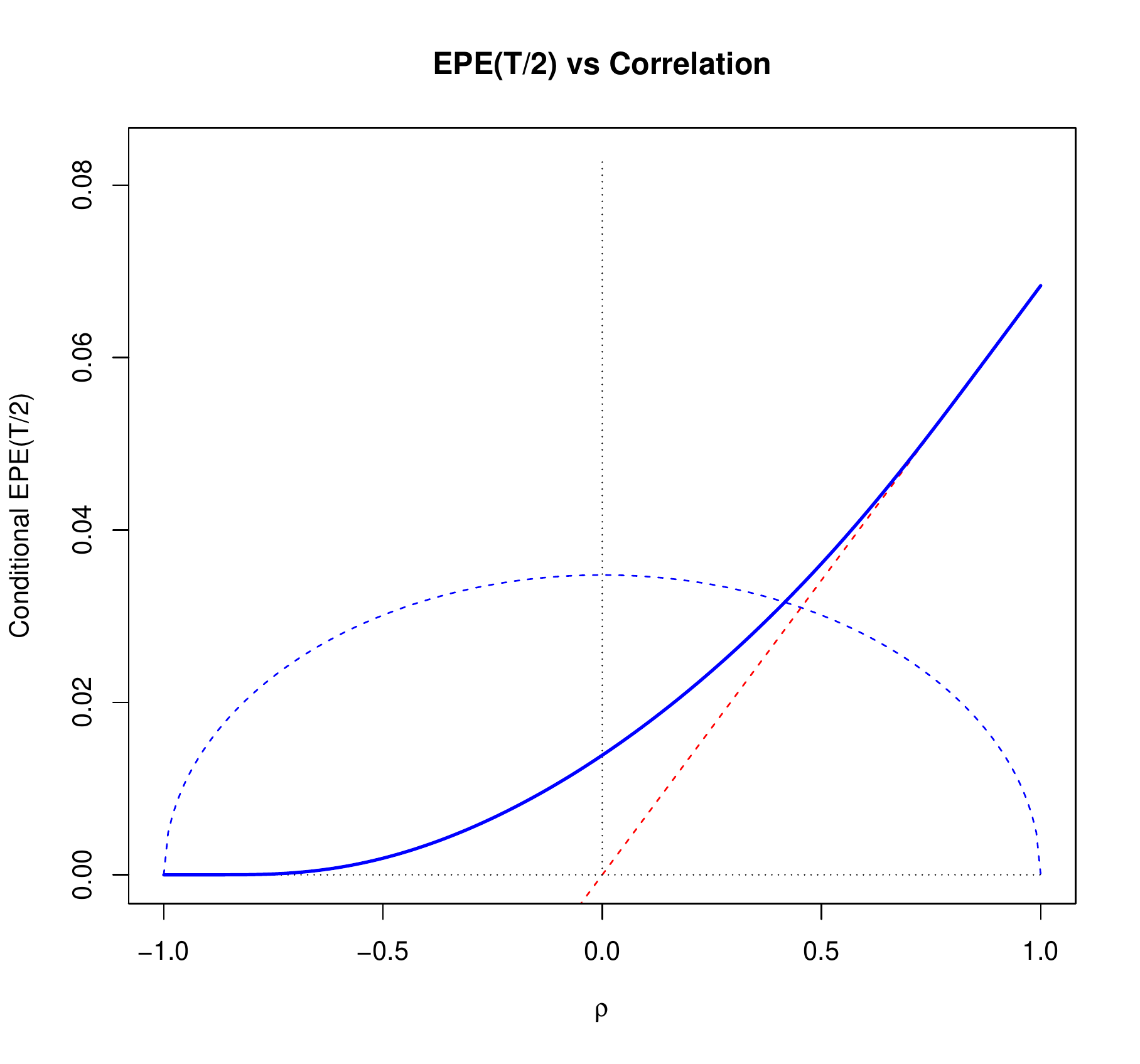}}\hspace{0.2cm}
\subfigure[IRS-type profile (h=1\%)]{\includegraphics[width=0.45\columnwidth]{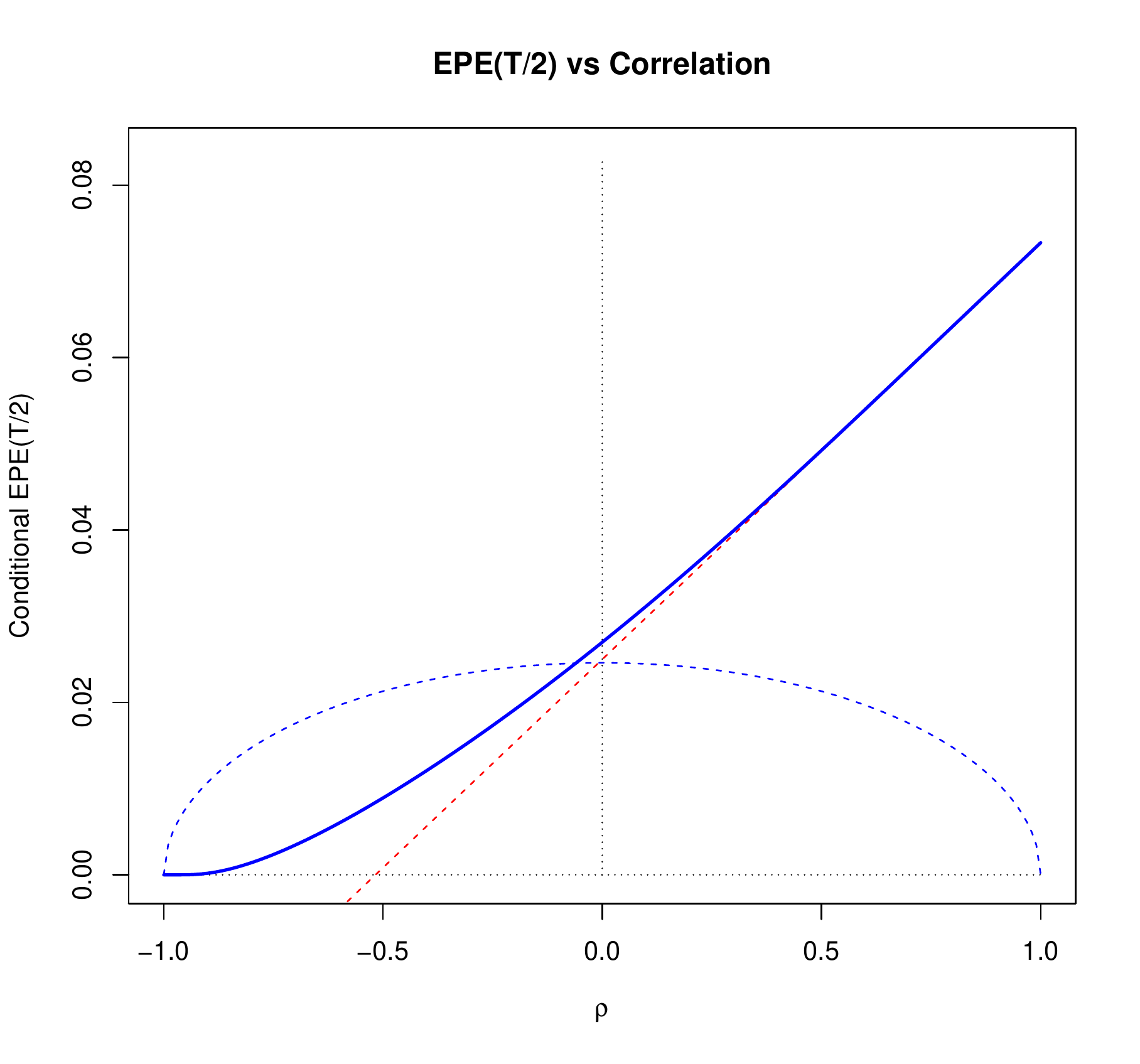}}\\
\subfigure[Forward-type profile (h=10\%)]{\includegraphics[width=0.45\columnwidth]{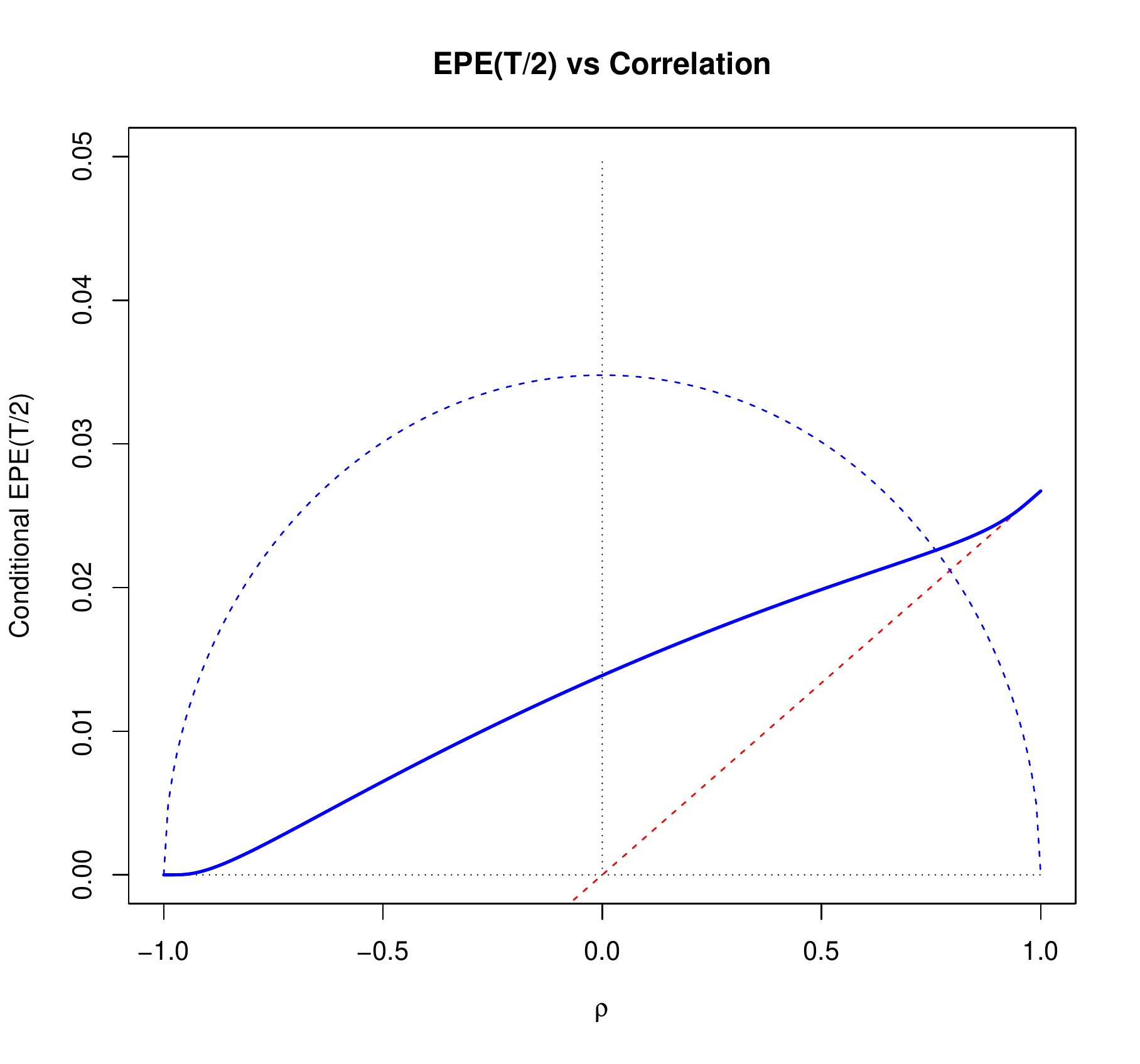}}\hspace{0.2cm}
\subfigure[IRS-type profile (h=10\%)]{\includegraphics[width=0.45\columnwidth]{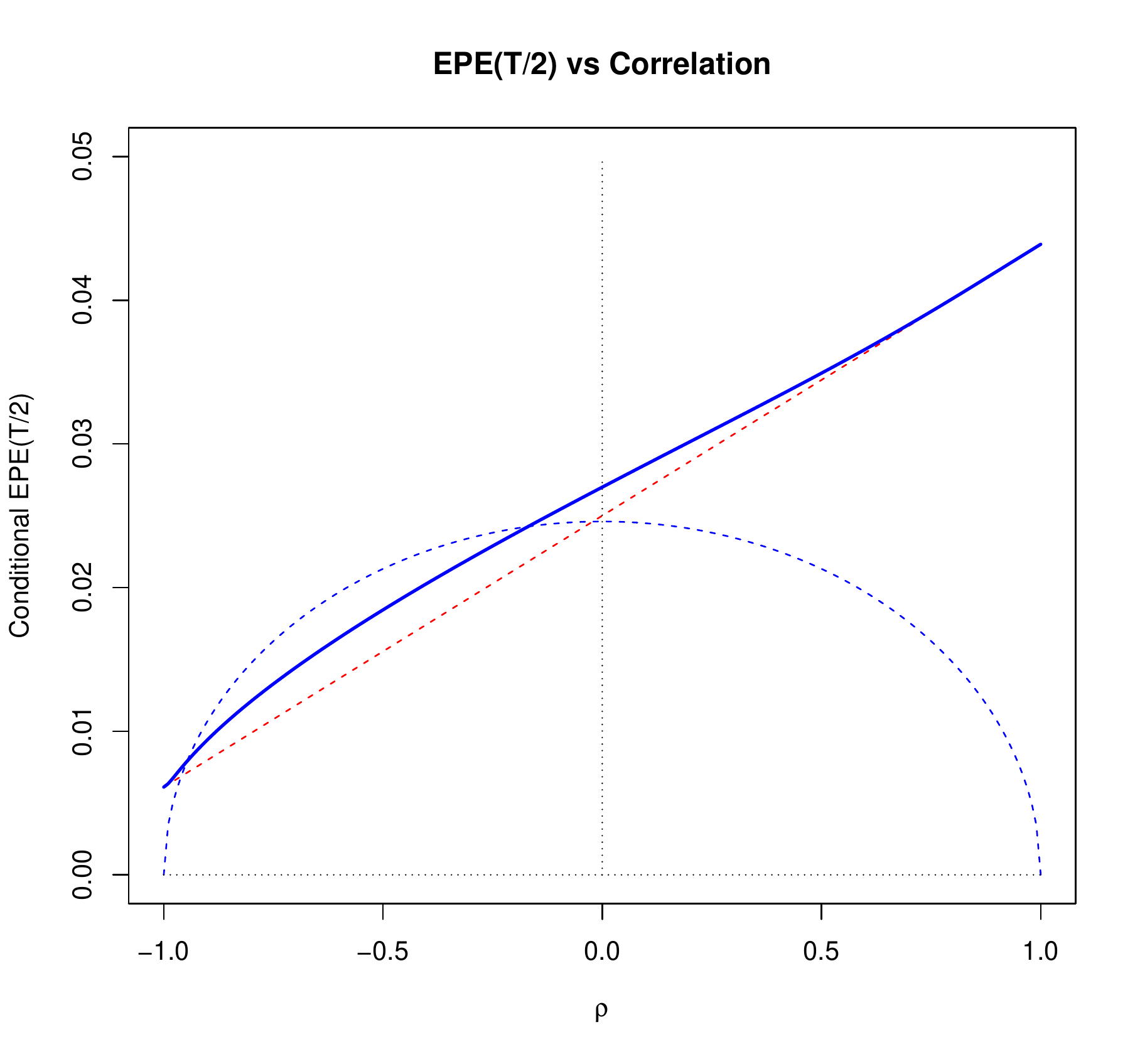}}\\
\caption{$\E[V^+_{t}(t)]$ wrt $\rho$ for $t=T/2$ and $T=5$, $\vartheta=2.2\%$,$\gamma=0.4\%$ (blue, solid), average of discounted conditional exposure $\delta\gamma t(1-t/T) + \vartheta\sqrt{t(1-\delta t/T)}\rho \Phi^{-1}(G(t))$ (red, dashed) and residual standard deviation $\vartheta\sqrt{t(1-\delta t/T)(1-\rho^2)}$ (blue, dashed) for Forward ($\delta=0$) and IRS ($\delta=1$) profiles.}\label{fig:WWRGCEPE}
\end{figure}

\subsection{Dynamic WWR CVA}
\label{sec:WWREPEDyn}

CVA is given by the integral of the WWR EPE with respect to the default probability $\bar{G}(t)=1-G(t)$ up to portfolio maturity; this is eq.~(\ref{eq:CVAdefint}). In the dynamic approach, this WWR EPE takes the form
\beq
\E[V_t^+|\tau=t]=\E[\zeta_tV_t^+]\;,
\eeq
where $\zeta_t=-\mu(t)/(h(t)G(t))$ is the wrong-way process defined according to the dynamics of the Az\'ema supermartingale implied by the model.

In this section, we analyze the intensity approach using the shifted Vasicek approach and the martingale approach. The reason why SSRD is not dealt with in details here is that we focus on cases that remain analytically tractable, which is not the case of SSRD (in particular, due to the difficulty to compute the correlation between $\lambda_t$ and $\Lambda_t=\int_0^t \lambda_s ds$). Moreover, the main advantage of the SRD (CIR) process over the Vasicek is (in our context) the positivity property. Unfortunately, the positivity property is not guaranteed due to the shift function $\phi$, which rends the model less attractive. Moreover, both intensity models have more or less the same behavior (in terms of WWR impact) when being far from zero (empirical evidences illustrate that point in Section~\ref{sec:Comp}). Therefore, we restrict ourselves to perform a detailed analysis of the Hull-White intensity approach, and will make some comments on the SSRD based on numerical results. The later will be based on Monte Carlo simulations using the Euler schemes proposed in~\cite{Brigo05}.

Denoting by $Z_i$ standard Normal variables with appropriate correlation matrix
\beqn
\lambda_t&\sim&A(t)+B(t) Z_1\\
A(t)&=&r_0\e^{-\kappa t}+\phi(t)\\
\phi(t)&=&h(t)+\frac{\sigma^2}{\kappa}(\xi(1,t)-\xi(2,t))-r_0\e^{-\kappa t}\\
B(t)&=&\sigma\sqrt{\xi(2,t)}\;.
\eeqn

On the other hand,
\beq
S_t=\e^{-\int_{0}^tr_s+\phi(s)ds}=:\e^{-\Lambda_t}
\eeq
where
\beqn
\Lambda_t&=&\int_{0}^t \lambda_s ds=\int_{0}^t (r_s+\phi(s)) ds\\
&\sim&\omega(t)+\Omega(t)Z_2\\
\omega(t)&=&r_0\xi(1,t)+\int_{0}^t \phi(s) ds\\
&=&\frac{\sigma^2}{2\kappa^2}(t-2\xi(1,t)+\xi(2,t))-\ln G(t)\\
\Omega(t)&=&\frac{\sigma}{\kappa}\sqrt{t-2\xi(1,t)+\xi(2,t)}
\eeqn
and recall that both profiles are Normally distributed, that is
\beq
V_t=a(t)+b(t)Z_3\label{eq:Vta+bZ}\;.
\eeq

For each time $t$, the intensity, survival probability and discounted exposure variables can be written as a function of iid standard Normal variables $X,Y,Z$ :
\beqn
\lambda_t&=&A(t)+B(t)X\\
S_t&=&\e^{-\omega(t)-\Omega(t)(r_{21}X+r_{22}Y)}\\
V_t&=&a(t)+b(t)(r_{31}X+r_{32}Y+r_{33}Z)
\eeqn
where $r_{ij}$ is the $(i,j)$ element of the lower triangular matrix $R$ obtained by Choleski decomposition of the $(\lambda_t, \Lambda_t, V_t)$ correlation matrix, and satisfy $||r_{1\cdot}||_2=||r_{2\cdot}||_2=1$. Disregarding the time-index for ease of reading,
\beq
\left[
\begin{array}{ccc}
	1 & \rho_{\lambda, \Lambda} & \rho_{\lambda, V}\\
	\rho_{\lambda, \Lambda} & 1 & \rho_{\Lambda, V}\\
	\rho_{\lambda, V} & \rho_{\Lambda, V} & 1
\end{array}
\right]
= RR^T
\eeq
with $R$ taking the form
\beq
R:=\left[
\begin{array}{ccc}
	1 & 0 & 0\\
	\rho_{\lambda, \Lambda} & \sqrt{1-\rho^2_{\lambda, \Lambda}} & 0\\
	\rho_{\lambda, V} & \frac{\rho_{\Lambda, V}-\rho_{\lambda, V}\rho_{\lambda, S}}{\sqrt{1-\rho^2_{\lambda, \Lambda}}} & \sqrt{1-\rho_{\lambda, V}^2 - \frac{\left(\rho_{\Lambda, V}-\rho_{\lambda, V}\rho_{\lambda, \Lambda}\right)^2}{1-\rho^2_{\lambda, \Lambda}}}
\end{array}
\right]\;.
\eeq

Using this formulation, the expression of $\E\left[\zeta_tV_t^+\right]$ becomes analytically available as $\E\left[\lambda_tS_tV_t^+\right]$ can be obtained in closed-form; this is performed in Appendix, Section~\ref{sec:App2:HW}. It remains to compute the expression of the correlation matrix $RR^T$ for each time index $t$, which we now derive. To that purpose, first recall that the covariance between two It\^o integrals $I_t$ and $J_t$ of deterministic integrands $a(t),b(t)$ driven by two Brownian motions $W^I$ and $W^J$
\beqn
I_t&=&\int_0^t a(s) dW^I_s\\
J_t&=&\int_0^t b(s) dW^J_s
\eeqn
is given by 
\beq
\cov(I_t,J_t)=\int_0^t a(s)b(s) d\langle W^I,W^J\rangle_s\;.
\eeq

With this expression in hand, we can compute the pairwise correlations between the involved processes.

In particular, the intensity-integrated intensity correlation $\rho_{\lambda,\Lambda}(t)$ does not depend on the portfolio process. It corresponds to the correlation between $r_t=\lambda_t-\phi(t)$ and $y_t:=\int_{0}^t r_s ds$, and is given by
\beq
\rho_{r,y}(t)=\rho_{\lambda,\Lambda}(t)=\frac{\xi(1,t)-\xi(2,t)}{\sqrt{\xi(2,t)}\sqrt{t-2\xi(1,t)+\xi(2,t)}}\;.
\eeq

We now compute the correlation with discounted exposure variable for both FRA and IRS prototypical exposure profiles. With regards to the forward-type contract, the correlation between $\lambda_t$ and $V_t$ is constant and given by
\beq
\rho_{V,\lambda}(t)=\frac{\rho\sigma\vartheta t}{\sqrt{\sigma^2 t}\sqrt{\vartheta^2 t}}=\rho
\eeq
and the correlation between $\Lambda_t$ and $V_t$ is given by
\beq
\rho_{V,\Lambda}(t)=\frac{\rho(t-\xi(1,t))}{\sqrt{t}\sqrt{(t-2\xi(1,t)+\xi(2,t)}}\;.
\eeq
For the IRS-type profiles, the correlation between $\lambda_t$ and $V_t$ is given by
\beq
\rho_{V,\lambda}(t)=\frac{\rho\sigma\vartheta (T-t)\int_{0}^t\frac{1}{T-s}ds}{\sqrt{\sigma^2 t}\sqrt{\vartheta^2\frac{t(T-t)}{T}}}=\frac{\rho \sqrt{(T-t)T}\ln(T/(T-t))}{t}
\eeq
and the correlation between $\Lambda_t$ and $V_t$ is given by
\beqn
\rho_{V,\Lambda}(t)&=&\frac{\rho\vartheta\sigma(T-t)\left(\ln(T/(T-t))+\e^{-\kappa t}\int_{-T}^{t-T}\frac{\e^{\kappa s}}{s}ds\right)}{\sqrt{\sigma^2 t}\sqrt{\vartheta^2\frac{t(T-t)}{T}}}\\
&=&\frac{\rho(T-t)\left(\ln(T/(T-t))+\e^{\kappa (T-t)}\int_{-T}^{t-T}\frac{\e^{\kappa s}}{s}ds\right)}{\sqrt{t-2\xi(1,t)+\xi(2,t)}\sqrt{\frac{t(T-t)}{T}}}\;.
\eeqn

The above results show that one can compute CVA analytically under a stochastic ``intensity'' model where both exposures and ``intensities'' are Gaussian processes. 

The analytical tractability is preserved when working with the $\Phi$-martingale, as we now show. The resulting expressions are simpler than those of the Hull-White intensity approach as there is no need to compute $\rho_{\lambda,\Lambda}(t)$. Indeed, eq.~(\ref{eq:ZetaConic}) reveals that $\zeta$ is given by a simple function (smooth bijection) of a Gaussian process, and not the product of two Gaussian processes:
\beqn
\zeta_t&=&\varphi(X_t)k(t)\\
k(t)&=&\e^{\mu t}/\varphi(\Phi^{-1}(G(t))
\eeqn
with $\mu=\sigma^2/2$ and where the latent (Vasicek) process is distributed as 
\beq
X_t\sim\underbrace{\Phi^{-1}(G(t))\e^{\mu t}}_{A(t)}+\underbrace{\sqrt{\e^{2\mu t}-1}}_{B(t)}Z\;.
\eeq

On the other hand, 
$(X_t,V_t)$ is jointly Gaussian with some correlation $\rho(t)$, that will be derived below.

Therefore, with $\brho(t)=\sqrt{1-\rho(t)^2}$ the time-$t$ contribution of WWR EPE to CVA $f(t):=\E[V^+_t\zeta_t]$ is given by
\beqn
f(t)&=&k(t)\int_{x,y}\varphi\Big(A(t)+B(t)x\Big)\Big(a(t)+ b(t)(\rho(t)x+\brho(t)y)\Big)^+\varphi(x)\varphi(y)dxdy\\
&=&k(t)\int_{x}\varphi\Big(A(t)+B(t)x\Big)I(x)\varphi(x)dx\label{eq:IntConic}\\
I(x)&:=&\int_{-\infty}^\infty\Big(a(t)+ b(t)(\rho(t)x+\brho(t)y)\Big)^+\varphi(y)dy\\
&=&\int_{-m(x)}\Big(a(t)+ b(t)(\rho(t)x+\brho(t)y)\Big)\varphi(y)dy\;.
\eeqn

Setting
\beq
m(x):=\frac{\frac{a(t)}{ b(t)}+\rho(t)x}{\brho(t)}
\eeq
the integral $I(x)$ above becomes
\beqn
I(x)&=&\Big(a(t)+ b(t)\rho(t)x\Big)\int_{-m(x)}^\infty\varphi(y)dy+ a(t)+ b(t)\brho(t)\int_{-m(x)}^\infty y\varphi(y)dy\\
&=&\Big(a(t)+ b(t)\rho(t)x\Big)\Phi(m(x))+ b(t)\brho(t)\varphi(m(x))\;.
\eeqn

Using this expression for $I(x)$, the integral~(\ref{eq:IntConic}) can be derived explicitly (see Appendix~\ref{sec:App2:CM}). The only missing piece at this stage is $\rho(t)$, a quantity that we now compute for both FRA and IRS profiles. In FRA-type contracts, the correlation between $X_t$ and $V_t$ is given by
\beq
\rho(t)=\frac{\sigma\vartheta\rho\e^{\mu t}\int_0^t\e^{-\mu s}ds}{\sqrt{\sigma^2\int_0^t\e^{2\mu(t-s)}ds}\sqrt{\vartheta^2 t}}=2\rho\frac{1-\e^{-\sigma^2/2 t}}{\sigma\sqrt{t(1-\e^{-\sigma^2 t})}}\;.
\eeq
For swap-types contract, the correlation between $X_t$ and $V_t$ is given by
\beqn
\rho(t)&=&\frac{\sigma\vartheta(T-t)\rho\e^{\mu t}\int_0^t\frac{\e^{-\mu s}}{T-s}ds}{\sqrt{\sigma^2\int_0^t\e^{2\mu(t-s)}ds}\sqrt{(\vartheta(T-t))^2\int_0^t\frac{1}{(T-s)^2}ds}}\\
&=&\frac{\sigma\vartheta(T-t)\rho\e^{\mu t}\int_{T-t}^T\frac{\e^{\mu s}}{s}ds}{\sqrt{\e^{2\mu t}-1}\vartheta\sqrt{\frac{t(T-t)}{T}}}\\
&=&\frac{\sigma\rho\int_{T-t}^T\frac{\e^{\mu s}}{s}ds}{\sqrt{1-\e^{-2\mu t}}\sqrt{\frac{t}{T(T-t)}}}\\
&=&\sigma\rho\sqrt{\frac{T(T-t)}{t(1-\e^{-\sigma^2 t})}}\int_{T-t}^T\frac{\e^{\sigma^2/2 s}}{s}ds\;.
\eeqn
%

\subsection{Impact of correlation $\rho$ on WWR EPE profiles and CVA}
\label{sec:Comp}

In this section, we shall compare the WWR EPE profiles $f(t)=\E[V_t^+|\tau=t]$ as a function of $\rho$ for both Forward and IRS prototypical examples. We handle those analytically or semi-analytically for the Gaussian Copula (GC), the Hull-White (HW) intensity and the Conic martingale (CM) setups.\footnote{The $\Phi$-martingale belongs to the class of \textit{conic martingales}, as defined in~\cite{Vrins14} or \cite{Vrins16}} Corresponding values for CVA, obtained by integrating the above with respect to the default distribution $\bar{G}(t)=1-G(t)$, are then analyzed for various risk profiles (hazard rate levels $h$). Notice that in order to facilitate the comparison exercise, we swapped the sign of $\rho$ in the Conic martingale approach.

\begin{landscape}
\begin{figure}
\centering
\subfigure[]{\includegraphics[width=0.2\columnwidth]{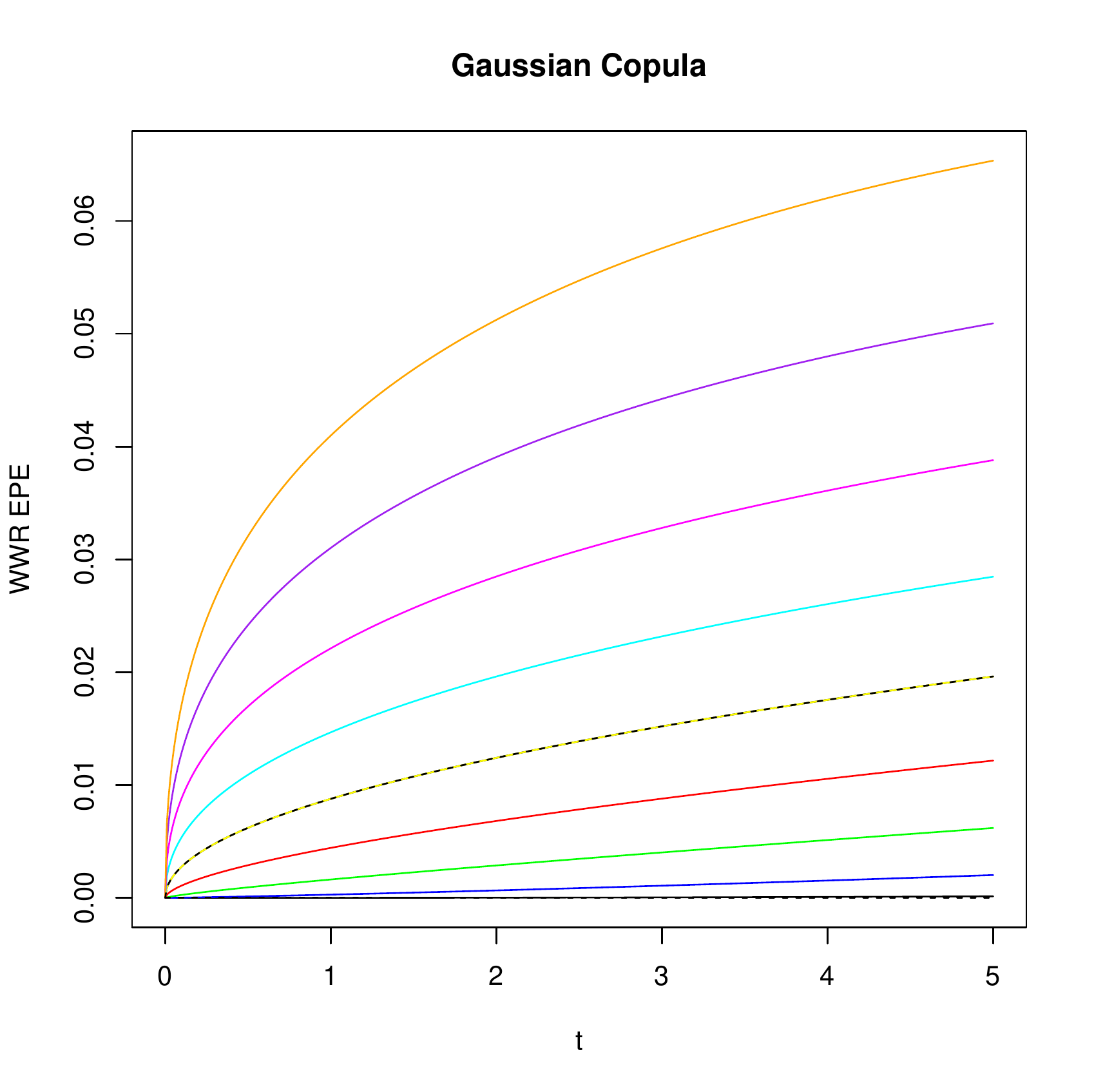}}
\subfigure[]{\includegraphics[width=0.2\columnwidth]{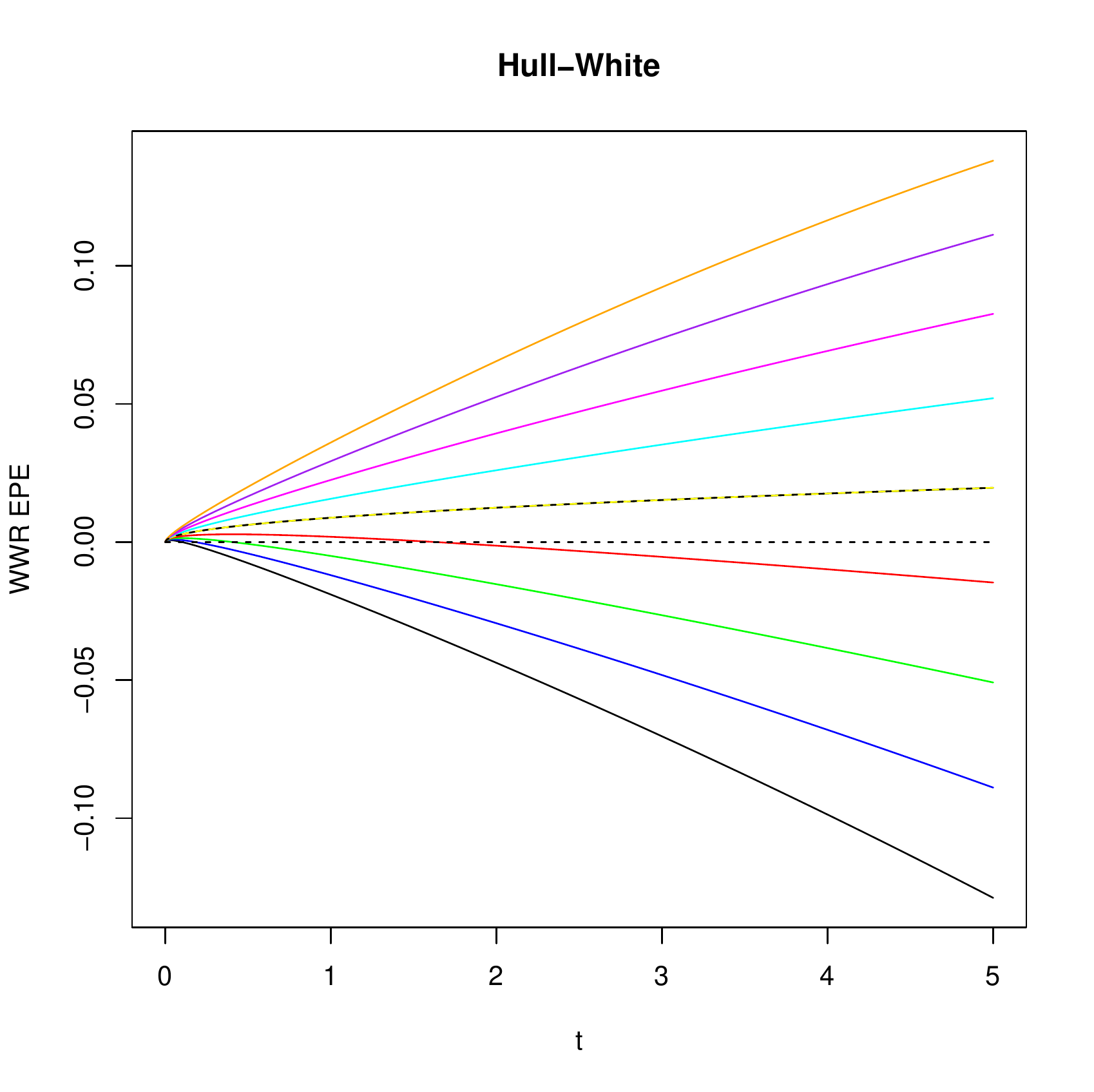}}
\subfigure[]{\includegraphics[width=0.2\columnwidth]{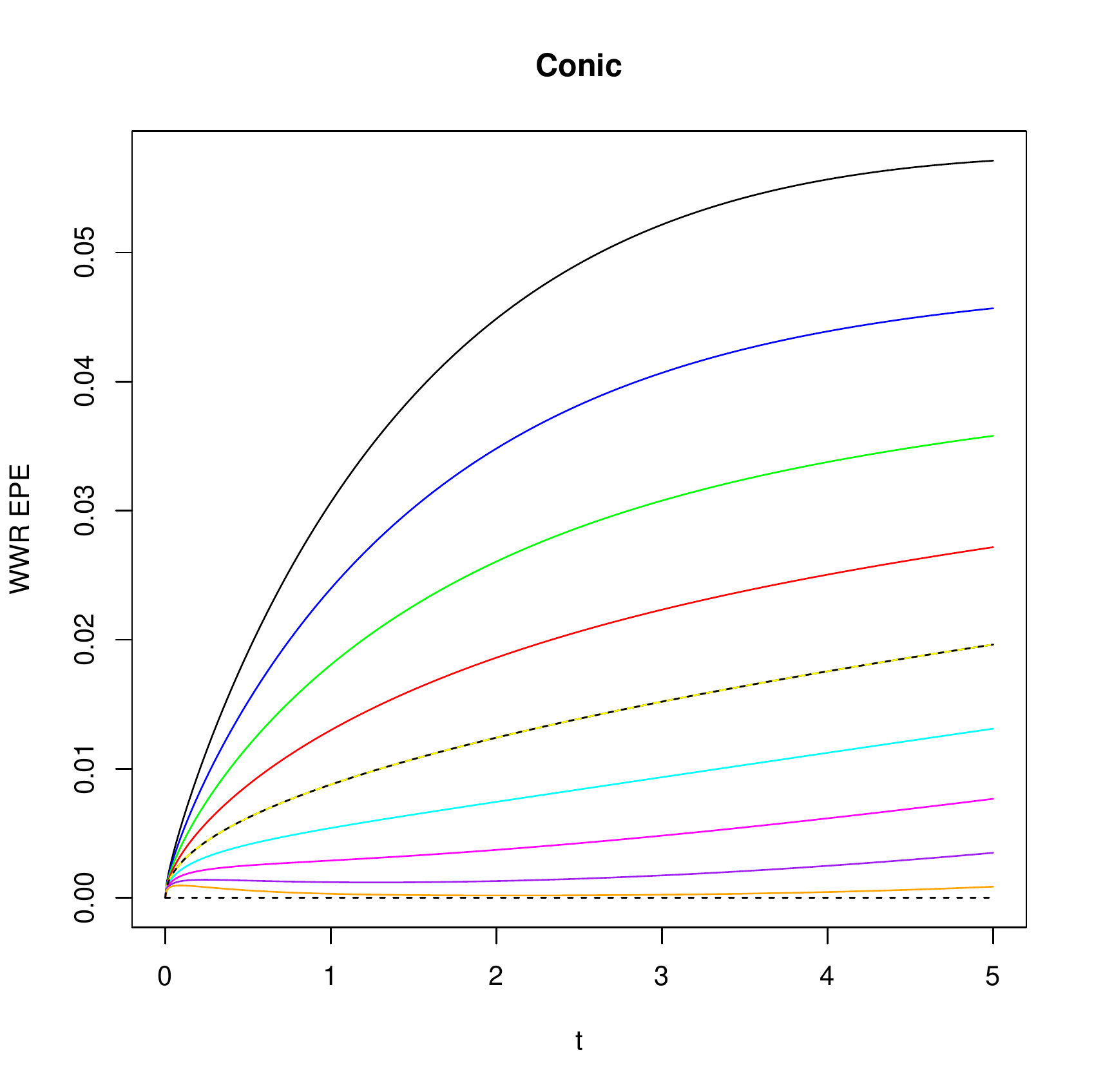}}
\subfigure[]{\includegraphics[width=0.2\columnwidth]{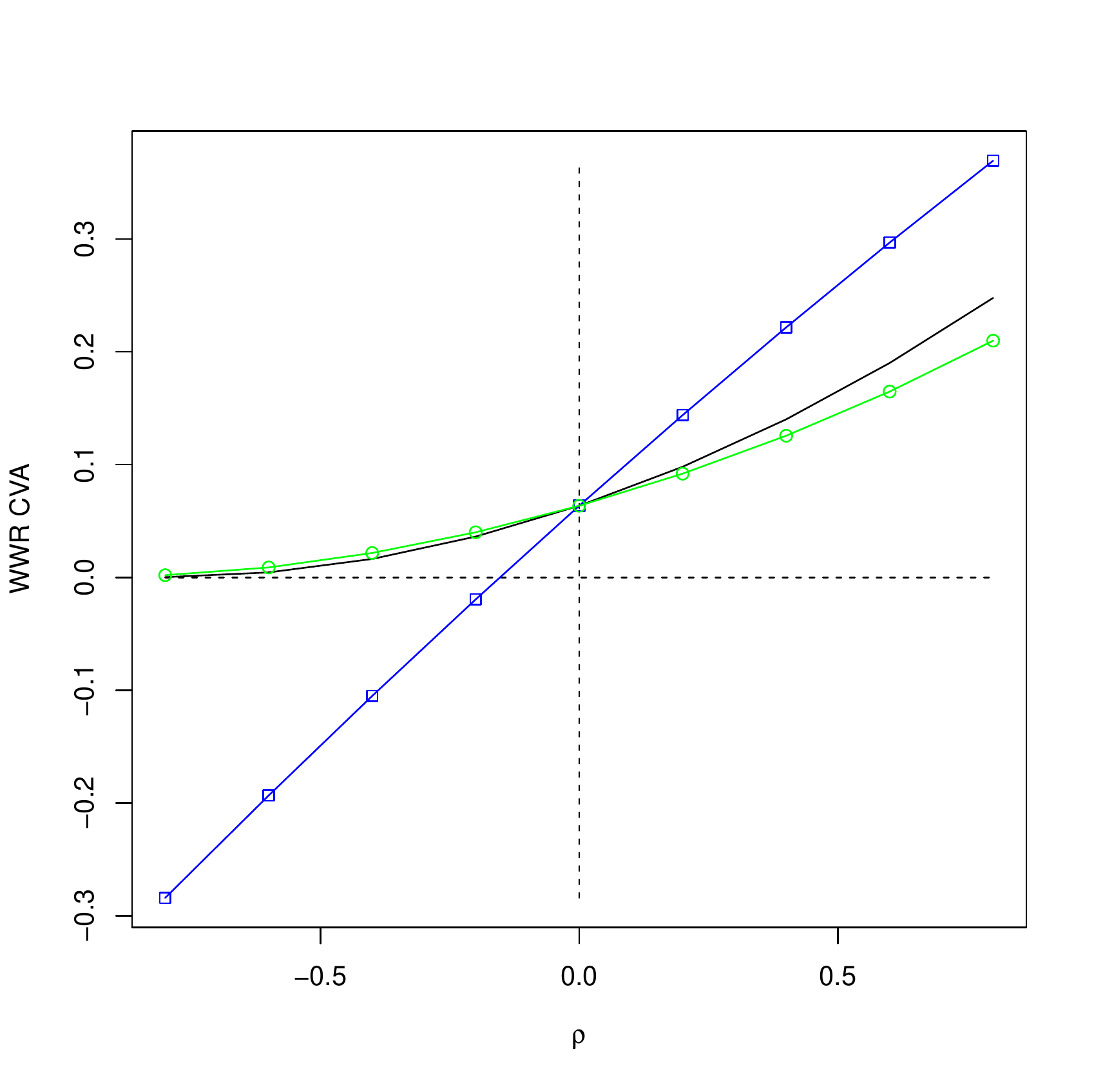}}\\
\subfigure[]{\includegraphics[width=0.2\columnwidth]{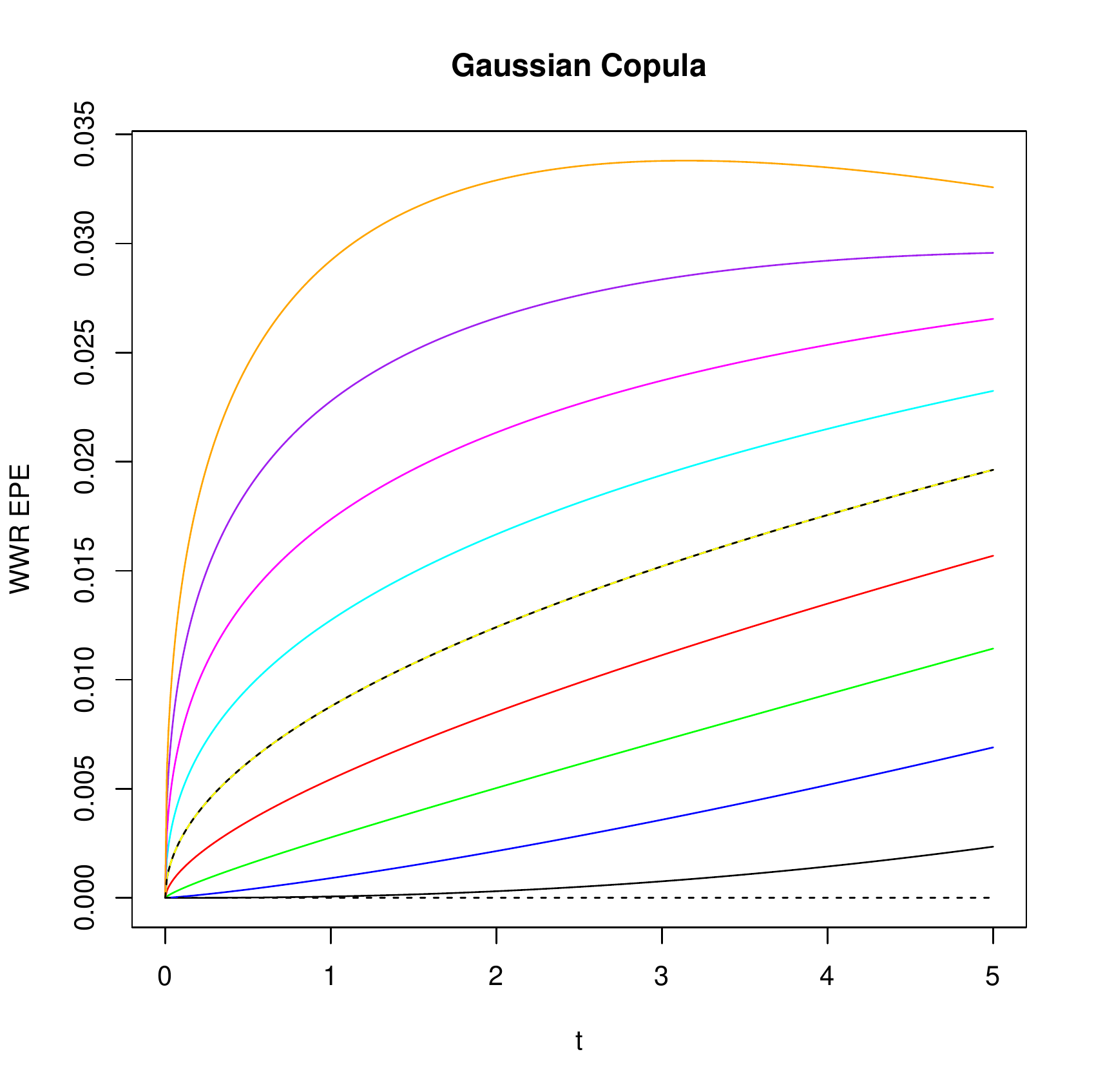}}
\subfigure[]{\includegraphics[width=0.2\columnwidth]{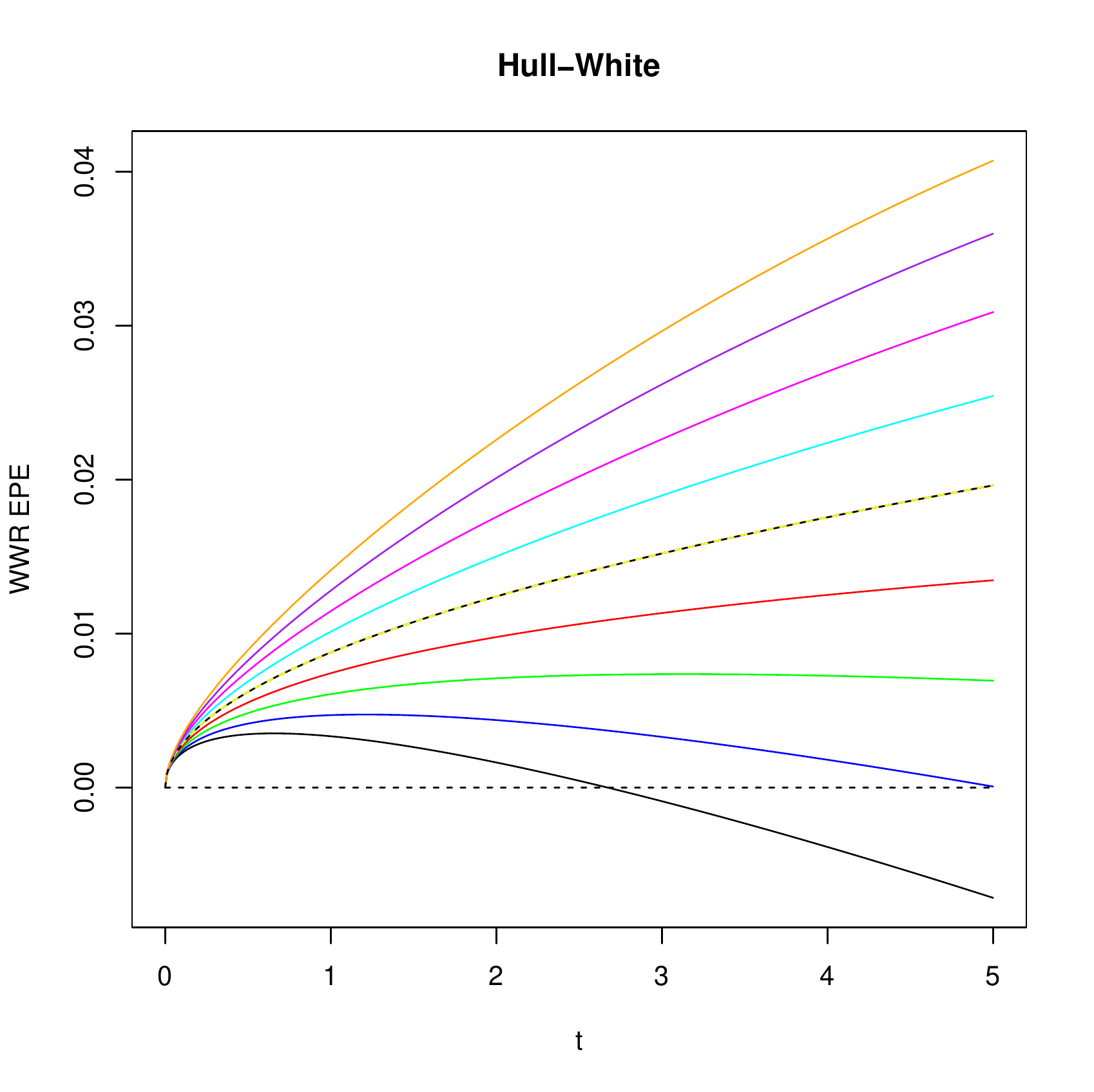}}
\subfigure[]{\includegraphics[width=0.2\columnwidth]{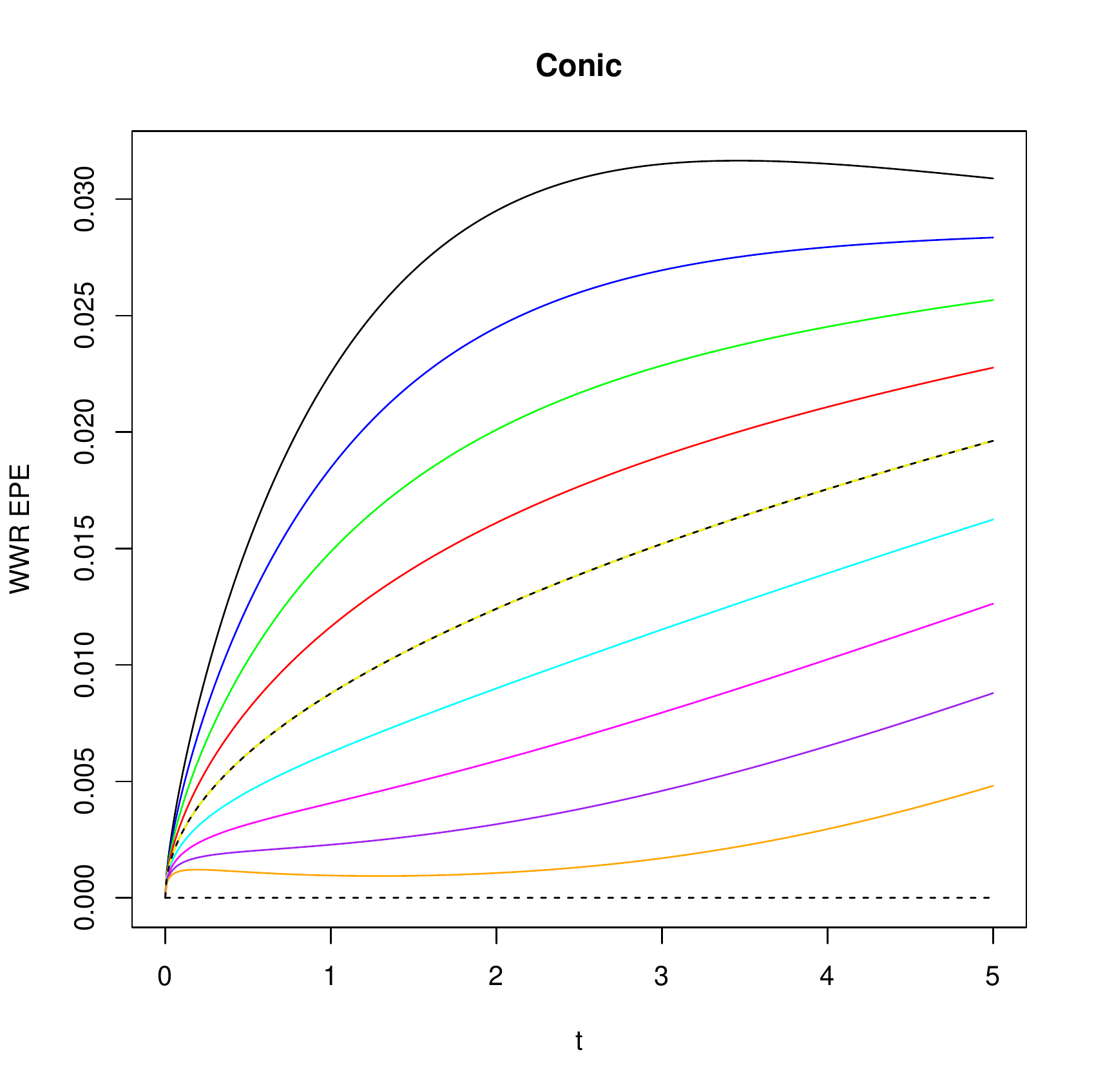}}
\subfigure[]{\includegraphics[width=0.2\columnwidth]{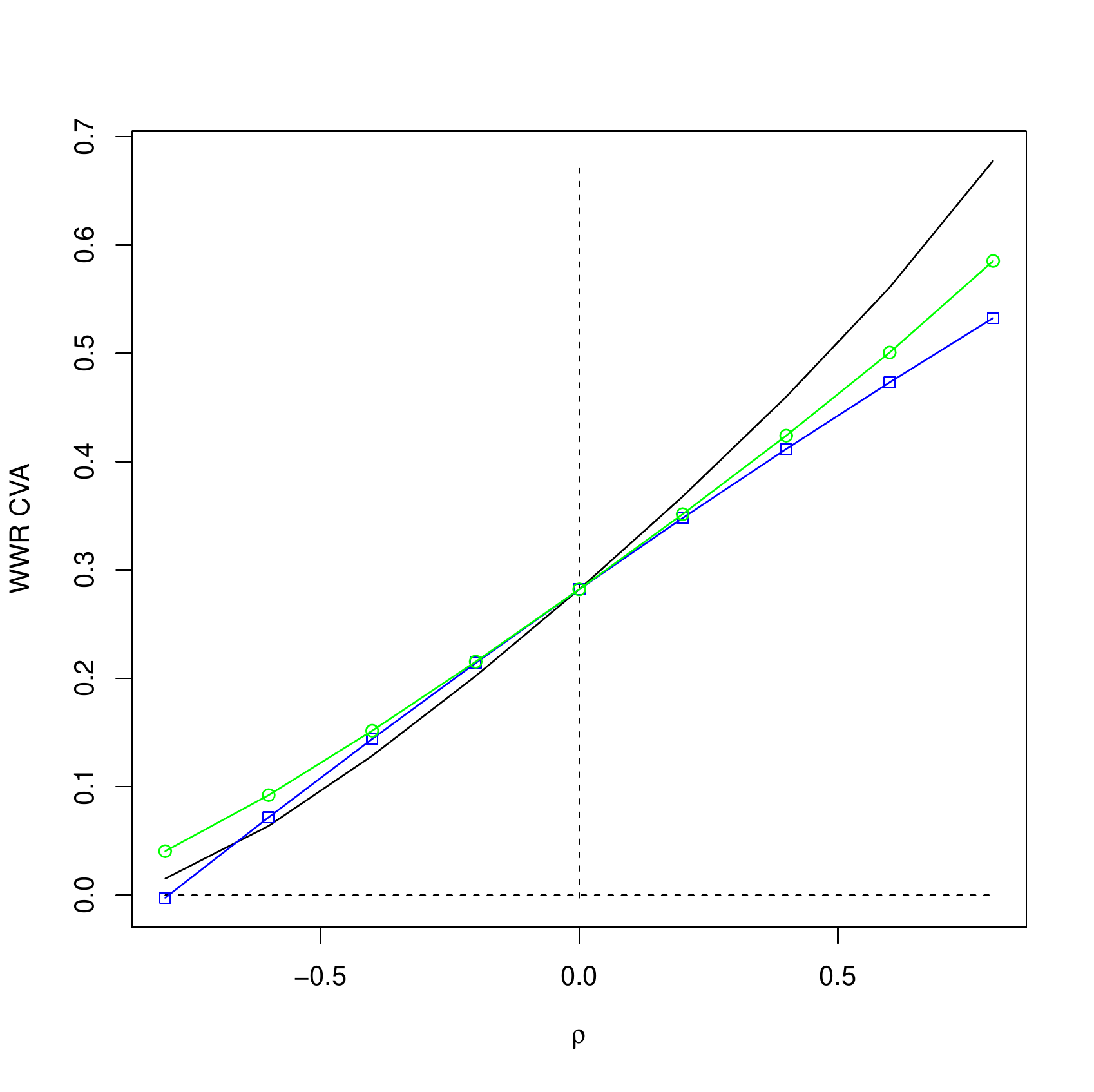}}\\
\subfigure[]{\includegraphics[width=0.2\columnwidth]{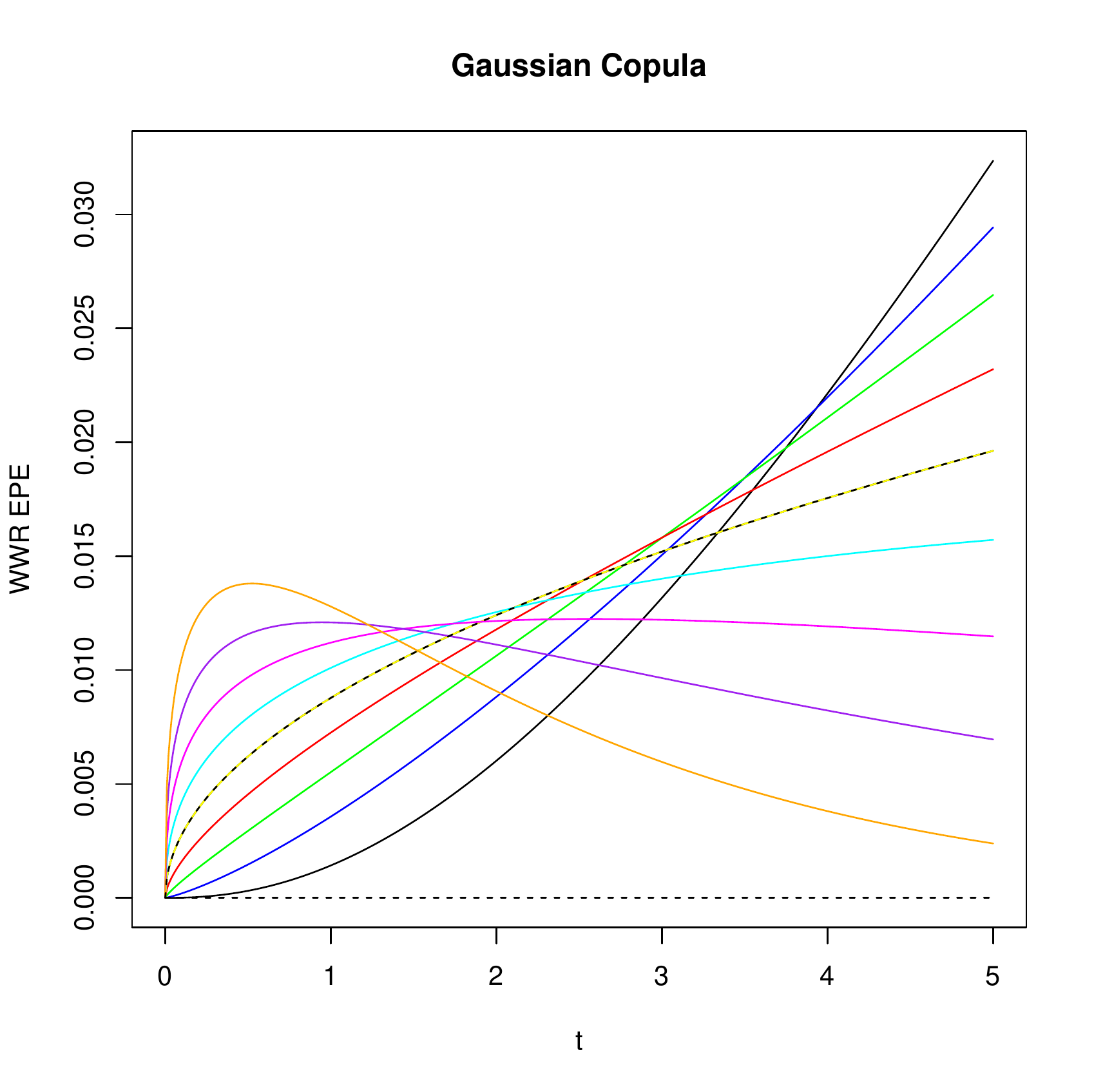}}
\subfigure[]{\includegraphics[width=0.2\columnwidth]{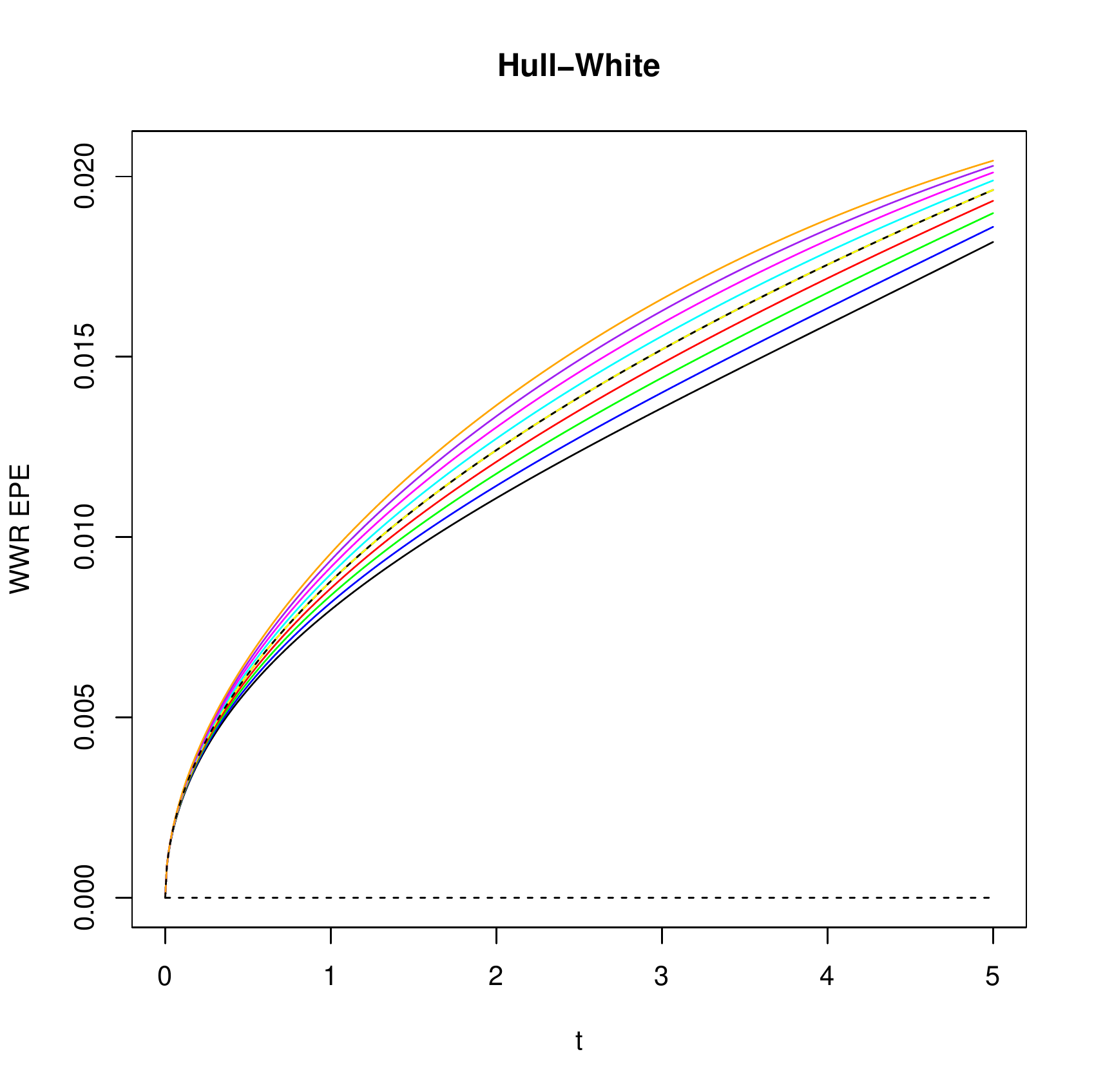}}
\subfigure[]{\includegraphics[width=0.2\columnwidth]{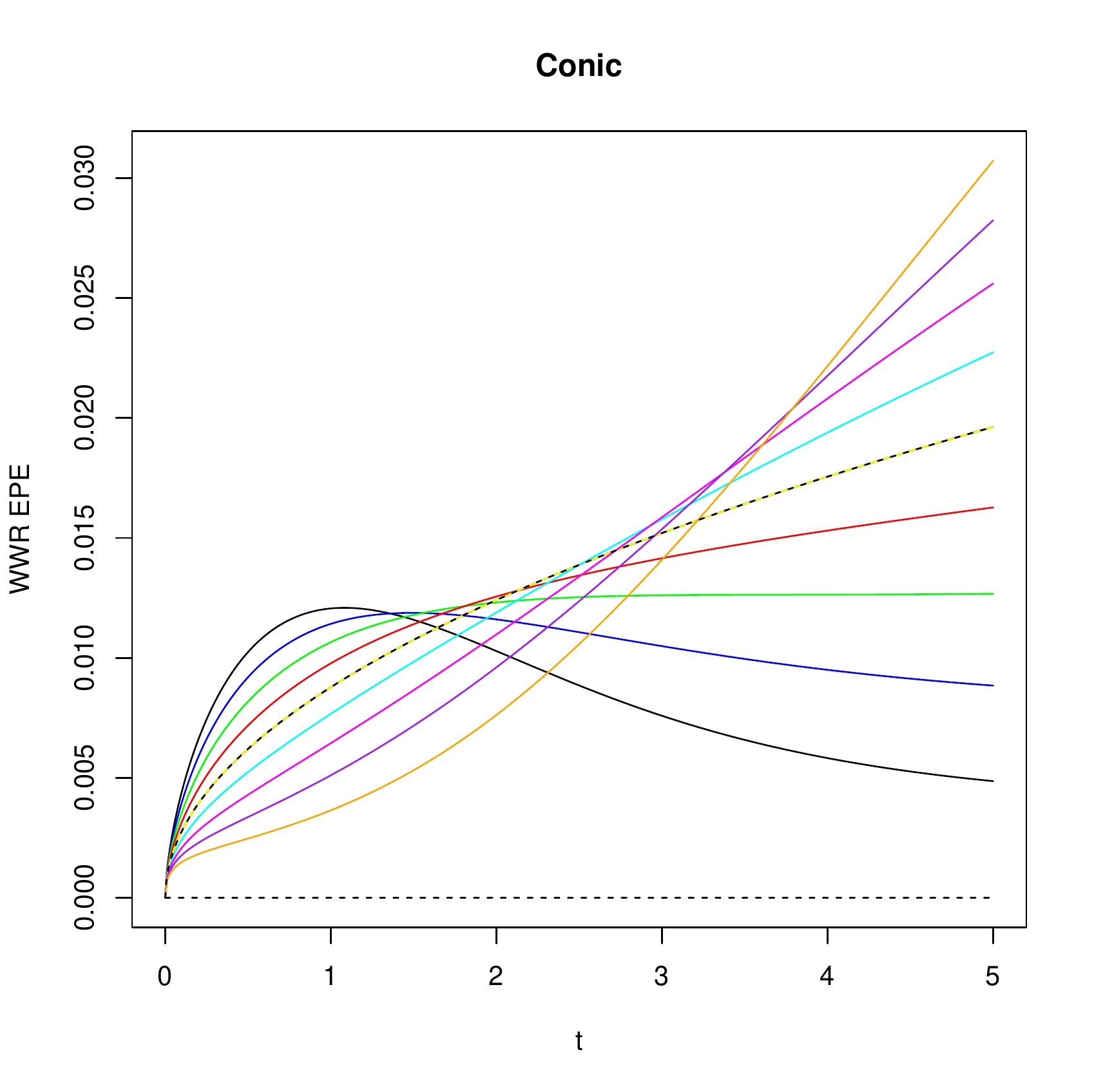}}
\subfigure[]{\includegraphics[width=0.2\columnwidth]{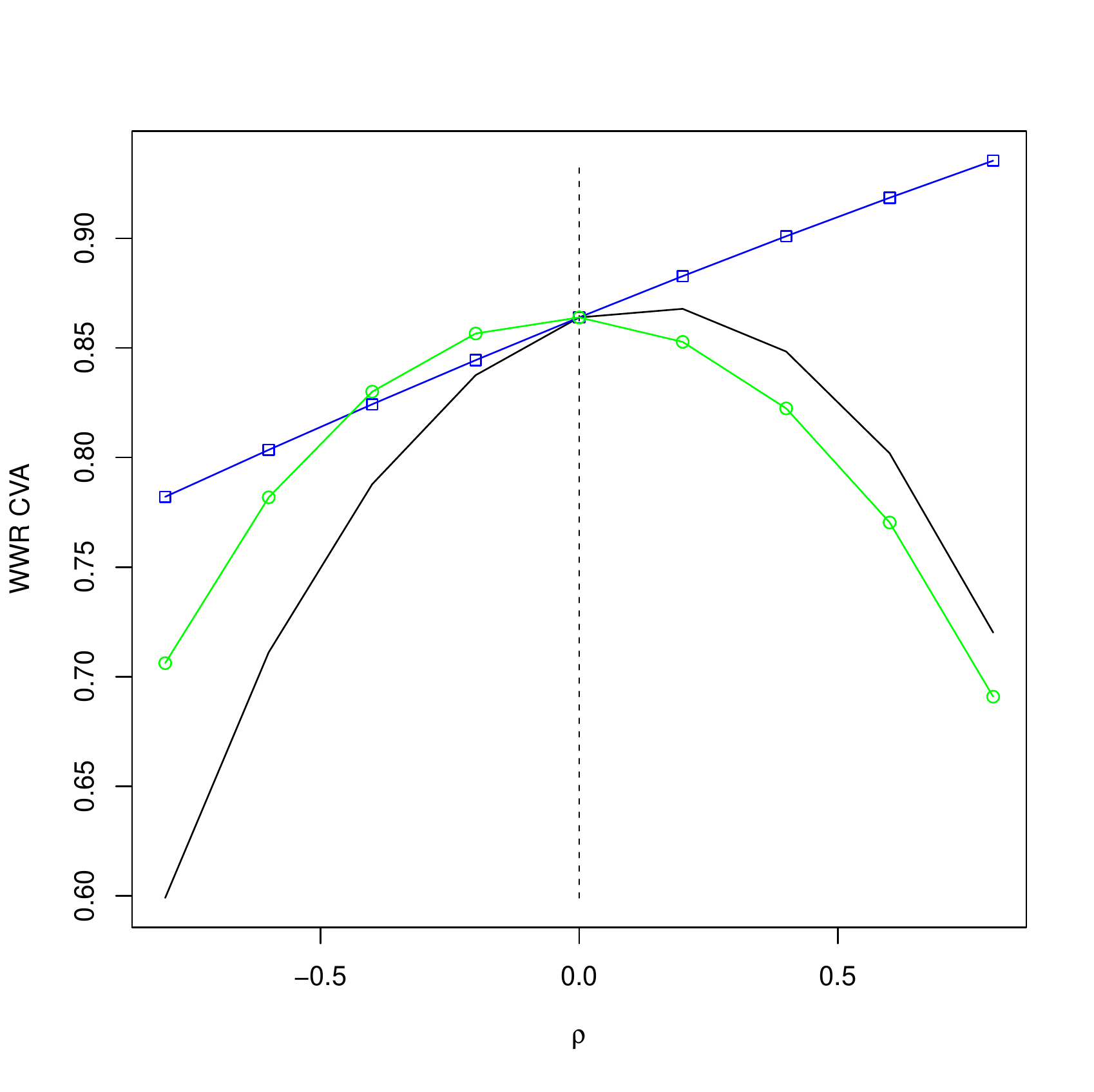}\label{fig:CVAFX}}\\
\caption{WWR EPE Forward profiles for the Gaussian Copula (1st col.), Hull-White (2nd col.; $\sigma=4\%,\kappa=0.5\%$) and Conic Martingale (3rd col.; $\sigma=0.9$), as well as WWR CVA (last col.). Top row ($h=1\%$), middle row ($h=5\%$), bottom row ($h=30\%$). Each curve corresponds to a level of $\rho\in\{-0.8,-0.6,\ldots,0.6,0.8\}$ with colormap (by ascending order) blue, green, red, yellow, cyan, magenta, purple, orange.}\label{fig:WWRFigFX}
\end{figure}
\end{landscape}

\begin{landscape}
\begin{figure}
\centering
\subfigure[]{\includegraphics[width=0.2\columnwidth]{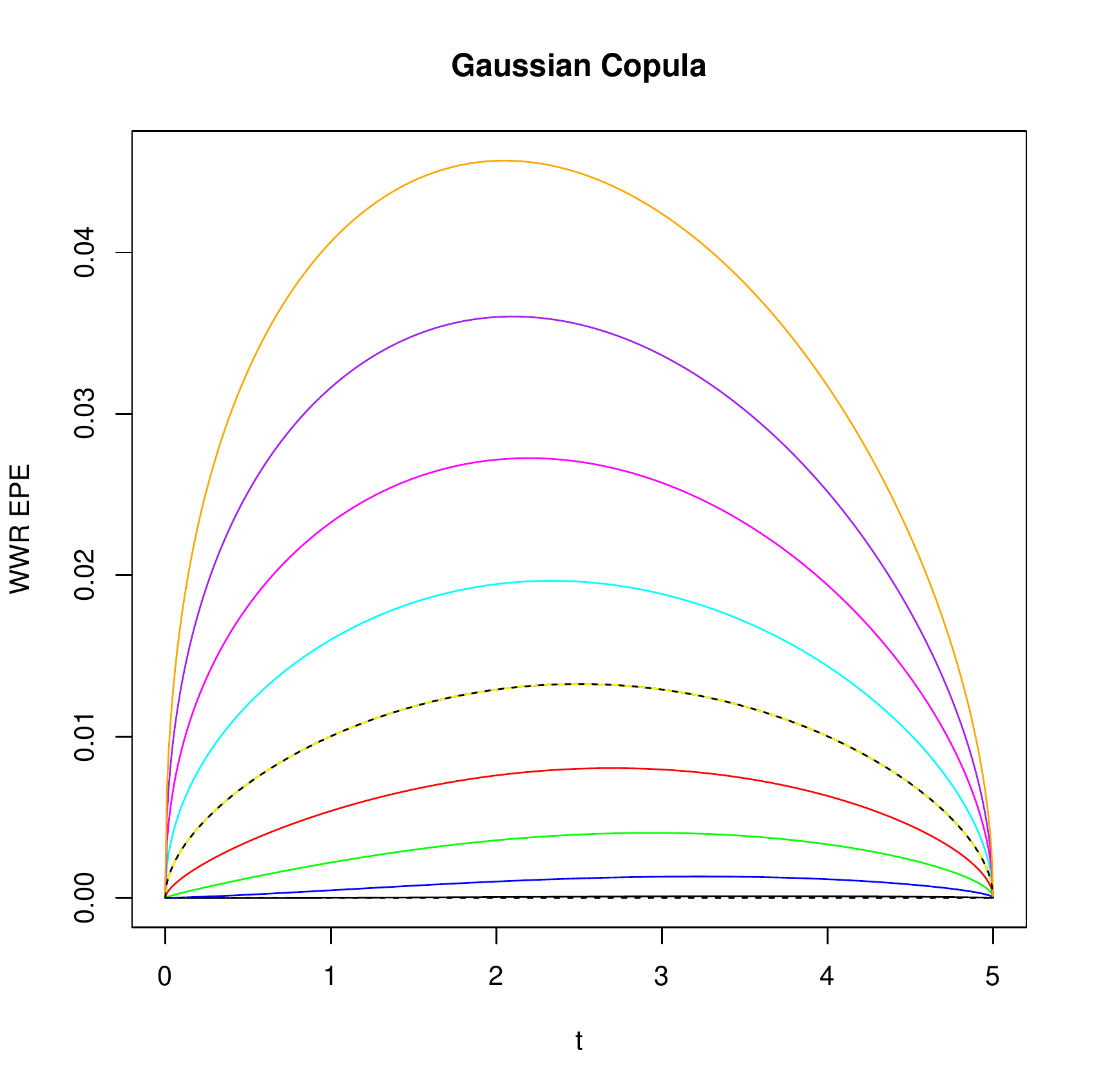}}
\subfigure[]{\includegraphics[width=0.2\columnwidth]{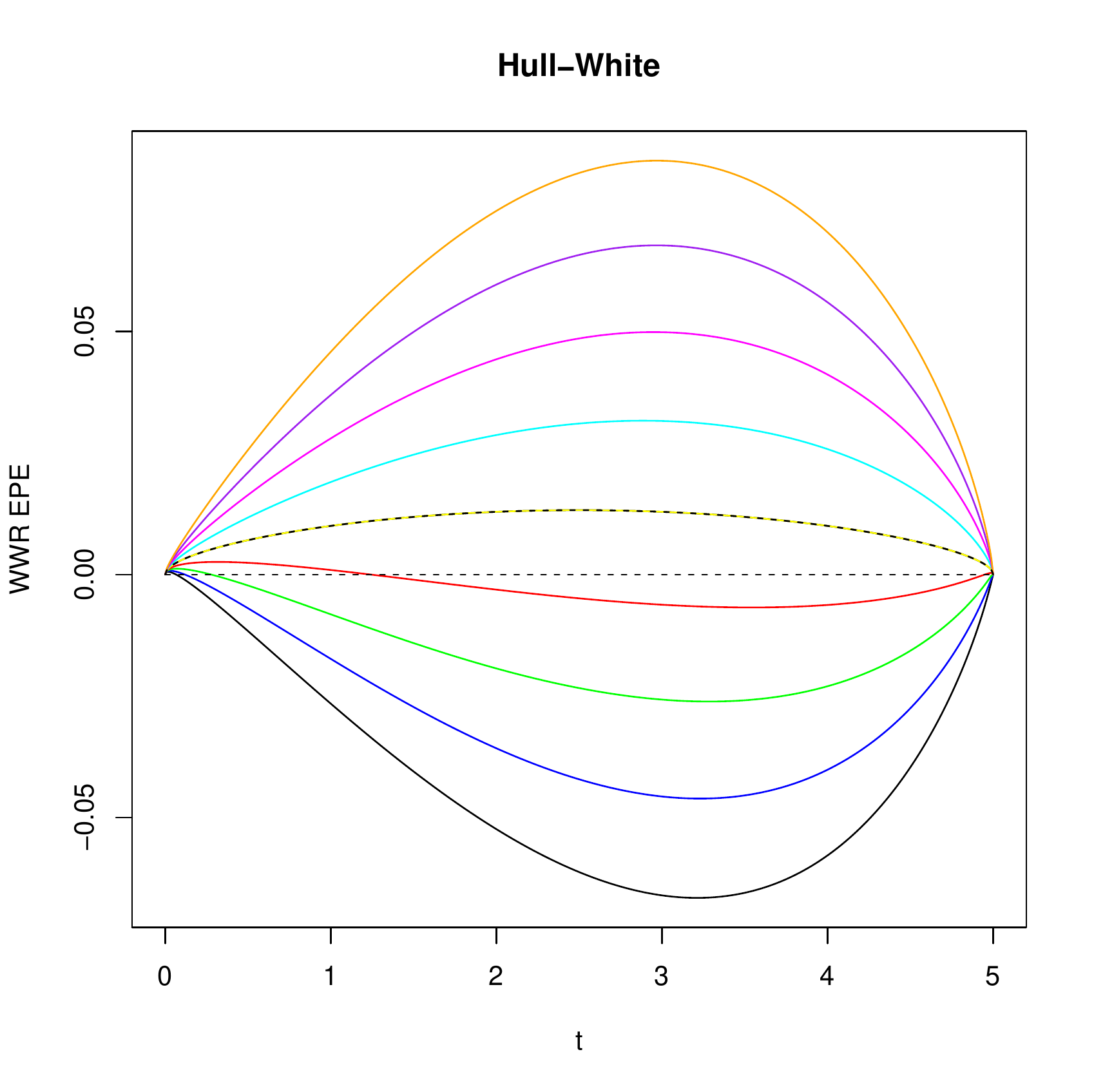}}
\subfigure[]{\includegraphics[width=0.2\columnwidth]{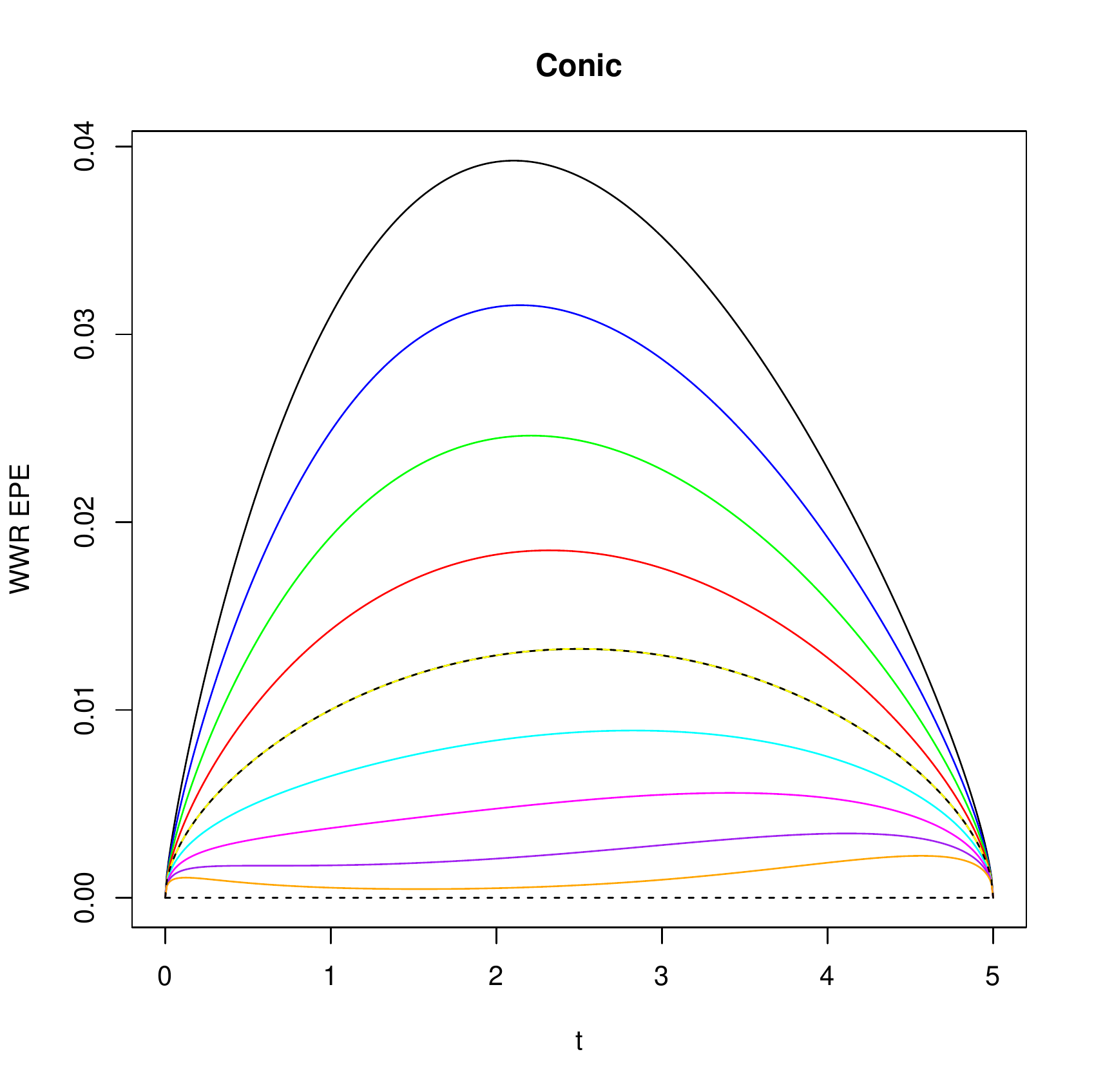}}
\subfigure[]{\includegraphics[width=0.2\columnwidth]{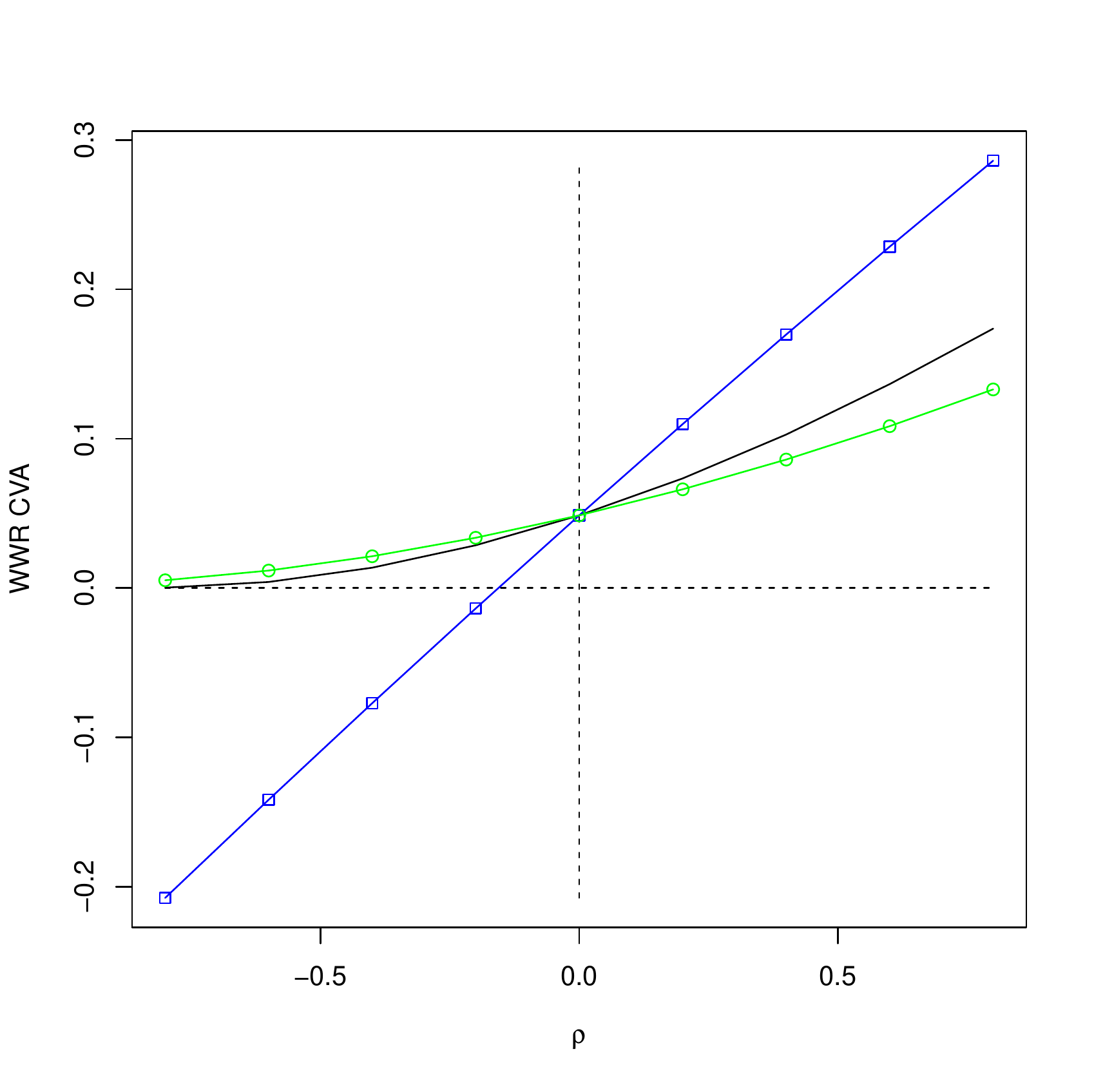}}\\
\subfigure[]{\includegraphics[width=0.2\columnwidth]{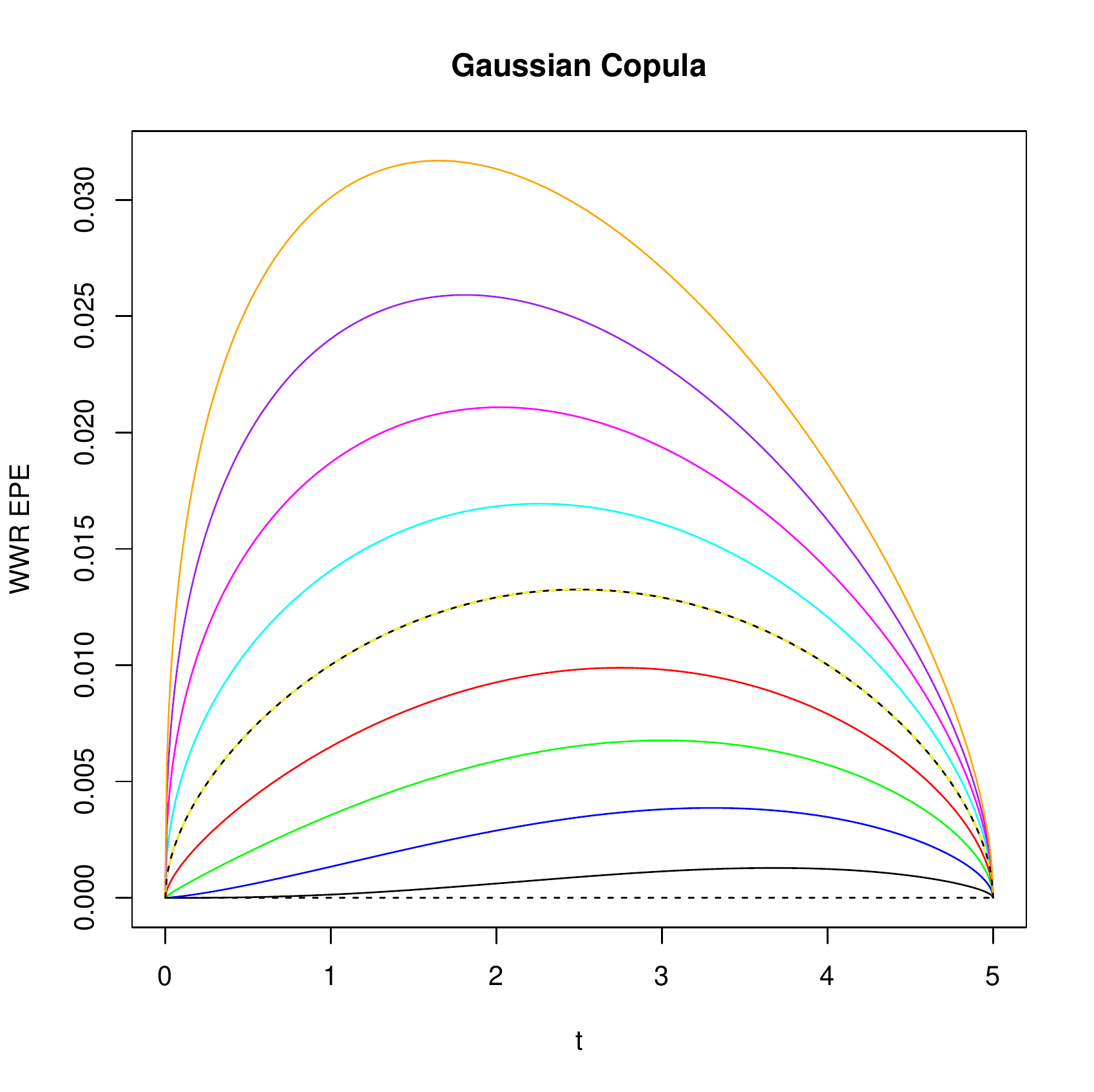}}
\subfigure[]{\includegraphics[width=0.2\columnwidth]{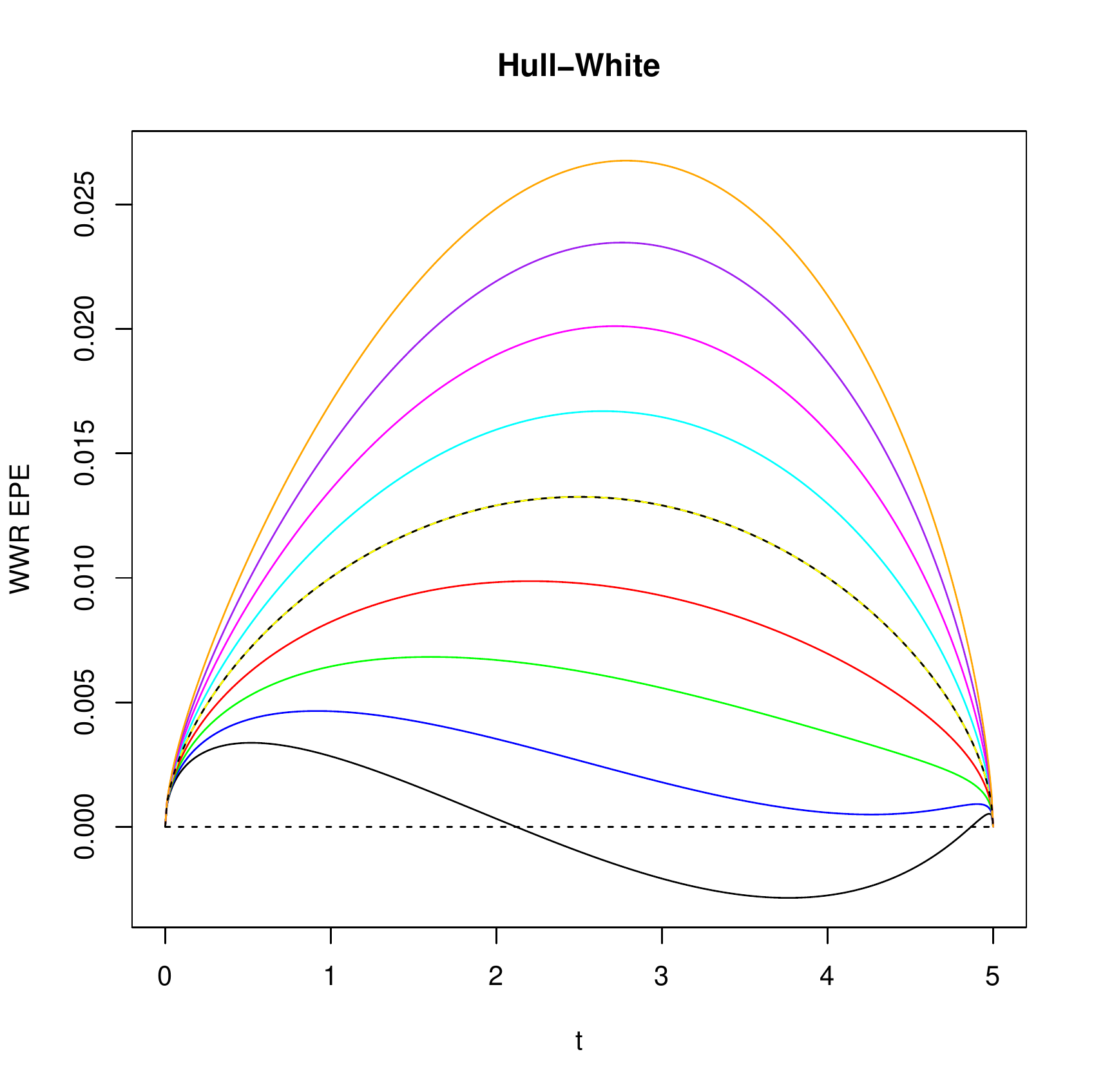}}
\subfigure[]{\includegraphics[width=0.2\columnwidth]{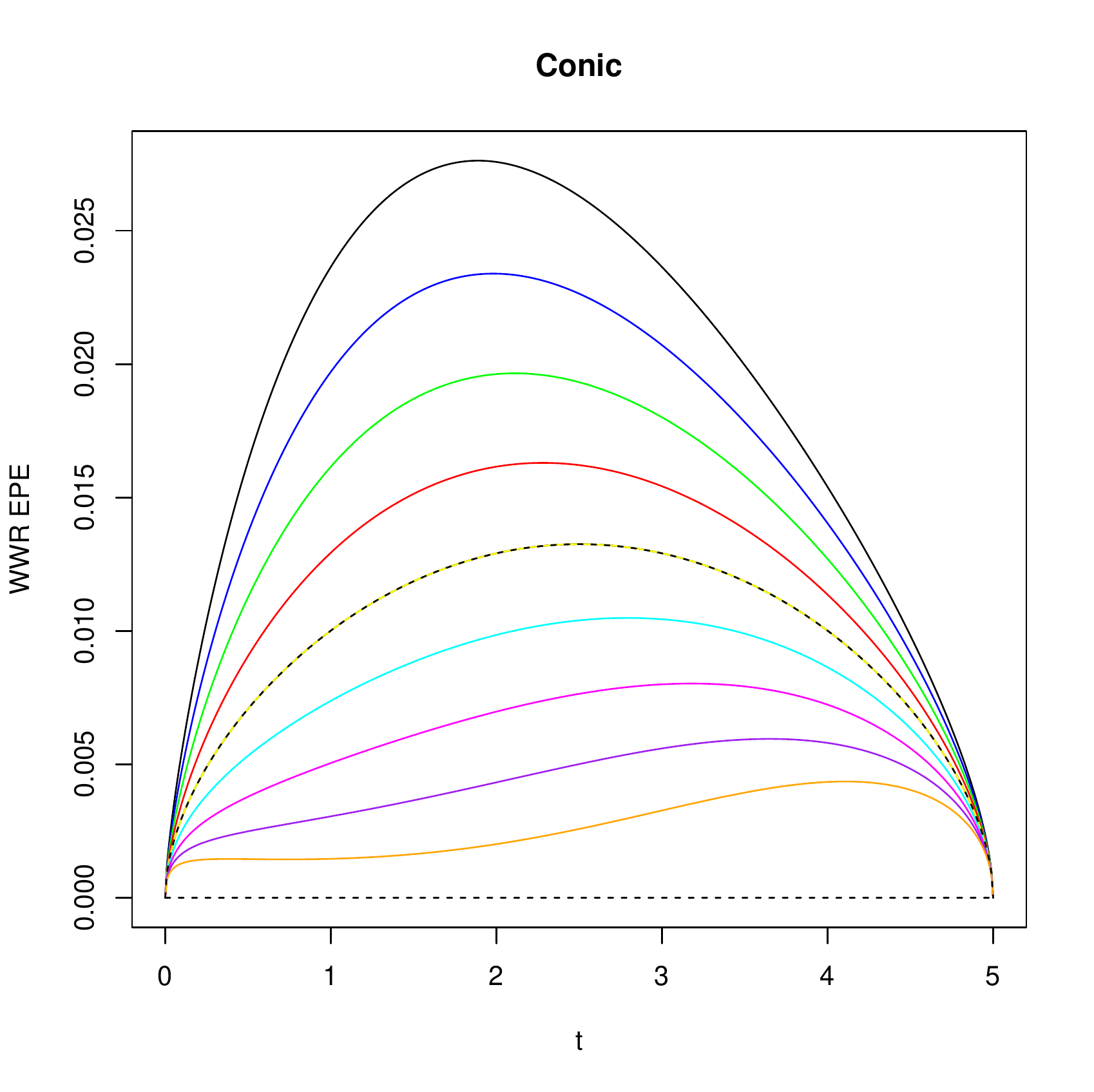}}
\subfigure[]{\includegraphics[width=0.2\columnwidth]{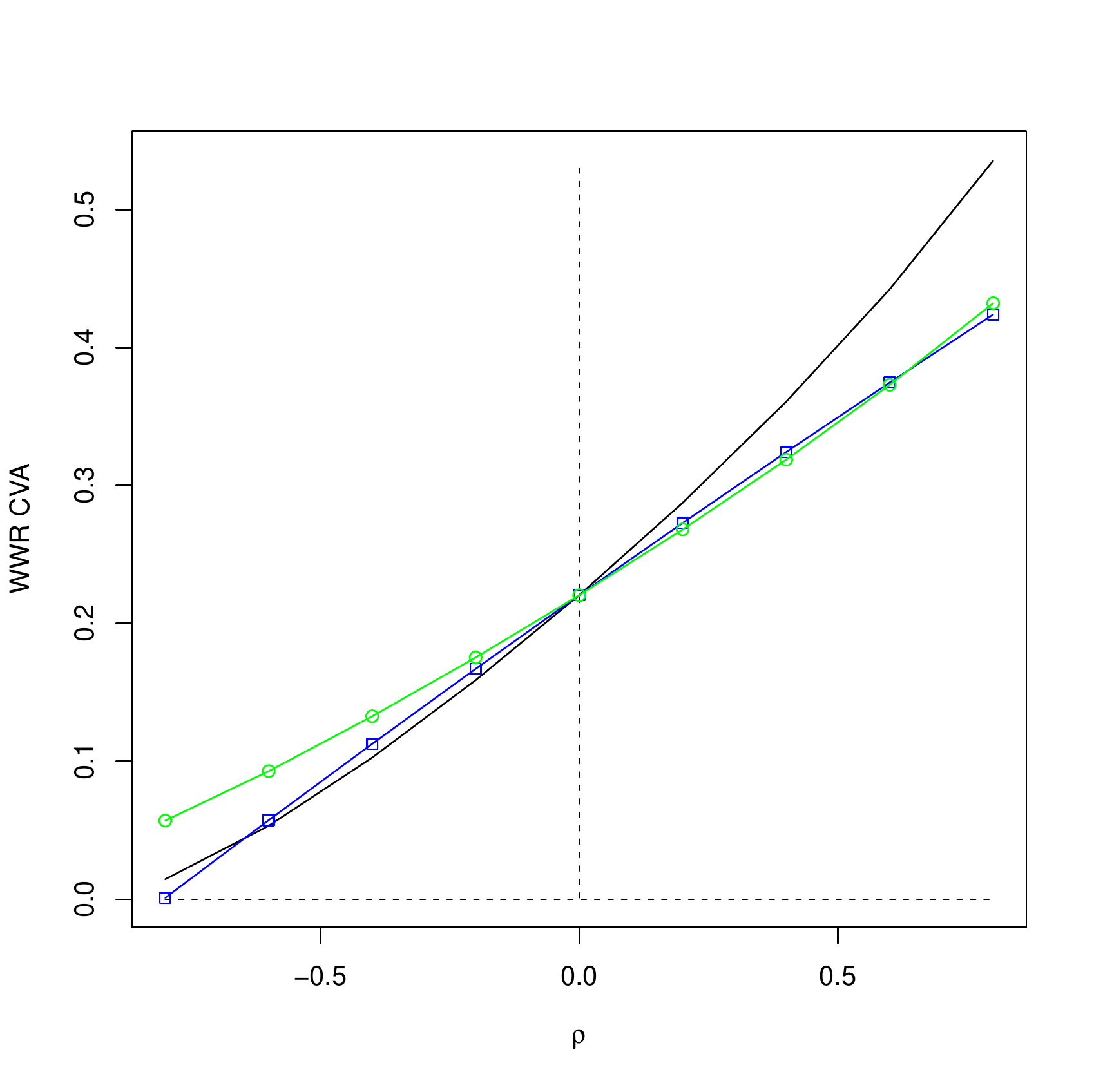}}\\
\subfigure[]{\includegraphics[width=0.2\columnwidth]{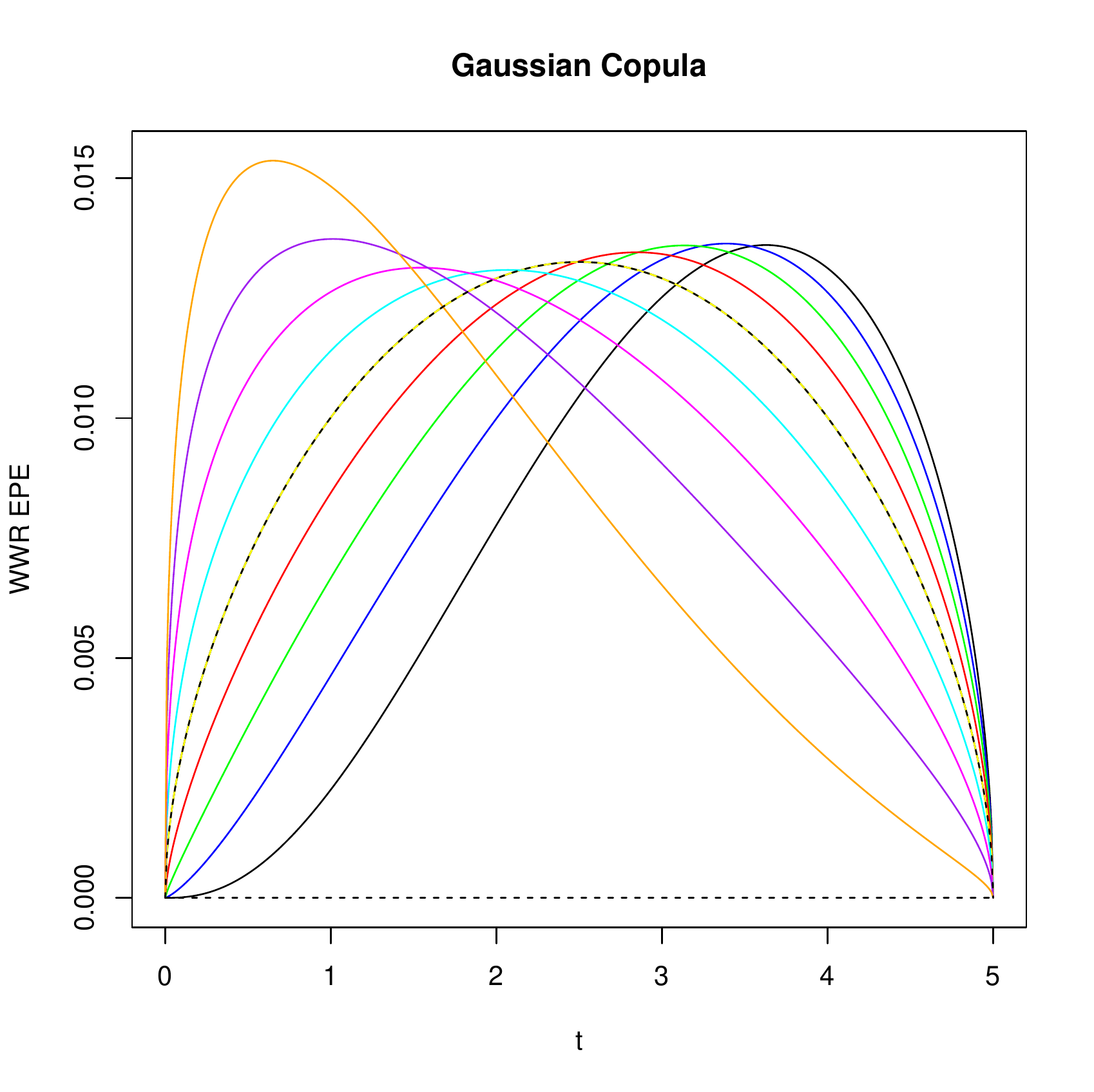}}
\subfigure[]{\includegraphics[width=0.2\columnwidth]{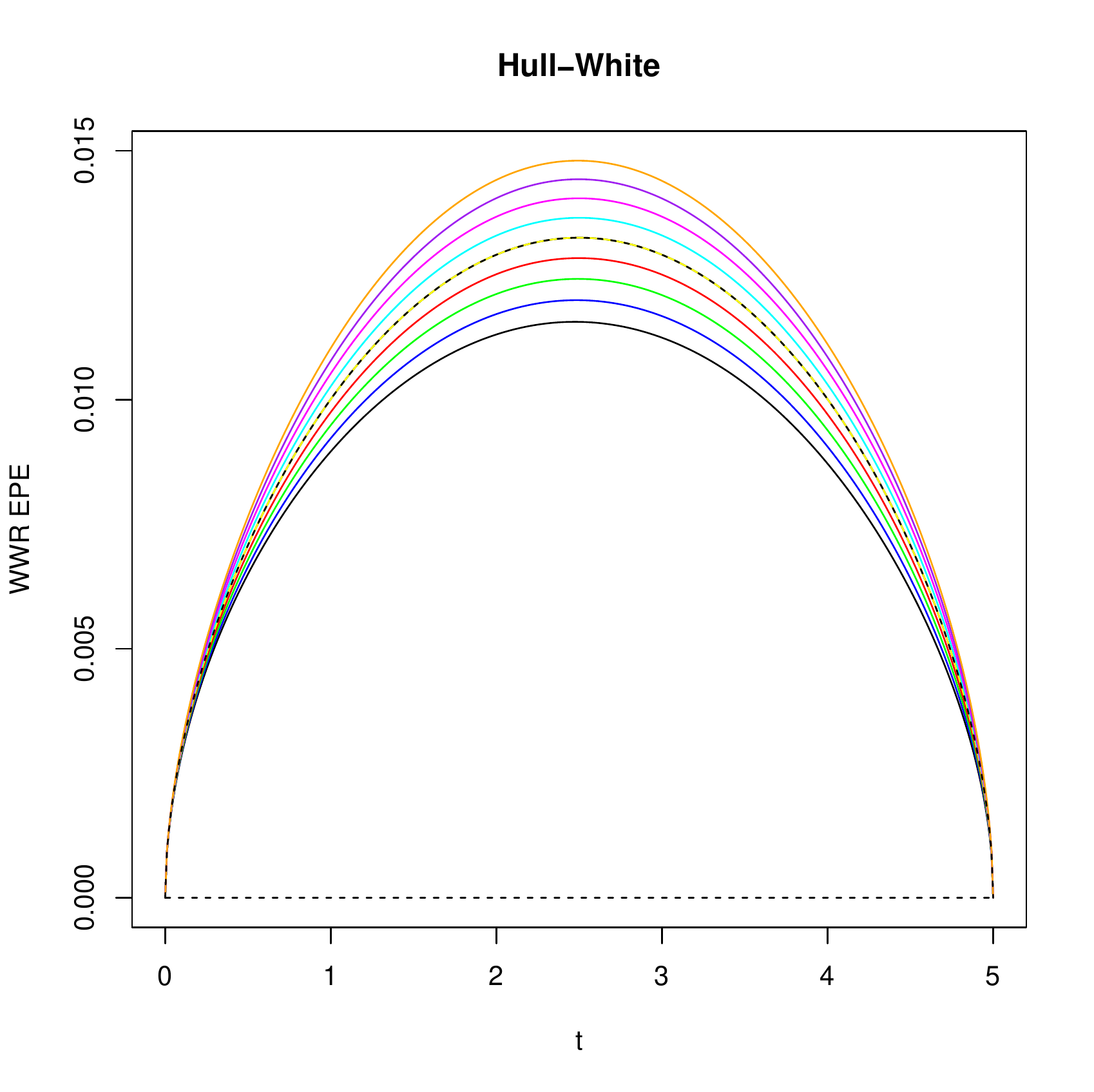}}
\subfigure[]{\includegraphics[width=0.2\columnwidth]{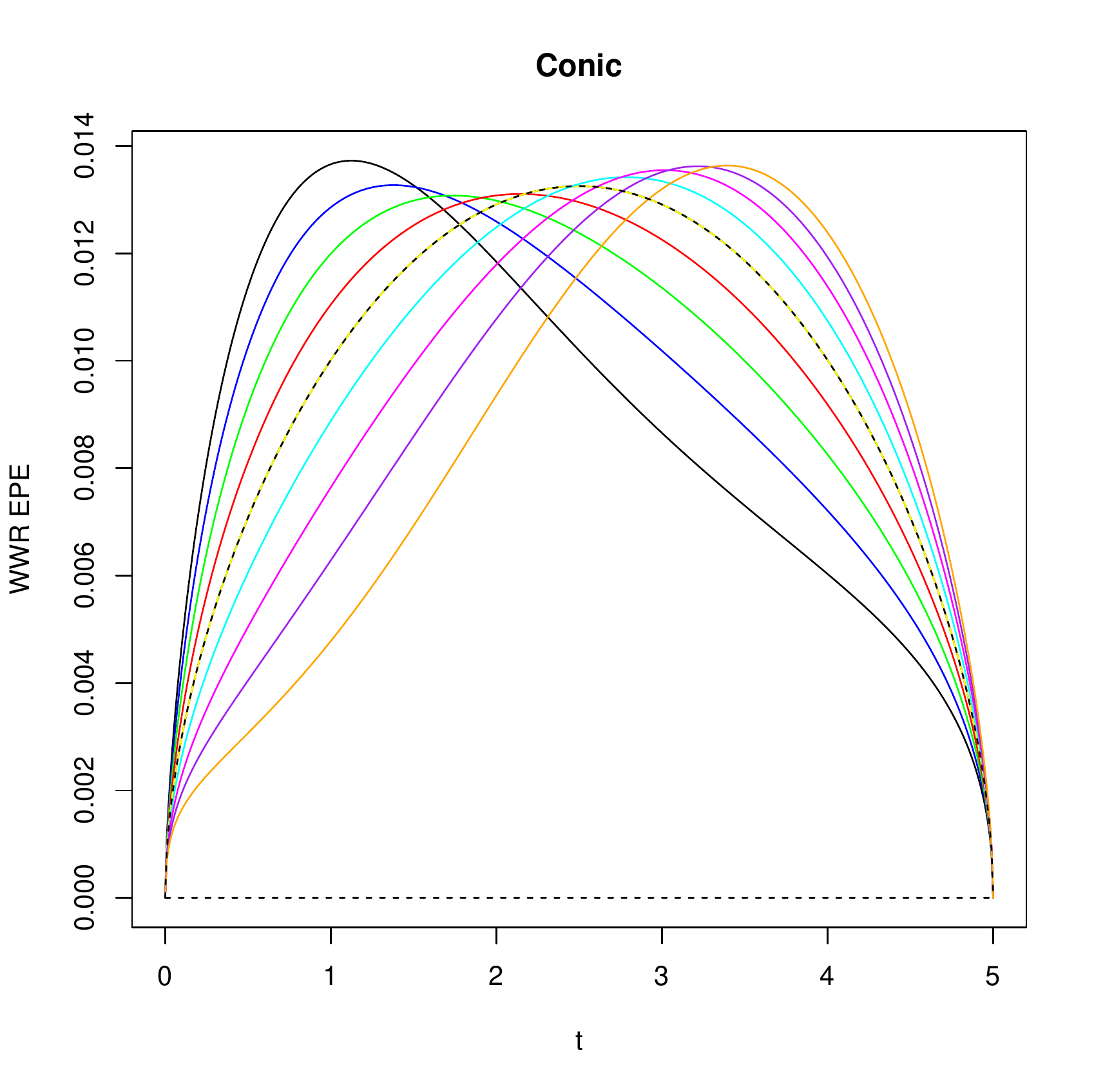}}
\subfigure[]{\includegraphics[width=0.2\columnwidth]{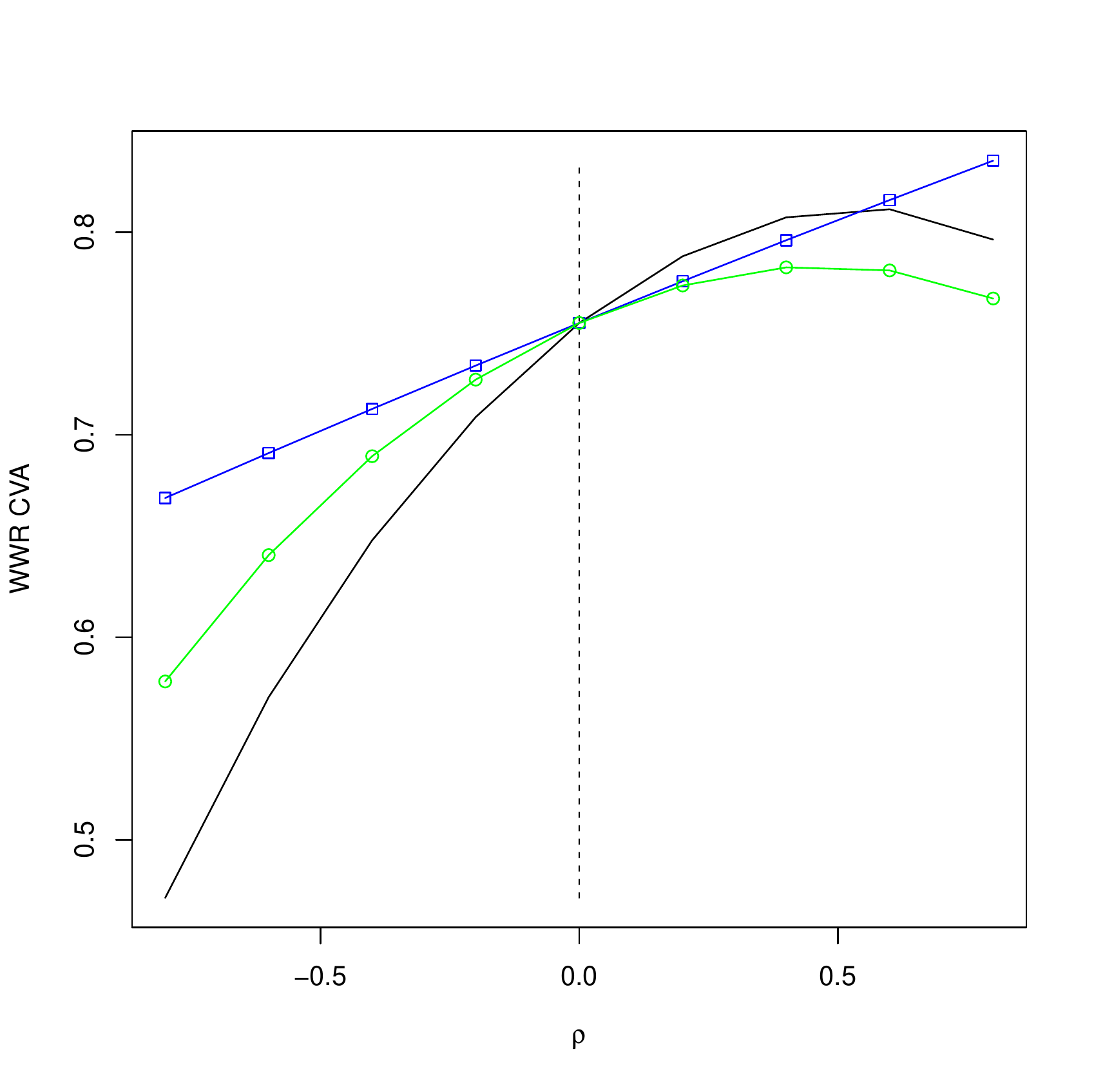}\label{fig:CVAIRS}}\\
\caption{WWR EPE IRS profiles for the Gaussian Copula (1st col.), Hull-White (2nd col.; $\sigma=3.15\%,\kappa=0.5\%$) and Conic Martingale (3rd col.; $\sigma=0.9$), as well as WWR CVA (last col.). Top row ($h=1\%$), middle row ($h=5\%$), bottom row ($h=30\%$). Each curve corresponds to a level of $\rho\in\{-0.8,-0.6,\ldots,0.6,0.8\}$ with colormap by (by ascending order) blue, green, red, yellow, cyan, magenta, purple, orange.}\label{fig:WWRFigIRS}
\end{figure}
\end{landscape}

The drawback of the HW approach (negative hazard rates) becomes specifically material in the EPE context for counterparty embedding little credit risk ($h$ small) as soon as decent volatility is plugged in the model. Indeed, as the intensity process is Normal, there is no skew in the model: the intensity process $\lambda$ can deviate in a quite significant way from the level $h(t)$. When $h(t)$ is small, this rapidly leads to both positive and negative values for the stochastic intensity. Consequently, negative $\zeta$ (proportional to $\lambda$) can be  observed and even the WWR EPE can become negative for negative $\rho$ at some time point $t$. This is clearly emphasized in panels (b) and (f) of Fig.~\ref{fig:WWRFigFX} and~\ref{fig:WWRFigIRS} for Forward and IRS profiles, respectively. Although ``negative EPEs'' are not observed in SSRD, the shift function can be negative, leading to negative values for $\lambda$ (we shall come back on this point in the next section). When credit risk increases, the intensity process moves around more positive values, rending the negative paths more rare. 

The HW and CM volatilities are quite large here. We decided to specify exogenously this parameter as (i) there is no liquid quotes for single-name CDS options and (ii) our main purpose here is to compare the capabilities of the models, justifying the analysis of limit behavior as well. For HW, the impact of $\rho$ on CVA seems to increase without bounds with $\sigma$ for large positive correlation. However, CVA becomes then negative for negative correlations. Therefore, for HW, we decided to set $\sigma$ to a value such that $CVA$ is 0 for $h=5\%$ and $\rho=-80\%$ for either profiles. For CM, CVA does not increase monotonously with $\sigma$: it first increases and then decrease; we have thus chosen the $\sigma$ allowing for maximum CVA at $\rho=80\%$.

For all three models, WWR EPE values generally increase pointwise with correlation $\rho$. This is always true for HW, and is valid as well as for GC and CM for counterparties with little credit risk (small $h$: see panels (a), (b), (c) and (e), (f), (g) of Fig.~\ref{fig:WWRFigFX} and \ref{fig:WWRFigIRS} for Forward and IRS profiles, respectively). This holds true for HW even for increasing $h$ but for the last two models however, the monotonicity of WWR EPE with respect to $\rho$ breaks down for the long-term part of WWR EPE for very risky counterparties. See panels (i), (j), (k) of the same figures. Again, it is worth noting that all these results are obtained in a analytical or semi-analytical way, there is no resampling, estimation or simulation error.

Another important remark is that compared to GC and CM, HW can only introduce small WWR effect on the short-end; WWR takes time to materialize. This is not a specificity of HW, but is a common feature of intensity-based approaches.

WWR EPE comparison revealed that CM and GC behave quite similarly in that they both allow for WWR impact on the  short-term profiles, and that in both cases, monotonicity of WWR EPE values with respect to $\rho$  may not hold for the long-term part of the profiles. A comparison in terms of CVA is also quite interesting. In particular, it allows to put in perspectives common thoughts. 

First, it is often thought that $\rho\to\pm 1$ provide upper and lower bounds on CVA in the WWR framework. Again, this is true here for HW (and in fact, seems valid for intensity models), but can fail to hold in general. In particular, CVA is not monotonic wrt $\rho$ for large $h$ when considering GC or CM models. This monotonicity is true only to some extend in terms of default rate: this behavior fails to hold for risky counterparties. We illustrate this using our Forward and IRS examples in panels (d), (h) and (l) of Fig.~\ref{fig:WWRFigFX} and~\ref{fig:WWRFigIRS}. As explained above, we have set the volatility parameter $\sigma$ of the HW model such that when setting $\rho=-80\%$, CVA is zero for $h=5\%$ (see panels (h)). 

Another common thought deserves to be clarified. It is often believed that intensity models typically fail to induce large correlation impact because dependency between exposure and default times is incorporated only via intensity of default events, but that conditional upon the intensity path, defaults are independent events. It appears in fact that in the context of CVA, the moderate ``correlation impact'' is in fact a \textit{covariance} effect, which is very low due to small instantaneous volatility of intensity processes. Increasing the later will enhance the ``correlation impact''. Consider for instance the Hull-White approach: any correlation impact level can be achieved by tweaking the instantaneous volatility. However, this is hard to achieve in a consistent way. The reason is that increasing $\sigma$ will increase the number of paths exceeding $1$ at some point. Same applies to the SSRD model: increasing the volatility will break the Feller condition $2\kappa\theta>\sigma^2$ which limits the levels of implied volatilities that can be achieved under positivity constraint. This problem can be partly circumvented by using jumps ($JCIR^{++}$), but the positivity constraint becomes more difficult to check. We refer to~\cite{Brigo06} and \cite{Brigo10} for a detailed discussion and to~\cite{Brigo13}[Sec. 3.3.6] for a summary.  

A last point worth mentioning is the CVA convexity implied by the model. Here again, HW exhibits no convexity: CVA increases quite linearly with $\rho$, and this is why negative CVA cannot be avoided for large negative correlation values and small $h$ (driving the intercept of the $\rho-CVA$ plot). Again, this is not the case of the other two models. Positive (negative) convexity for small (large) correlation value $\rho$ can be observed, provided that $h$ is small (large) enough.

\begin{figure}
\centering
\subfigure[$(h,35\%,0.12\%,2\%)$]{\includegraphics[width=0.8\columnwidth]{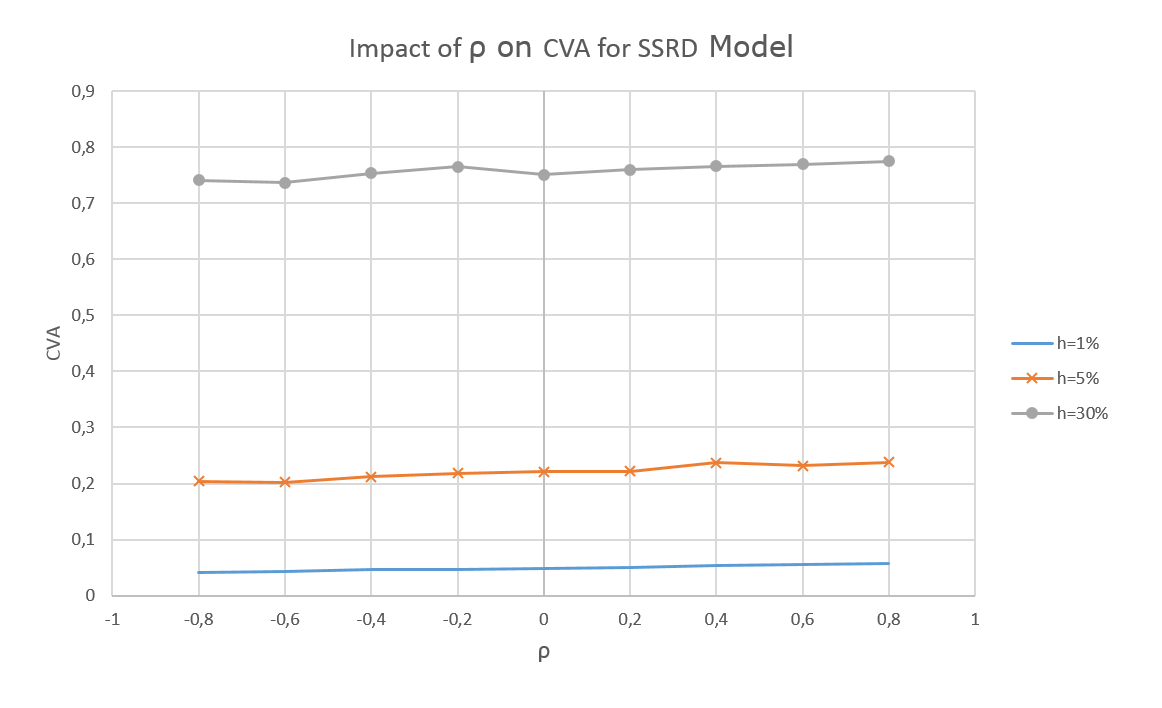}}\\
\subfigure[$(h,35\%,12\%,20\%)$]{\includegraphics[width=0.8\columnwidth]{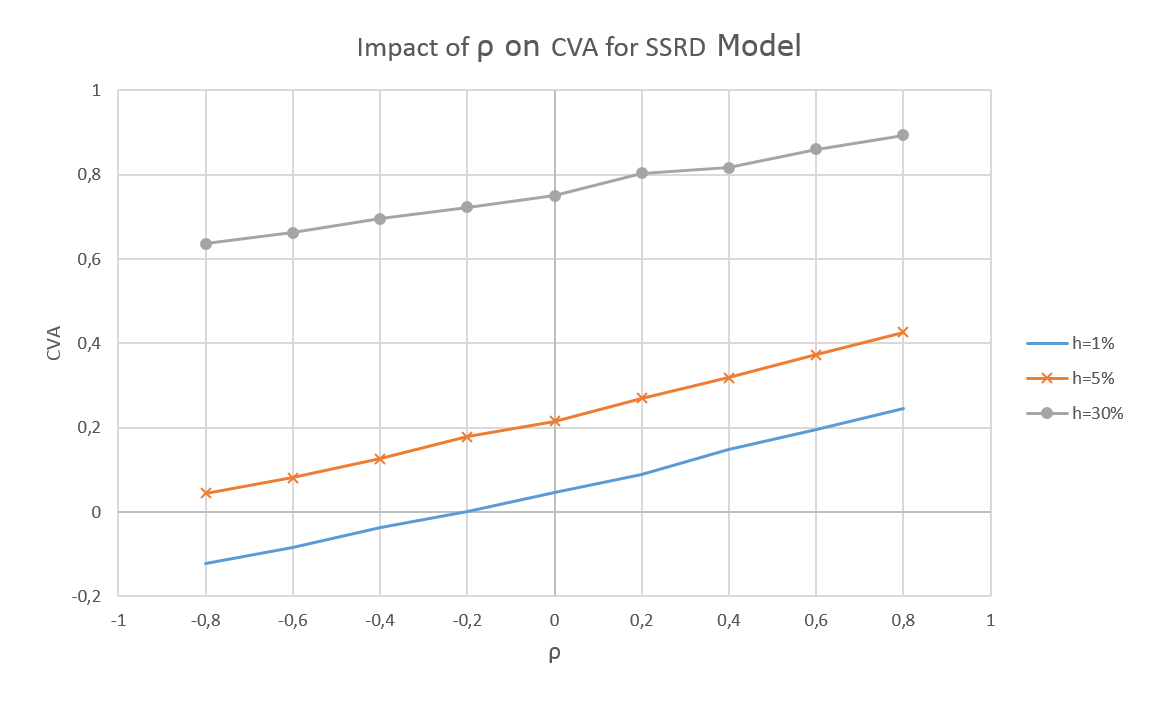}}
\caption{Impact of Brownian correlation $\rho$ on CVA on prototypical Payer swap $(\gamma,\vartheta)=(0.5\%,2.2\%)$ for SSRD model for standard and bumped values of $(r_0,\kappa,\theta,\sigma)$. The impact is pretty linear in both cases, and CVA can become negative as a result of the deterministic shift required for calibration purposes when counterparty features little credit risk (small $h$) and intensity has large volatility. These values have been computed based on 10,000 Monte Carlo simulations with time step $\Delta=0.01$.}\label{fig:CVASSRD}
\end{figure}

The similarity between HW and SSRD models in terms of CVA profiles (more specifically, the almost perfect linearity of CVA with respect to $\rho$) can be seen on Fig.~\ref{fig:CVASSRD}. Left panel uses values close to those in ~\cite{Brigo05} whilst in the right panel, more extreme values are used to emphasize the impact of larger volatilities (right panel). Just like Hull-White, the SSRD model does not really exhibit a skew, and relaxing the perfect fit assumption does not really help obtaining convexity. This can be understood as follows. Assume we drop the perfect calibration constraint so that we rely on a ``pure'' (i.e. unshifted) CIR model. We will then optimize the parameters such that $\E[S_t]\approx G(t)$ subject to the Feller constraint. The process $\lambda$ will thus remain strictly positive, ensuring $CVA(\rho)>0$ for all $\rho$. In such circumstances, this zero lower bound should imply a convexity in $\rho$-CVA$(\rho)$ profile when $CVA(0)$ is small (that is when $h$ is small)\footnote{If $CVA(0)$ is large (and thus $h$ is large), there is no reason to have convexity because the curve is shifted in the large positive values, so that the zero lower bound can be achieved without any convexity effect.}. However, one cannot observe such a con,vex curve. For a given exposure process, small $CVA(0)$ implies small $h$ and means a relatively small $\kappa\theta$. Therefore, only little values for $\sigma$ can be chosen in order to comply with the Feller condition. Hence, the impact of $\rho $ (proportional to $\sigma$) will be very limited and the profile $CVA(\rho)$ stays above zero, but is essentially flat (and thus rather linear). Using a shifted CIR does not fix the issue. Indeed, one can plug larger values for $\sigma$ (magnifying the impact of $\rho$ on CVA) and tweaking the other SRD parameters so that Feller condition is met, and then play with the deterministic shift to get the approximate (or perfect) calibration. Nonetheless, the later will then be typically negative, leading to potentially negative CVA. Summarizing, CVA$(\rho)$ profile remains quite linear in both intensity models. It seems there is no way to obtain large ratio CVA$(\rho)$/CVA($0)$ values for large $\rho$ while preventing CVA$(\rho)$ to be negative for quite negative $\rho$ when using intensities.

\begin{landscape}
\begin{figure}
\centering
\subfigure[Hull-White: $\sigma(\%)\in\{0.1,0.5,1,2,3,4,5\}$]{\includegraphics[width=0.495\columnwidth]{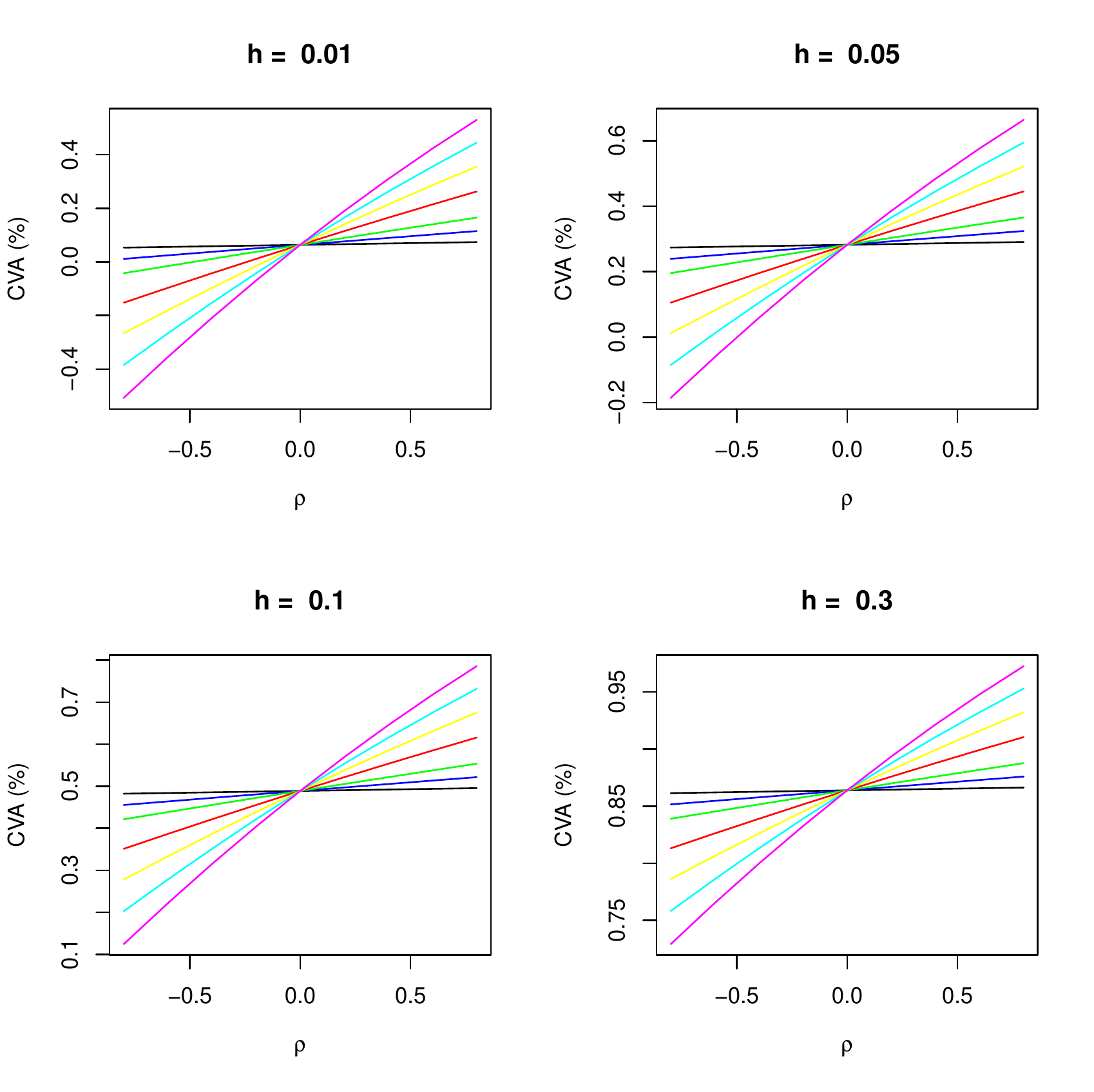}}
\subfigure[Conic Martingale: $\sigma(\%)\in\{0.5,1,5,10,40,100,200\}$]{\includegraphics[width=0.495\columnwidth]{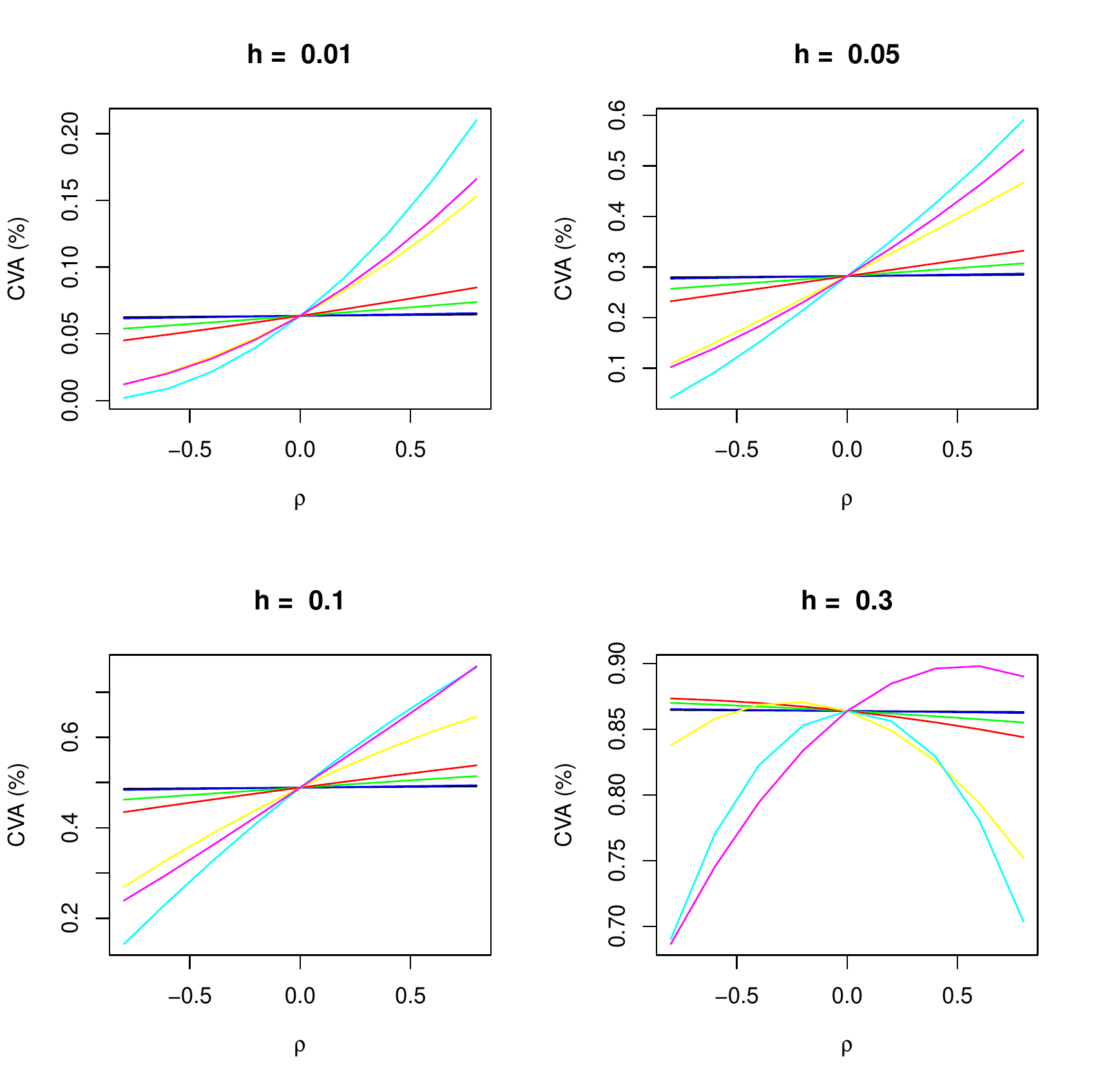}}
\caption{Impact of $\rho$ on the CVA (FW profiles) for increasing values of $\sigma$ (color map: black, blue, green, red, yellow, cyan, magenta). For HW, $\kappa=0.5\%$.}\label{fig:CVAImpactSigma_FW}
\end{figure}
\end{landscape}

\begin{landscape}
\begin{figure}
\centering
\subfigure[Hull-White: $\sigma(\%)\in\{0.1,0.5,1,2,3,4,5\}$]{\includegraphics[width=0.495\columnwidth]{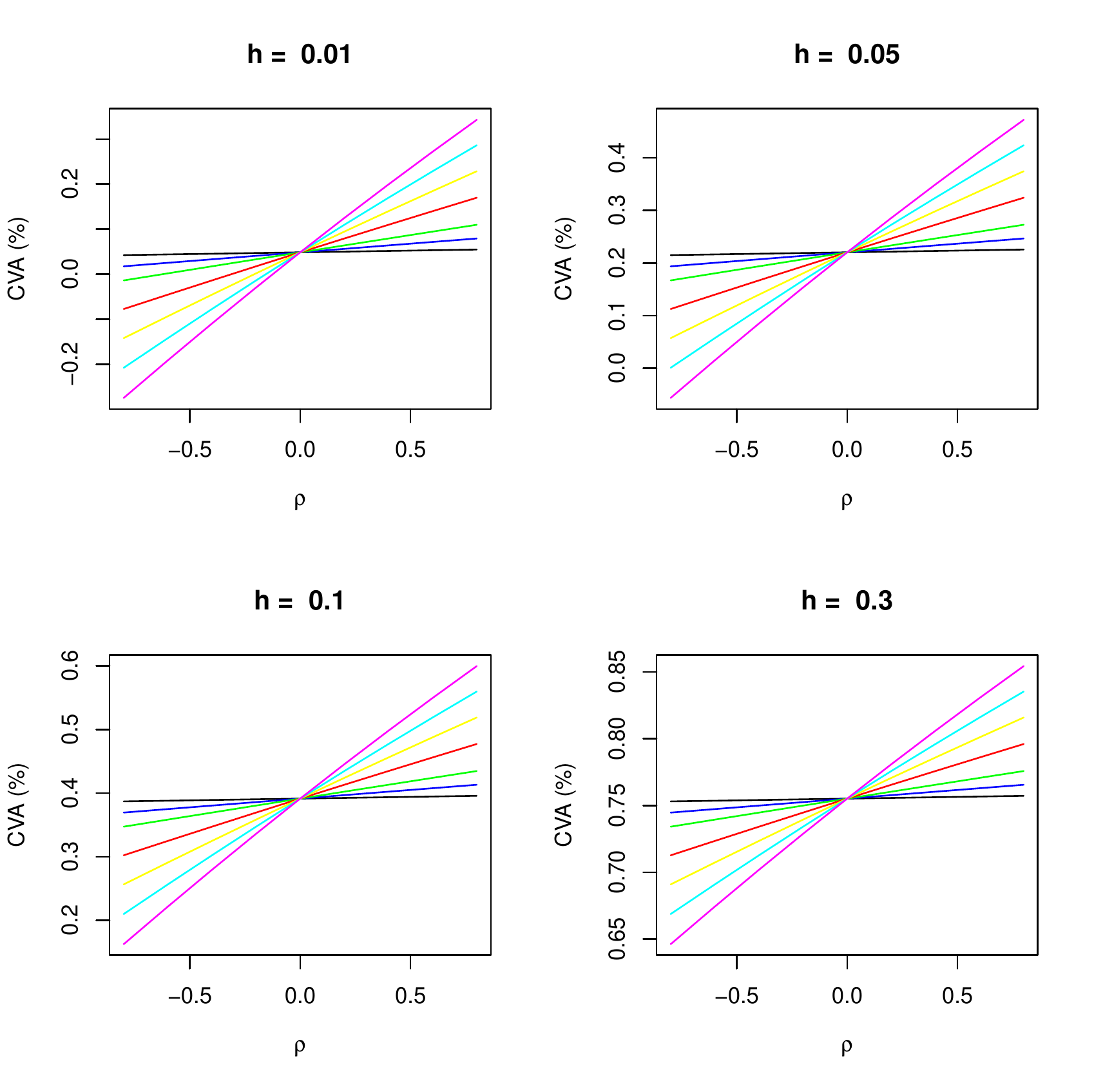}}
\subfigure[Conic Martingale: $\sigma(\%)\in\{0.5,1,5,10,40,100,200\}$]{\includegraphics[width=0.495\columnwidth]{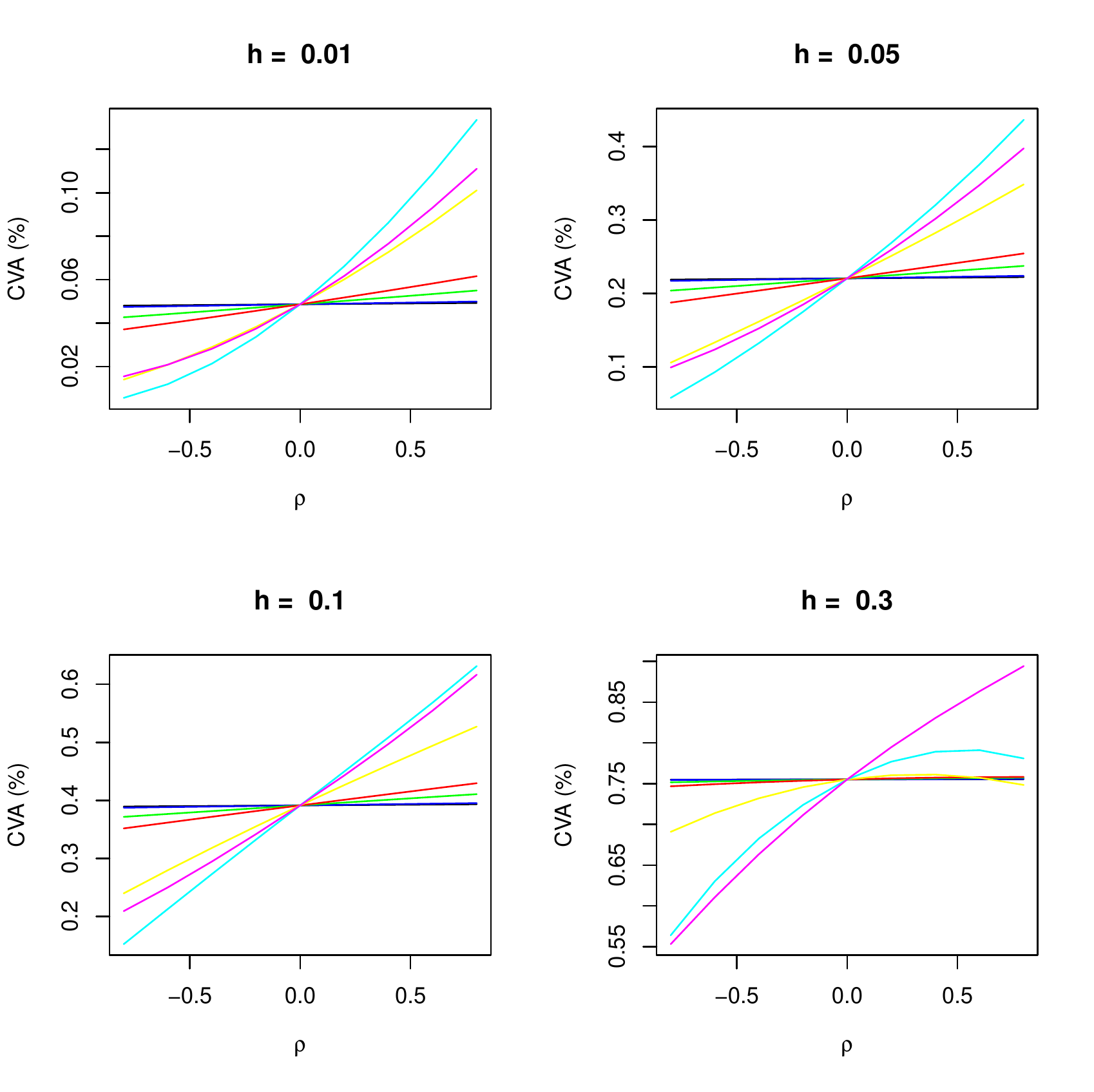}}
\caption{Impact of $\rho$ on the CVA (IRS profiles) for increasing values of $\sigma$ (color map: black, blue, green, red, yellow, cyan, magenta). For HW, $\kappa=0.5\%$.}\label{fig:CVAImpactSigma_IRS}
\end{figure}
\end{landscape}

We conclude this section by stressing the fact that correlation in the static approach GC represents in fact a set of terminal correlations, and embeds both instantaneous correlation and volatility effects featured in dynamic setups. This explains why large short-term correlation impact can be observed in static models. It appears from our results that large GC correlation $\rho$ corresponds to extremely large (that is to say, unrealistic) volatilities in dynamic setups. Indeed, as explained above, panels (d), (g) and (l) corresponds to results where dynamic volatilities are set at their (reasonable) ``maximum''. The impact of $\sigma$, and in particular the monotonic and convexity behaviors of $CVA(\rho)$ can be seen on Fig.~\ref{fig:CVAImpactSigma_FW} and~\ref{fig:CVAImpactSigma_IRS}.

\section{Discussion and wrong-way measures}
\label{sec:Discussion}\label{sec:WWRCoM}

A central point is whether these wrong-way risk models imply an equivalent measure $\tilde{\Q}$ under which one could simulate $V_t$ to incorporate the wrong-way risk effect. If so, CVA given by the integral of conditional $\Q$-expectations would become an integral of unconditional $\tilde{\Q}$-expectations:
\beqn
CVA&=&-\int_{0}^T\E[V^+_t|\tau=t]dG(t)\label{eq:CondExp}\\
&=&-\int_{0}^T\tilde{\E}[V^+_t]dG(t)\;.\label{eq:CoM}
\eeqn

If this proves to be the case, one could incorporate wrong-way risk by adopting the independent (no-WWR) setup, but adjusting the dynamics of price processes according to Girsanov's theorem. This would hopefully allow us to disentangle the credit and market variables, and lower the dimensionality of the problem. 

Such a measure $\tilde{\Q}$ would exist if there exists a positive $\Q$-martingale $\mathcal{L}$ with unit expectation such that
\beq
CVA=-\int_{0}^T\E\left[V^+_t\mathcal{L}_t\right]dG(t)\;.\label{eq:CVAzeta}
\eeq

Comparing eq.~(\ref{eq:CVAzeta}) with eq.~(\ref{eq:EdS}) we see that in the above dynamic models, $\zeta_t$ seems to play the role of $\mathcal{L}_t$. From eq.~(\ref{eq:CondExp}) and (\ref{eq:CVAzeta}), we have that $\E\left[V^+_t\zeta_t\right]=\E\left[V^+_t|\tau=t\right]$. 
The variable $\zeta_t$ controls the ratio of the conditional and unconditional probability densities. The condition in the former however depends on time in a way that is not ``merely'' depending on the information available at time $t$, i.e. not only through the filtration $\mathcal{F}_t$, but also through the ``$\tau=t$'' condition:
\beqn
\E\left[V^+_t\zeta_t\right]&=&\E[V^+_t\E[\zeta_t|V_t]]\\
&=&\int_{v=0}^\infty v \E[\zeta_t|V_t=v] f_{V_t}(v)dv\\
\E\left[V^+_t|\tau=t\right]&=&\int_{v=0}^\infty v f_{V_t|\tau}(v,t)dv
\eeqn
\beq
\E[\zeta_t|V_t=v]=\frac{f_{V_t|\tau}(v,t)}{f_{V_t}(v)}
\eeq
where $f_{V_t|\tau}(v,t)$ is the (conditional) density of $V_t$ given $\tau=t$ evaluated at $V_t=v$. 

Therefore, dynamic WWR models would imply a change of measure provided that $\zeta_t\geq 0$ $\Q$-a.s., $\E[\zeta_t]=1$ and $\E[\zeta_t|\mathcal{F}_s]=\zeta_s$. Clearly, the wrong-way process in both Cox and Conic Martingale setups satisfies the first two conditions. With regards to the positivity requirement, it is clear from eq.~(\ref{eq:zetalambda}) that $\zeta$ is non-negative provided that $\lambda>0$ which is the case in Cox setup (recall that this is not ensured when modeling $\lambda$ using standard interest rates processes). Similarly from eq.~(\ref{eq:ZetaConic}), same holds true in the $\Phi$-martingale case, where $\zeta$ is well-defined provided that $S\in [0,1]$, which is the case by construction.

The unit-expectation constraint is trivially met: this is eq.~(\ref{eq:calzeta}) which results from eq.~(\ref{eq:cal}). We can check that this is indeed the case in the considered models. In Cox setup for instance, we have
\beq
\E[\zeta_t]=\frac{\E[\lambda_tS_t]}{h(t)G(t)}=\frac{-\E[dS_t/dt]}{h(t)G(t)}=\frac{-d\E[S_t]/dt}{h(t)G(t)}=1\;.
\eeq

In the Conic Martingale case, the calibration equation~(\ref{eq:calCM}) leads to
\beqn
\E[\zeta_t]&=&\frac{\e^{\frac{\sigma^2}{2}t}}{\varphi(\Phi^{-1}(G(t)))}\E[\varphi(\Phi^{-1}(S_t))]\\
&=&\frac{\e^{\sigma^2/2t}}{\varphi(\Phi^{-1}(G(t)))}\E[\varphi(X_t)]\\
&=&\frac{\e^{\sigma^2/2t}}{\varphi(\Phi^{-1}(G(t)))}\E[\varphi(\Phi^{-1}(S_{0,t})\e^{\sigma^{2}/2t}+\sqrt{\e^{\sigma^{2}t}-1}Z)]\\
&=&\frac{\e^{\sigma^2/2t}}{\varphi(\Phi^{-1}(G(t)))}\frac{\varphi(\Phi^{-1}(S_{0,t}))}{\e^{\sigma^{2}/2t}}\\
&=&1\;.
\eeqn
Unfortunately however, $\zeta$ does generally not satisfy the martingale property of Radon-Nikodym derivative processes. This implies that $\zeta$ does not defines an equivalent measure $\tilde{\Q}$ but instead, defines a \textit{set} of measures. Indeed,  $\zeta$ is a unit expectation random variable which is strictly positive in Cox and CM approaches. Therefore, each ``term of the integral'' can be computed using a ``local'' change of measure~\cite{Shrev04}. The problem is that we need to determine the dynamics of the exposure process in each of these measures, so that this perspective is of little use in practice. 

We conclude this analysis with the comparison of the paths of $S$ and $\zeta$ for 4 dynamic approaches: Hull-White, SSRD, Gaussian martingale and $\Phi$-martingale. Simulation schemes for the first three approaches are well known in the literature. With regards to the last case, the exact scheme for simulating $S_{t_i}$ can be obtained by setting $S_0=1$ and noting that $S_{t}=\Phi(X_{t})$ where $X_{t}$ is a Vasicek process, which can also be sampled in an error-free way. 

As explained above, modeling stochastic intensities using short-rate processes possibly lead to negative paths for $\lambda$, that is, for $\zeta$. The only way to prevent this (especially for safe counterparties, that is with small hazard rate function $h(t)$) is to have a small volatility $\sigma$ which, may not be consistent with option quotes, but more importantly in this context, will cancel the effect of the instantaneous correlation $\rho$ between Brownian drivers $W$ and $B$ (no impact on terminal correlations). The Conic martingale approach, however, does not suffer from this drawback. It allows for fast and easy (analytical, and in fact automatic) calibration without facing the problem of generating probabilities out of [0,1]. In particular, a volatile wrong-way process $\zeta$ can be obtained in all cases.

\begin{figure}
\centering
\includegraphics[width=0.95\columnwidth]{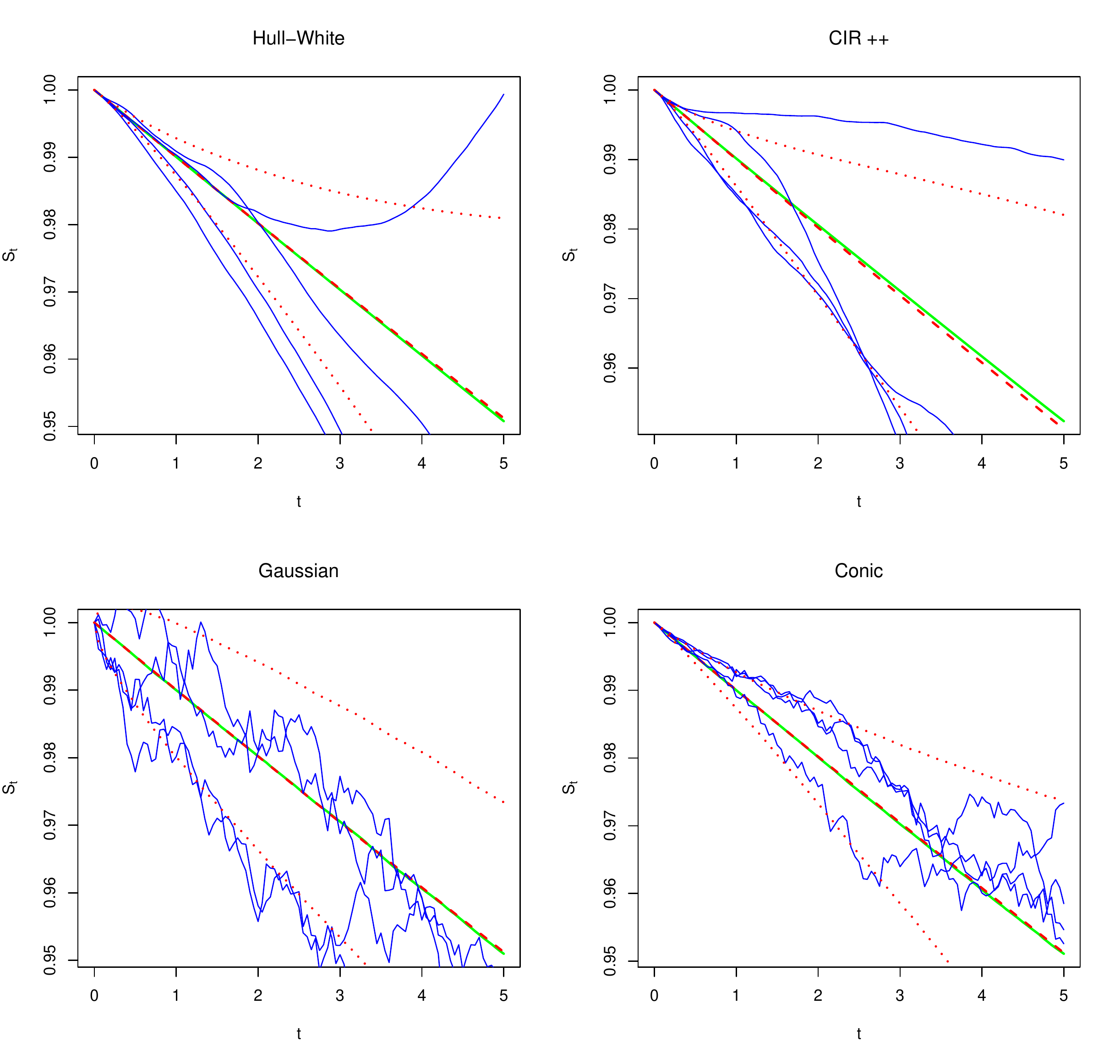}
\caption{Sample paths of the Az\'ema supermartingale $S_t$ for the four models. The corresponding parameters are (clockwise, in \%): $(\kappa,\theta,\sigma)=(0.5,1,1)$ (Hull-White), $(\kappa,\theta,\sigma)=(8,30,1.1)$ (SSRD or CIR$^{++}$), $\sigma=10$ (Conic) and $\sigma=1$ (Gaussian). We have used $h(t)=1\%$ and $T=5$.}\label{fig:StPaths}
\end{figure}

\begin{figure}
\centering
\includegraphics[width=0.95\columnwidth]{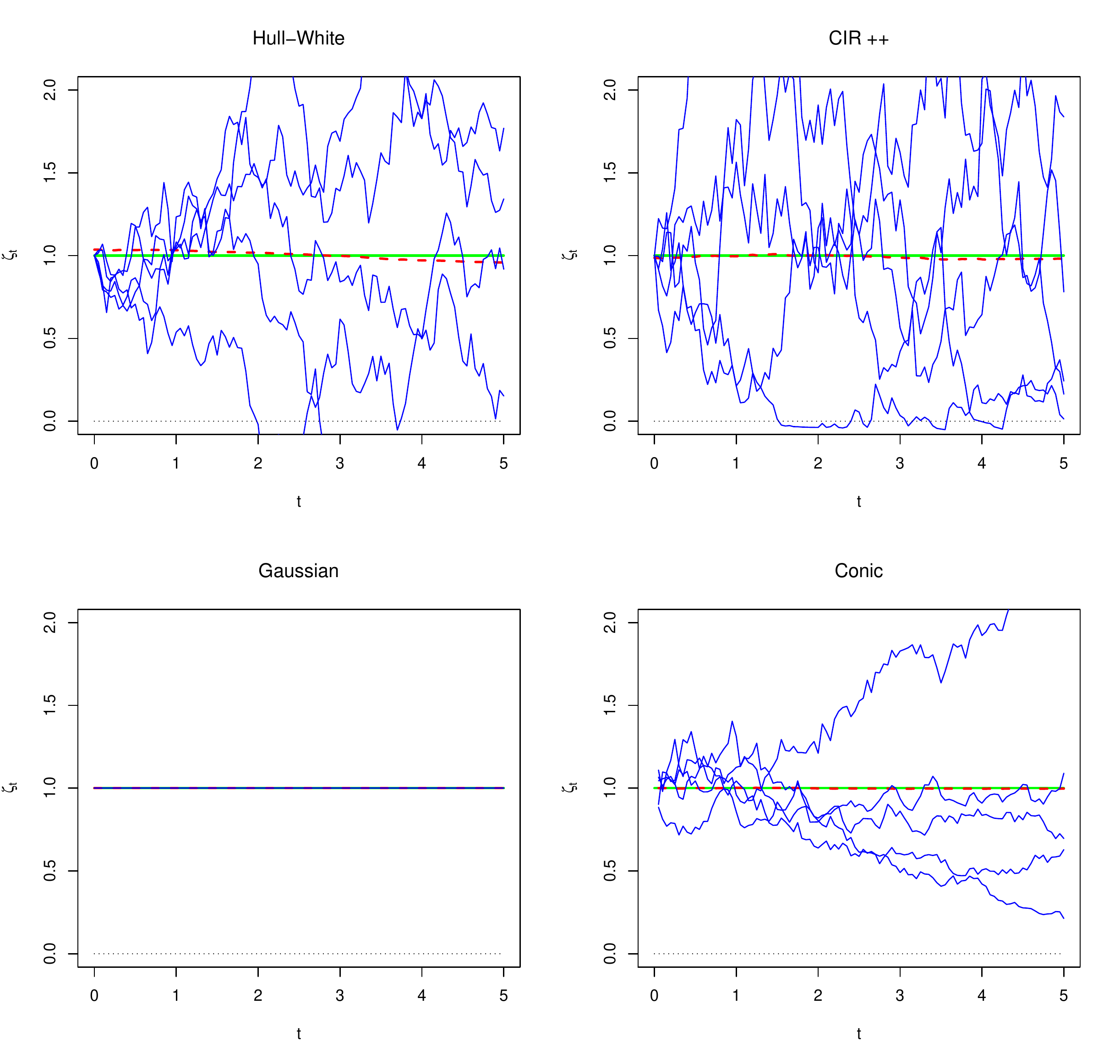}
\caption{Sample paths of the unit-expectation wrong-way process $\zeta_t$ (parameters are as in caption of Fig.~\ref{fig:StPaths}).}\label{fig:ZetatPaths}
\end{figure}

Figure~\ref{fig:StPaths} shows sample paths of the Az\'ema supermartingale for both intensity models (Hull-White and SSRD) as well as for the Gaussian and $\Phi$-martingale with constant diffusion coefficients. Among those, only the second and the fourth scheme exhibit sample paths in $[0,1]$. In theory, only the fourth one is ensured to share this property in all circumstances, as the shift function in the SSRD can become negative and there are in fact SSRD paths exceeding 1. This is clear from the wrong-way process paths drawn in Fig.~\ref{fig:ZetatPaths}, where negative values are visible (and only $\lambda$ can be responsible for that); the volatility of $\zeta$ decreases when going close to zero, but vanishes only when $r_t=0$ (and not when $\lambda_t=0$). The first two sets of paths are smoother as the intensity paths are smoothed through the integral operator, as opposed to the last two cases. All of them, however, lead to perfect calibration to the initial survival probability curve $G(t)$.

\section{Conclusion}
\label{sec:Conclusion}

In this paper, we have reviewed both static (copula-based resampling) and dynamic (stochastic intensity) approaches for modeling Wrong-Way Risk in the context of CVA. We have introduced another approach for dynamic modeling based on so-called \textit{Conic martingales}. We have made the link between WWR and change-of-measure techniques. It has been shown that in the static setup with Gaussian exposures, the conditional exposure at any point in time can indeed be computed by disregarding the default condition, but tweaking the distribution of the portfolio process. 

We have compared three WWR models by focusing on prototypical exposure paths depicting Forward and Swap profiles; analytical expressions for WWR EPE and semi-analytical expressions for CVA are obtained, throwing out any simulation issues in this exercise. Intensity models prove here to be quite specific. In particular WWR EPE profiles and CVA are monotonic with regards to the hybrid correlation $\rho$, and to some extend, WWR EPE profiles (and even CVA) can become negative. 

The conic martingale approach reveals interesting in that it ensures $S_t\in(0,1]$ for all parameter values (and thus guarantees a positive CVA), is sparse (one parameter, $\sigma$ which is a nice feature here as there is no quotes for CDS options), allows for automatic calibration to CDS quotes. Moreover it behaves quite similarly to the simple static approach. In particular, it exhibits short-term impact and convexity. Comparing dynamic and static approaches, one conclude that large correlation parameters of Gaussian Copula lead to CVA values implied by dynamic models with large correlation, but also extreme (probably unrealistic) volatility values.

In this paper, we restricted the study of stochastic intensity models to the most popular approaches resulting from the transposition of short rates models to credit: Hull-White and, to some extend, the shifted square-root diffusion. In particular, we did not cover the approaches that do not admit analytical calibration, like those associated to lognormal intensity models (\cite{Hull12},\cite{Willemen14}). This needs to be studied relying on well-chosen simulation and discretization schemes. Finally, it is worth stressing that potential arbitrage problems exist in all the considered approaches. This is rather clear for the artificial resampling technique. It is also the case for the standard intensity models because of the negative intensities, which violate Cox setup and may lead the Az\'ema supermartingale to be larger than 1. Depending on the assets traded on the considered market, arbitrage opportunities may also affect the Conic martingale in spite of the fact that $S_t$ belongs to $[0,1]$ almost surely. This is a delicate question linked to enlargement of filtrations and more precisely, linked to the \textit{immersion} property, also known as the $H$-hypothesis~(\cite{Biel11}\cite{Blan04}). This will be the topic of a specific research work. Although this may not be relevant in practice since CVA is computed at netting set level, an interesting theoretical question to answer is whether stochastic intensity models do indeed generates EPE profiles ordered pointwise with respect to correlation when only one risk factor is involved since this would imply a similar monotonic behavior for CVA as well. At this stage, this seems to be specific to intensity models.

\section{Appendix}
\label{sec:App}

In this section, we derive the analytical expression of the wrong-way risk expected positive exposures 
\beq
f(t)=\E[\zeta_tV_t^+]
\eeq
for both Hull-White stochastic intensity and Conic Martingale approaches.

In both cases, the below relationship obtained by completing the squares is helpful:
\beq
\varphi(a+bx)\varphi(c+dx)=\varphi\left(\frac{ab+cd}{\sqrt{b^2+d^2}}+x\sqrt{b^2+d^2}\right)\varphi\left(\frac{ad-bc}{\sqrt{b^2+d^2}}\right)
\eeq

\subsubsection{Hull-White}
\label{sec:App2:HW}

Let us consider the following three random variables $(\lambda,S,V)$ built from $(X,Y,Z)$, a vector of independent standard Normal variables:
\beqn
\lambda&\sim&A+B X\\
S&\sim&ke^{-(\alpha X+\beta Y+\gamma Z)}\\
V&\sim&a+\talpha X+\tbeta Y+\tgamma Z
\eeqn

We want to evaluate the below expression
\beqn
\E\left[\lambda SV^+\right]
&=&k\E\left[(A+BX)\e^{-(\alpha X+\beta Y+\gamma Z)}\left(a+\talpha X+\tbeta Y+\tgamma Z\right)^+\right]\\
&=&k\e^{(\alpha^2+\beta^2+\gamma^2)/2}\left(A I_1+B I_2\right)\label{eq:IntCM}\\
I_1&=&\int_{x,y} J(x,y)\varphi(x+\alpha)\varphi(y+\beta)dxdy\\
I_2&=&\int_{x,y} xJ(x,y)\varphi(x+\alpha)\varphi(y+\beta)dxdy\\
J(x,y)&=&\int (k(x,y)+\tgamma z)^+\varphi(z+\gamma)dz\\
&=&(k(x,y)-\gamma \tgamma)\Phi\left(\frac{k(x,y)-\gamma\tgamma}{|\tgamma|}\right)+|\tgamma|\varphi\left(\frac{k(x,y)-\gamma\tgamma}{|\tgamma|}\right)\\
k(x,y)&=&a+\talpha x+\tbeta y
\eeqn

To evaluate $I_1$ and $I_2$, set $v=(\mu,\sigma,\delta,a,b,c,d)$ and define
\beqn
s(v)&=&\sign(bd)\\
A(v)&=&\frac{\delta b-\mu a}{\sqrt{b^2+\mu^2}}\\
B(v)&=&\frac{\sigma b}{\sqrt{b^2+\mu^2}}\\
C(v)&=&A(v)d-B(v)c\\
\beta(v)&=&\sqrt{d^2+B(v)^2}\\
\alpha(v)&=&(A(v)B(v)+cd)/\beta(v)\\
\hat{A}(v)&=&s(v)(\delta bd-\mu ad-\sigma cb)/(d\sqrt{b^2+\mu^2})\\
\tilde{A}(v)&=&\mu^2+b^2+(\delta\mu+ab)^2\\
\tilde{B}(v)&=&2\sigma(ab\mu+\delta\mu^2)\\
\tilde{C}(v)&=&(\sigma\mu)^2\\
\varphi(x,k)&=&\varphi(x/k)/|k|\\
\varphi(v)&=&\varphi\left(C(v),\beta(v)\right)
\eeqn

Equipped with these notations, one can evaluate the 6 integrals below:
\beqn
I_A(v)&=&\int_{x,y}\varphi(\mu x+\sigma y +\delta)\varphi(a+b x)\varphi(c+dy)dxdy\\
&=&\varphi\left(v\right)/\sqrt{b^2+\mu^2}\\
I_B(v)&=&\int_{x,y}\Phi(\mu x+\sigma y +\delta)\varphi(a+b x)\varphi(c+dy)dxdy\\
&=&\frac{s(v)}{bd}\Phi\left(\frac{\hat{A}(v)d}{\beta(v)}\right)\\
I_C(v)&=&\int_{x,y}x\Phi(\mu x+\sigma y +\delta)\varphi(a+b x)\varphi(c+dy)dxdy\\
&=&\frac{\mu}{b^2}\left(\frac{\varphi\left(\hat{A}(v)d,\beta(v)\right)}{\sqrt{\mu^2+b^2}}-\frac{a}{b}I_B(v)\right)\\
I_D(v)&=&\int_{x,y}x\varphi(\mu x+\sigma y +\delta)\varphi(a+b x)\varphi(c+dy)dxdy\\
&=&\frac{-\varphi\left(v\right)}{b\sqrt{b^2+\mu^2}}\left(\frac{\mu A(v)}{\sqrt{\mu^2+b^2}}+a-\frac{\mu B(v)}{\sqrt{\mu^2+b^2}}\left(\frac{A(v)B(v)+cd}{\beta(v)^2}\right)\right)\\
I_E(v)&=&\int_{x,y}x^2\Phi(\mu x+\sigma y +\delta)\varphi(a+ x)\varphi(c+y)dxdy\\
&=&\frac{1}{b^3}\left((1+a^2)I_{E,1}(v)-(\frac{2a\mu}{\sqrt{b^2+\mu^2}}+\frac{A(v)\mu^2}{b^2+\mu^2})\varphi(v)-\frac{B(v)\mu^2}{b^2+\mu^2}I_{E,2}(v)\right)\\
I_F(v)&=&\int_{x,y}xy\Phi(\mu x+\sigma y +\delta)\varphi(a+ x)\varphi(c+y)dxdy\\
&=&\frac{\mu}{b^2\sqrt{b^2+\mu^2}}I_{F,1}(v)-\frac{a}{b^2}I_{F,2}(v)
\eeqn
where
\beqn
I_{E,1}(v)&=&\frac{s(v)}{d}\Phi\left(s(v)C(v)/\beta(v)\right)\\
I_{E,2}(v)&=&-\varphi\left(v\right)\frac{A(v)B(v)+cd}{\beta(v)^2}\\
I_{F,1}(v)&=&I_{E,2}(v)\\
I_{F,2}(v)&=&\frac{B(v)}{d^2}\varphi\left(v\right)-\frac{c}{d}I_{E,1}(v)
\eeqn

Let us define $v_1=\left(\frac{\talpha}{\tgamma},\frac{\tbeta}{\tgamma},\frac{a}{\tgamma}-\gamma,\alpha,1,\beta,1\right)$ and $v_2=\left(\frac{\tbeta}{\tgamma},\frac{\talpha}{\tgamma},\frac{a}{\tgamma}-\gamma,\beta,1,\alpha,1\right)$. 
Then,
\beqn
I_1&=&(a-\gamma\tgamma)I_B(v_1)+\talpha I_C(v_1)+\tbeta I_C(v_2)+|\tgamma| I_A(v_1)\label{eq:I1}\\
I_2&=&(a-\gamma\tgamma)I_C(v_1)+\talpha I_E(v_1)+\tbeta I_F(v_1)+|\tgamma| I_D(v_1)\label{eq:I2}
\eeqn

Finally, the computation of the time-$t$ expected positive exposure is given by 
\beq
f(t)=\E\left[\lambda_tS_tV_t^+\right]/(h(t)G(t))
\eeq
where the expectation can be evaluated by plugging $I_1$ and $I_2$ given in eq.~(\ref{eq:I1}) and~(\ref{eq:I2}) in~(\ref{eq:IntCM}) using $A=A(t)$, $B=B(t)$, $\alpha=\Omega(t)r_{21}(t)$, $\beta=\Omega(t)r_{22}(t)$, $\gamma=0$, $k=\e^{-\omega(t)}$, $a=a(t)$, $\talpha=b(t)r_{31}(t)$, $\tbeta=b(t)r_{32}(t)$, $\tgamma=b(t)r_{33}(t)$.

\subsubsection{Conic Martingale}
\label{sec:App2:CM}

The expected positive exposure $f(t)$ given in~(\ref{eq:IntConic}) takes the form
\beqn
f(t)&=&K(t)\int_{-\infty}^\infty I(x)\varphi\left(\frac{A(t)B(t)}{\sqrt{B(t)^2+1}}+\sqrt{B(t)^2+1}x\right)dx\\
K(t)&=&k(t)\varphi\left(\frac{A(t)}{\sqrt{B(t)^2+1}}\right)
\eeqn
Let us fix $t$ and set
\beqn
\alpha&=&\frac{A(t)B(t)}{\sqrt{B(t)^2+1}}\\
\beta&=&\sqrt{B(t)^2+1}\\
\talpha&=&\frac{ a(t)}{ b(t)\brho(t)}\\
\tbeta&=&\frac{\rho(t)}{\brho(t)}
\eeqn

With these notations, $f(t)=K(t)\int_{-\infty}^\infty I(x)\varphi(\alpha+\beta x)dx$ and $m(t,x)=\talpha+\tbeta x$
\beqn
f(t)&=&K(t)( a(t)I_1(x)+ b(t)\rho(t)I_2(x)+ b(t)\brho(t)I_3(x))\\
I_1(x)&=&\int_{-\infty}^\infty  \Phi(\talpha+\tbeta x)\varphi(\alpha+\beta x) dx\\
I_2(x)&=&\int_{-\infty}^\infty  x\Phi(\talpha+\tbeta x)\varphi(\alpha+\beta x) dx\\
I_3(x)&=&\int_{-\infty}^\infty  \varphi(\talpha+\tbeta x)\varphi(\alpha+\beta x) dx
\eeqn

Setting
\beqn
\mu&=&\talpha-\alpha\sigma\\
\sigma&=&\tbeta/\beta\\
\frac{\mu}{\sqrt{1+\sigma^2}}&=&\frac{\beta\talpha-\alpha\tbeta}{\sqrt{\beta^2+\tbeta^2}}
\eeqn
a simple change of variable $y=\alpha+\beta x$ yields
\beqn
I_1(x)&=&\frac{1}{\beta}\int \Phi\left(\mu+y\sigma\right)\varphi(y)dy\\
&=&\frac{1}{\beta}\Phi\left(\frac{\mu}{\sqrt{1+\sigma^2}}\right)\\
I_2(x)&=&\frac{\sigma}{\beta^2\sqrt{1+\sigma^2}}\varphi\left(\frac{\mu}{\sqrt{1+\sigma^2}}\right)-\frac{\alpha}{\beta^2}\Phi\left(\frac{\mu}{\sqrt{1+\sigma^2}}\right)\\
I_3(x)&=&\varphi\left(\frac{\mu}{\sqrt{1+\sigma^2}}\right)\int \varphi\left(\frac{\alpha\tbeta+\beta\talpha}{\sqrt{\beta^2+\tbeta^2}}+x\sqrt{\beta^2+\tbeta^2}\right)dx\\
&=&\varphi\left(\frac{\mu}{\sqrt{1+\sigma^2}}\right)/\sqrt{\beta^2+\tbeta^2}
\eeqn

Finally, with
\beqn
\beta&=&\sqrt{B(t)^2+1}\\
\frac{\mu}{\sqrt{1+\sigma^2}}&=&\frac{\frac{ a(t)}{ b(t)}(B(t)^2+1)-\rho(t) A(t)B(t)}{\sqrt{B(t)^2+1}\sqrt{1-\rho(t)^2B(t)^2+B(t)^2}}
\eeqn
the expression of the expected positive exposure in the CM model is given by
\beq
f(t)=K(t)\left(\left(\frac{\beta a(t)-\alpha b(t)\rho(t)}{\beta^2}\right)\Phi\left(\frac{\mu}{\sqrt{1+\sigma^2}}\right)+\frac{ b(t)}{\sqrt{\beta^2+\tbeta^2}}\left(\frac{\tbeta\rho(t)}{\beta^2}+\brho(t)\right)\varphi\left(\frac{\mu}{\sqrt{1+\sigma^2}}\right)\right)
\eeq

It is easy to check that this expression agrees with the no-WWR EPE when $\rho=0$, $\tbeta=0$. In this case indeed, and one gets
\beqn
f(t)&=&\frac{K(t)}{\sqrt{B(t)^2+1}}\left(\left( a(t)\right)\Phi\left(\frac{ a(t)}{ b(t)}\right)+ b(t)\varphi\left(\frac{ a(t)}{ b(t)}\right)\right)
\eeqn
where $K(t)/\sqrt{B(t)^2+1}\equiv 1$:
\beqn
\frac{K(t)}{\sqrt{B(t)^2+1}}&=&\frac{k(t)}{\sqrt{B(t)^2+1}}\varphi\left(\frac{A(t)}{\sqrt{B(t)^2+1}}\right)\\
&=&\frac{\e^{\sigma^2/2 t}/\varphi(\Phi^{-1}(G(t))}{\sqrt{B(t)^2+1}}\varphi\left(\frac{A(t)}{\sqrt{B(t)^2+1}}\right)\\
&=&\e^{\sigma^2/2 t}/\varphi(\Phi^{-1}(G(t))\int x \varphi\left(A(t)+B(t)x\right)dx\\
&=&\e^{\sigma^2/2 t}/\varphi(\Phi^{-1}(G(t))\E\left[\varphi(X_t)\right]\\
&=&1
\eeqn
%

    \bibliography{\MyBib}
  \bibliographystyle{unsrtnat}

\end{document}